\documentclass[showpacs,twocolumn,pre,amssymb,amsmath,nobibnotes,aps]{revtex4-1}
\usepackage{bm}

\usepackage{color}

\newcommand{\nh}{\hat{n}}
\newcommand{\xv}{{\bf x}}
\newcommand{\rv}{{\bf r}}
\newcommand{\qv}{{\bf q}}

\newcommand{\bfq}{\mathbf{q}}
\newcommand{\be}{\begin{equation}}
\newcommand{\ee}{\end{equation}}
\newcommand{\bea}{\begin{eqnarray}}
\newcommand{\eea}{\end{eqnarray}}

\newcommand{\half}{\frac{1}{2}}

\newcommand{\gv}{{\bf g}}

\def\rf#1{(\ref{#1})}

\usepackage[dvips]{graphicx}
\begin{document}

\title{Smectic order, pinning, and phase transition in a smectic liquid crystal cell with a random substrate}

\author{Quan Zhang}
\author{Leo Radzihovsky}
\affiliation{Department of Physics, University of Colorado,
   Boulder, CO 80309, USA}
\date{\today}

\begin{abstract}
We study smectic-liquid-crystal order in a cell with a heterogeneous substrate imposing surface random positional and orientational
pinnings. Proposing a minimal random elastic model, we demonstrate that, for a thick cell, the smectic state without a rubbed substrate is
always unstable at long scales and, for weak random pinning, is replaced by a smectic glass state. We compute the statistics of the
associated substrate-driven distortions and the characteristic smectic domain size on the heterogeneous substrate and in the bulk. We find
that for weak disorder, the system exhibits a three-dimensional temperature-controlled phase transition between a weakly and strongly pinned
smectic glass states akin to the Cardy-Ostlund phase transition. We explore experimental implications of the predicted phenomenology and suggest
that it provides a plausible explanation for the experimental observations on polarized light microscopy and x-ray scattering.

\end{abstract}

\maketitle

\section{Introduction}
\label{sec:introduction}

\subsection{Motivation and background}
\label{sec:motivation}

Research on ordered condensed matter systems subject to bulk random heterogeneities
has been an active field.
Considerable progress has been made, providing a better understanding of real materials,
where quench disorder is always present \cite{FisherPhysicsToday,reviewRandom_MPV,ChargeDensityWave,disorderSC,
randomHelium,RTaerogelPRL,RTaerogelPRB,BelliniScience}.
Recently, attention has turned to
systems where the heterogeneity is confined to a surface, e.g., nematic
\cite{FeldmanVinokurPRL,usFRGPRL,usFRGPRE} and smectic-liquid-crystal cells \cite{usSmecticEPL} with dirty substrates. These surface disordered systems
are of considerable interest and exhibit phenomenology qualitatively distinct
from their bulk disordered counterparts.

The commonly observed Schlieren texture \cite{Schlieren_texture} is a
manifestation of such surface pinning in nematic cells. Recent studies
also include photo alignment and dynamics in self-assembled liquid
crystalline monolayers \cite{LeeLinkClarkSAMs,LeeClarkSAMs}, as well as memory effects and
multistability in the alignment of nematic cells with a heterogeneous
random anchoring substrate \cite{Aryasova2004}. The existence of the
corresponding phenomena in smectic liquid crystals has been recently
revealed in ferroelectric smectic-$C$ cells in a book shelf
geometry \cite{ClarkSmC,CDJonesThesis}. This latter system exhibits long-scale
smectic layer distortions, driven by collective
random surface pinning, and awaits a detailed theoretical description.

\begin{figure}[htbp]
\centering
\includegraphics[height=8 cm]{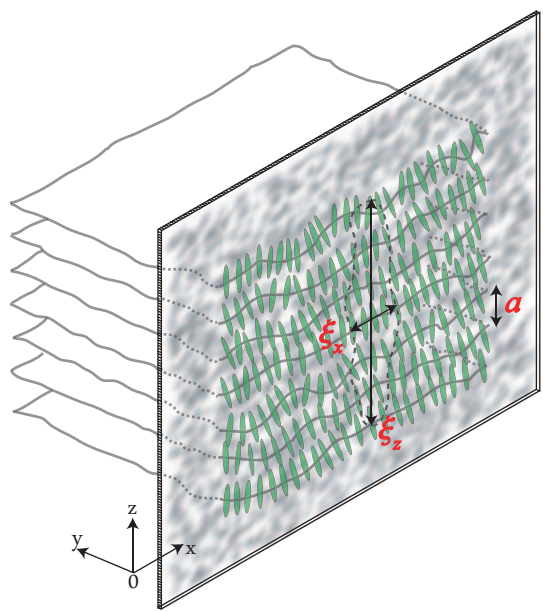}\\
\caption{(Color online) A schematic of a half-infinite smectic-liquid-crystal cell in the presence of
surface random pinning, with the region enclosed by dashed line indicating a domain of area $\xi_x\times\xi_z$ ($\xi_z\sim\xi_x^2/a$)
within which the perturbative treatment of surface disorder is valid. The smectic
long-range order is destroyed by an arbitrarily weak surface disorder beyond the domain size.
}
\label{fig:smecticCartoon}
\end{figure}

A schematic of such a smectic-liquid-crystal cell is illustrated in Fig.~\ref{fig:smecticCartoon},
with a ``dirty'' front substrate imposing two types of surface disorders:
surface random orientational pinning of local nematic order and surface random
positional pinning of smectic layers.
Generically, such surface pinning leads to elastic and plastic smectic disordering, latter
characterized by proliferation of topological defects, e.g., dislocations.
In this paper, we focus on the simpler limit of weak disorder, where the topological
defects are either not present at long scales or are
sufficiently dilute, whereby they can be neglected for a range of experimentally relevant length
scales.  Clearly this leaves a rich and challenging regime of strong pinning to future studies.

\subsection{Summary of the results}
\label{sec:summary_introduction}
We explore surface pinned smectic cell, focusing on the bookshelf geometry [with the coordinate system shown in Fig.~\ref{fig:smecticCartoon}, where $y$ is perpendicular to the substrate and points into the cell, $x$ (along smectic layer) and $z$ (along smectic layer normal) are parallel to the substrate]. As we will show in Sec.~\ref{sec:parallel_cell_smectic} in the homeotropic geometry, the effects of random substrate reduce to that of a random-field $xy$ model in $(d-1)$ dimensions, and are significantly less strong than in this more common bookshelf geometry.

Using a harmonic elastic description of smectic liquid crystal order with surface quenched disorder, which
characterizes a thick smectic cell subject to surface heterogeneities,
we studied the distortion of the smectic layers on the surface, characterized by Larkin (Imry-Ma)-like \cite{Larkin,ImryMa} length scales
given by $\xi_x$, $\xi_z\sim\xi_x^2/\lambda$ that relates the
distortion along the layer normal
($z$) and parallel to the layer ($x$) from either random orientational or random positional pinning, where
$\lambda=\sqrt{K/B}$ is the standard smectic-liquid-crystal length,
typically comparable to the layer
thickness $a$ \cite{deGennes}.
In the regime of dominant surface \textit{orientational} pinning we find (with $\Delta_f$ the associated mean-squared pinning strength)
\begin{equation}
\xi_x^f=
c\frac{B^2\lambda^3a^2}{\Delta_f}\approx \sqrt{\lambda\xi_z^f}.
\label{LLrandomTilt}
\end{equation}
For the dominant surface \textit{positional} disorder (of mean-squared strength $\Delta_v$) we instead find
\begin{equation}
\xi_x^v=
\left(3c\frac{B^2\lambda^3a^2}{\Delta_v}\right)^{1/3}\approx \sqrt{\lambda\xi_z^v},
\label{LLrandomPosition}
\end{equation}
where $c=\frac{4\pi^2}{\pi-2}\approx 34.6$. In the presence of both types of surface disorders,
the domain size is determined by a combination of these two lengths as discussed later in Sec.~\ref{sec:larkinLength_smectic}.

Adapting the standard Imry-Ma-Larkin analysis to the problem at hand, we find
that long-range smectic order on the heterogeneous substrate is
unstable against surface random orientational pinning in a $d$-dimensional
system with $d<d^f_{lc}$, where
\begin{equation}
d^f_{lc}=4,
\end{equation}
and for the surface random positional pinning in
a $d$-dimensional system with $d<d^v_{lc}$, where
\begin{equation}
d^v_{lc}=6.
\end{equation}
Thus, on sufficiently long scales, long-range smectic order in a three-dimensional cell is unstable to arbitrarily weak random pinning of statistically isotropic (nonrubbed) substrate.
Below these lower critical dimensions
arbitrarily weak pinning destabilizes long-range smectic order.
Within the resulting finite smectic domains
the correlations of smectic layer distortions grow as a power law of in-plane separation and decay exponentially into the bulk of the cell, with the characteristic length at depth $y > \xi_x$ set by $y$ itself.

We employed the functional renormalization group (FRG) \cite{DSFisherFRG,GiamarchiLedoussalFluxLattice_PRB,GiamarchiLedoussalFluxLattice_PRL,LedoussalWiese,BalentsFisher} to assess the effectively strong
nonlinear pinning physics on scales beyond the smectic domain size.
The most interesting and potentially experimentally relevant prediction of this analysis
is a three-dimensional (3D) Cardy-Ostlund-like (CO) \cite{CardyOstlund} phase transition at
a temperature $T_g$ from a weakly disordered smectic for $T > T_g$ (where
on long scales the surface positional pinning is averaged away by thermal fluctuations)
to a low-temperature disorder-dominated smectic-glass for $T < T_g$. In the latter phase,
the positional disorder flows to a CO fixed line at which the correlations of layer
distortions asymptotically are characterized by the orientational pinning alone, but
with an effective strength additively enhanced with $(T_g - T)^2$.
The high- and low-temperature phases are distinguished by (among other
features) the effective temperature-dependent orientational pinning strength
\begin{equation}
\delta_f(T)\equiv\Delta_f(T)/\Delta_f,
\label{delta_f_T_definition}
\end{equation}
given by
\begin{equation}
\delta_f(T)=\left\{\begin{array}{ll}
1,&\mbox{ for }T>T_g,\\
1+
\frac{9}{8\pi^2}\frac{\xi^f_x}{a}\left(1-\frac{T}{T_g}\right)^2,
& \mbox{ for } T\leq T_g.
\end{array}\right.
\label{delta_f_T}
\end{equation}
The long-scale smectic layer correlations on the substrate (at $y=0$) are
characterized by $C(x,z)=\overline{\langle [u_0(x,z)-u_0(0,0)]^2 \rangle}$, the full behavior of
which is given in Eq.~(\ref{Cxz_long_full}) with limits:
\begin{equation}
C(x,z)\approx  a^2\left\{\begin{array}{ll}
\pi x/\xi_x (T),& \mbox{ for } x\gg\sqrt{\lambda z},\\
\sqrt{2\pi}\sqrt{z/\xi_z(T)},& \mbox{ for } x\ll\sqrt{\lambda z},
\end{array}\right.
\label{large_scale_correlation}
\end{equation}
with the temperature dependent sizes
\begin{eqnarray}
\xi_x(T)&=&\xi_x^f/\delta_f(T),\\
\xi_z(T)&=&\xi_z^f/\delta_f^2(T).
\end{eqnarray}
These predictions remain valid only on scales shorter than the
distance between unbound dislocations and the scale $\xi^{NL}_x$ ($\gg \xi^f_x$ for weak pinning), beyond which the
nonlinear elasticity may become important.

\begin{figure}[tbhp]
\centering
\includegraphics[width=9 cm]{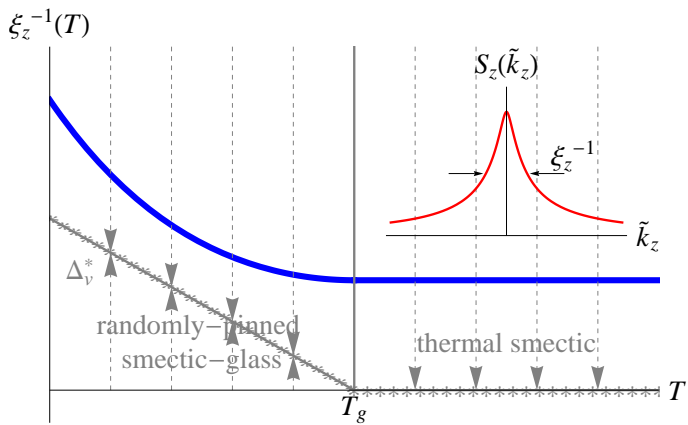}\\
\caption{(Color online) X-ray scattering peak width (thick smooth curve, as indicated by the top right inset) as a function of temperature. The peak broadens
below CO transition temperature $T_g$. The phase transition is indicated by the flow of random positional disorder strength to a nonzero fixed line value $\Delta_v^*(T)$ for $T<T_g$.}
\label{fig:Peakwidth}
\end{figure}

We also study various experimental features of our predictions for the polarized light microscopy.
We suggest that the highly anisotropic domain (as shown in Fig.~\ref{fig:smecticCartoon})
induced by surface disorder is a plausible
explanation for the commonly observed stripes in thin
smectic cells~\cite{CDJonesThesis}.
We further argue that the 3D
smectic glass transition predicted above (illustrated in Fig.~\ref{fig:DeltavFlowmap3D_2})
may have already been observed as the aforementioned precipitous x-ray
peak broadening in cooled smectic liquid crystal cells with a random
substrate~\cite{CDJonesThesis,ClarkSmC}, which is expected to have a peak width
proportional to $\xi_z^{-1}(T)$, as shown in Fig.~\ref{fig:Peakwidth}.
Although further systematic
detailed studies are necessary to test this conjecture, based on the
robustness of our theoretical prediction, we expect this transition to
be quite generic in smectic liquid crystal cells with heterogeneous
substrates.

The remainder of this paper is organized as follows. In Sec. \ref{sec:smecticmodel} we present
a simple model of a thick (half-infinite) bookshelf smectic cell with
surface heterogeneity and derive the effective surface
theory by dimensional reduction where bulk modes are integrated out.
In Sec.~\ref{sec:Imary-Ma}, the Imry-Ma argument is used to estimate the
domain size at which the influence of random pinnings becomes large.
In Sec.~\ref{sec:smectic_IntermediateResults}, the surface random positional pinning is
treated within the Larkin, perturbative approximation, which allows us to compute exactly
the short-scale correlation functions.
In Sec.~\ref{sec:smectic_FRG}
we present a functional renormalization group
analysis to treat fully nonlinear positional pinning necessary for characterization of long scale correlations. We demonstrate that, at finite $T$, this analysis predicts a 3D Cardy-Ostlund-like phase transition.
Possible features of observation under polarized microscopy and x-ray scattering
are analyzed in Sec.~\ref{sec:Experiments}.
In Sec.~\ref{sec:validity_of_theory}, the importance of
nonlinear elasticity is studied using a renormalization group method to show
that it is less relevant, thereby justifying the validity of the harmonic elastic treatment, and the stability of the
orientational order and relevant length scales are also discussed.
In Sec.~\ref{sec:parallel_cell_smectic} we derive a theory for a surface disordered
smectic cell with layers parallel to the substrates and show that it mimics a familiar bulk
disordered $xy$ model in one lower dimension.
We conclude in Sec.~\ref{sec:conclusion} with a brief summary of the work and future directions.  The technical details of our calculations are relegated to Appendices \ref{app:solution_complexIntegral}-\ref{app:full_long_scale_correlation}.

\section{Model}
\label{sec:smecticmodel}

\subsection{Bulk elasticity}
We will focus on a smectic cell in the experimentally and theoretically more interesting bookshelf geometry. We choose the coordinate system as illustrated in Fig.~\ref{fig:smecticCartoon}, so the smectic layers lie parallel to the $(x,z)$ plane, with the average layer normal and director $\hat{n}$ along the $z$ axis.
The random substrate is located at $y=0$ and running perpendicular to the $y$ axis.

Neglecting elastic nonlinearities \cite{GrinsteinPelcovits}, we model a half-infinite bookshelf
geometry smectic cell with
a heterogeneous substrate by an energy functional
\begin{equation}
 H_{bulk}=\int d^{d-1}x\int_0^{\infty} dy
\left[\frac{K}{2}(\nabla^2_\perp u)^2 +\frac{B}{2}(\partial_z  u)^2\right]+H_{pin},
\label{Hsmectic}
\end{equation}
in which $u(\xv,y)$ is the distortion of smectic layer at position $\rv=(\xv,y)$ and
in 3D $\xv=(x,z)$ spans the 2D random substrate and $y$ is the axis into the bulk of the cell.

\subsection{Surface pinnings}
The surface pinning energy of the substrate is given by
\begin{equation}
H_{pin}=H_y+H_{dn}+H_{d\rho},
\end{equation}
including uniform pinning of the layers, coupling of the substrate with the directors $\hat{n}$ and smectic density $\rho(\xv)$ on the $y=0$ substrate, respectively.

To stabilize the bookshelf cell geometry, a uniform component of the pinning is realized on the $y=0$ substrate via
\begin{equation}
H_y=\int d^{d-1}x \frac{W'}{2}(\hat{n}\cdot\hat{y})^2,\label{Hpiny}
\end{equation}
which dictates the layers to be perpendicular to the substrate. In experiment,
this is imposed by a uniform structureless treatment of the substrate \cite{pinningByStructurelessWall,Sluckin_pinningMechanism}.

Generically, the interaction between the substrate and the liquid crystal also includes the coupling between the nematic director $\hat{n}$
and the local random pinning axis, ${\bf g}(\xv)=(g_x,g_y,g_z)$, as well as the
coupling between the smectic density $\rho(\xv)$ and the local
random scalar potential $U(\xv)$. These are determined
by the substrate's local chemical and physical structure
(composition, roughness, rubbing, etc.).
The nematic director $\hat{n}$ is pinned by the local random pinning axis
${\bf g}(\xv)$, with
\begin{equation}
H_{dn}=-\int d^{d-1}x \left[\nh\cdot{\bf g}(\xv)\right]^2.
\end{equation}
We approximated this weak pinning by specializing to the smectic state, with the layer normal $\hat{n}$ taken along $\hat{z}\sqrt{1-\delta n^2} + \delta{{\bf n}}$ and $\delta{{\bf n}} \approx -\nabla_{\perp} u$ \cite{RTaerogelPRB,usSmecticEPL}.
With this, $H_{dn}$ becomes
\begin{eqnarray}
H_{dn}&\approx& \int d^{d-1}x \bigg\{\left[g_z^2(\xv)-g_{x}^2(\xv)\right](\partial_x u)^2\nonumber\\
&&+\left[g_z^2(\xv)-g_y^2(\xv)\right](\partial_y u)^2-2g_x(\xv)g_y(\xv)(\partial_x u)(\partial_y u)\nonumber\\
 &&+2g_x(\xv)g_z(\xv)(\partial_x u)+2g_y(\xv)g_z(\xv)(\partial_y u)\bigg\}.
\label{H_dn_full}
\end{eqnarray}
For a rubbed substrate used in the experiment \cite{CDJonesThesis,Sluckin_pinningMechanism}, the alignment is highly anisotropic such that $|g_z|\gg |g_x|\sim |g_y|$ and pinning of layer normal along $\hat{z}$ is imposed.
Combining with Eq.~(\ref{Hpiny}) and averaging over microscopic scales, we find
\begin{equation}
H_y+H_{dn}\approx \int d^{d-1}x\left[\frac{W}{2}(\partial_yu)^2+\frac{W_Q}{2}(\partial_xu)^2-h(\xv)(\partial_x u)\right],
\label{H_yandHdn}
\end{equation}
in which
\begin{equation}
\left\{\begin{array}{l}
W=W'+2\left[\overline{g_z^2(\xv)-g_y^2(\xv)}\right],\\
W_Q=2\left[\overline{g_z^2(\xv)-g_{x}^2(\xv)}\right],\\
h(\xv)=-2g_x(\xv)g_z(\xv).
\end{array}\right.
\end{equation}
For strong pinning $\frac{W}{2}(\partial_y u)^2$, the linear terms in $(\partial_y u)$ in
Eq.~(\ref{H_dn_full}) can be neglected and $\partial_y u$ is taken to be $0$.
Determined by the local chemical and physical structure, $h(\xv)$ is expected to be short ranged, with variance
\begin{equation}
\overline{h(\xv)h(\xv')}=\Delta_f\delta^{d-1}_a(\xv-\xv'),
\label{h_variance}
\end{equation}
where $\overline{\cdots}$ is the ensemble average and $\delta^{(d-1)}_a(\xv)$ is a $(d-1)$-dimensional short-ranged function set by the scale on
the order of the molecular size $a$. The precise form of $\delta^{d-1}_a(\xv)$ has no qualitative effect on the long-scale (longer than its range) behavior which is our focus here, and the effectiveness of field description of the system already averages out details on scales shorter than $a$.
Thus, it can be replaced by a $\delta$ function $\delta^{d-1}(\xv)$ without loss of generality.

Similarly, the local random scalar potential $U(\xv)$
is also expected to be short-ranged, and we thereby take it to be Gaussian characterized by variance
\begin{equation}
\overline{U(\xv)U(\xv')}=\Delta_U\delta^{d-1}_a(\xv-\xv')\approx \Delta_U\delta^{d-1}(\xv-\xv'),
\end{equation}
with strength $\Delta_U$. As studied in the context of bulk disorder for smectics in, e.g., aerogel \cite{RTaerogelPRL,RTaerogelPRB}, it couples to the smectic density on the $y=0$ substrate [with $u_0(\xv)=u(\xv,y=0)$ the
surface layer displacement]
\begin{equation}
\rho(\xv)=\rho_0+\sum_{n=1}^{\infty}\mathrm{Re}\{\rho_ne^{inq_0\left[z+u_0(\xv)\right]}\}
\end{equation}
through
\begin{equation}
H_{d\rho}=-\int d^{d-1}x \rho(\xv)U(\xv)\approx -\int d^{d-1}x V[u_0(\xv),\xv].
\end{equation}
We also take the random positional pinning potential $V[u_0,\xv]$ to be Gaussian with short-ranged variance
\begin{eqnarray}
&&\hspace{-0.6 cm}\overline{V[u_0(\xv),\xv]V[u'_0(\xv'),\xv']}\nonumber\\
&=&\left[\rho_0^2+\sum_{n=1}^{\infty}\frac{\rho_n^2}{2}
 \mathrm{Re}\left(\overline{e^{inq_0[z-z'+u_0(\xv)-u_0(\xv')]}}\right)\right]\nonumber\\
&&\times\Delta_U\delta^{d-1}(\xv-\xv')\nonumber\\
&=&R_v(u_0-u'_0)\delta^{d-1}(\xv-\xv'),
\label{R_v_variance}
\end{eqnarray}
in which the nonlinear variance function $R_v(u_0-u'_0)$ is given by
\begin{equation}
R_v(u_0-u'_0)=\Delta_U\left(\rho_0^2+\sum_{n=1}^{\infty}\frac{\rho_n^2}{2} \cos{\left\{nq_0\left[u_0(\xv)-u_0(\xv')\right]\right\}}\right),
\label{R_v_full_form}
\end{equation}
with the first dominant term corresponding to $q_0$ as studied in Sec.~\ref{sec:simple_Rf_smectic}.

Putting these terms together, the surface pinning energy is given by
\begin{equation}
H_{pin}\approx\int d^{d-1}x \left\{\frac{W_Q}{2}(\partial_x u)^2
-h(\xv)\partial_x u-V[u_0(\xv),\xv]\right\},
\label{H_pin}
\end{equation}
where we take the random fields $h(\xv)$ and $V[u_0(\xv),\xv]$ to be characterized by a Gaussian zero-mean distribution with variances given in Eqs.~(\ref{h_variance}) and (\ref{R_v_variance}) and
impose $\partial_yu=0$ on the $y=0$ substrate to satisfy the strong pinning in Eq.~(\ref{H_yandHdn}).
More generically, the random surface torque $h$ may also be a function the layer distortion $u_0$, with a periodic variance $\Delta_f(u_0-u'_0)$. However, as we will show in Sec.~\ref{sec:simple_Rf_smectic}, such a generalization is unnecessary at long scales.

\subsection{Effective surface model}
\label{sec:smecticDimensionalReduction}
Since the surface disorder is confined to the front substrate
at $y=0$, no nonlinearities (within harmonic elastic treatment) appear in the bulk
($y>0$) of the smectic cell. Consequently,
it is convenient to exactly eliminate the bulk degrees of freedom $u(\xv,y)$ in favor of the
layer distortion field on the random substrate, $u_0(\xv)\equiv u(\xv,y=0)$.
This can be done via a
constrained path-integral method by integrating out $u(\xv,y)$ with a
constraint $u(\xv,y=0)=u_0(\xv)$, thereby obtaining an
effective $(d-1)$-dimensional Hamiltonian for
$u_0(\xv)$ \cite{tiltLR}. Equivalently (for $T=0$ properties), we
can eliminate $u(\xv,y)$ by solving the Euler-Lagrange
equation
\begin{equation}
K\nabla_{\perp}^4u-B\partial_z^2u=f(\xv)\delta(y),
\label{smectic_eom_bulk}
\end{equation}
with $f(\xv)\delta(y)$ representing
the boundary condition that imposes $u_0(\xv)$ on the substrate.
To this end, we Fourier transform
$u(\xv,y)$ over $(\xv,y)$, obtaining an algebraic equation for
\begin{equation}
u(\qv,q_y)=\int d^{d-1}x dy u(\xv,y)e^{-i\qv\cdot\xv-i q_yy},
\end{equation}
whose solution after Fourier transform over $q_y$ becomes (as shown
in Appendix \ref{app:solution_complexIntegral})
\begin{subequations}
\begin{eqnarray}
&&\hspace{-0.6 cm}u(\qv,y) = u_0(q_x,q_z)
 e^{-\frac{y}{\sqrt{2\lambda}}\sqrt{\sqrt{\lambda^2q_x^4+q_z^2}+\lambda q_x^2}}\nonumber\\
 &&\times\Bigg[\frac{\sqrt{\sqrt{\lambda^2q_x^4+q_z^2}+\lambda q_x^2}}{\sqrt{\sqrt{\lambda^2q_x^4+q_z^2}-\lambda q_x^2}}
 \sin{\Big(\frac{y}{\sqrt{2\lambda}}\sqrt{\sqrt{\lambda^2q_x^4+q_z^2}-\lambda
q_x^2}\Big)} \nonumber\\
&&
+\cos{\Big(\frac{y}{\sqrt{2\lambda}}\sqrt{\sqrt{\lambda^2q_x^4+q_z^2}-\lambda q_x^2}\Big)}\Bigg]\\
&\equiv& u_0(q_x,q_z)\phi({\qv},y),
\end{eqnarray}
\label{uqy}
\end{subequations}
where $\lambda=\sqrt{K/B}$ is the length that, deep in the smectic phase, is comparable to the
microscopic length $a$ (determined by the layer thickness or molecular size)
and we define a shape function $\phi({\qv},y)$ that characterizes the extension of surface-imposed distortions into the bulk ($y>0$).
In obtaining Eq.~(\ref{uqy}), for strong surface pinning ($W\rightarrow \infty$) in Eq.~(\ref{H_yandHdn}), we also impose a constraint
$\partial_y u = 0$ at $y=0$. More generally, the relative importance
of the sine and cosine contributions depends on the system and the homogeneous transverse pinning strength $W$.

After substituting the above solution into (\ref{Hsmectic}) and integrating out
the $y$ (bulk) degrees of freedom in a half-infinite system ($0\leq y<\infty$), the energy functional
simplifies to an effective surface energy,
\begin{eqnarray}
H_{surface}&=&\int\frac{d^{d-2}q_x dq_z}{(2\pi)^{d-1}}\frac{\Gamma_{\qv}+W_Q q_x^2}{2}
 \left|u_0(q_x,q_z)\right|^2\nonumber\\
&&-\int d^{d-1}x \left[h(\xv)\partial_x u_0+V(u_0,\xv)\right],
\label{effectiveH0}
\end{eqnarray}
confined to the random substrate at $y=0$, with
\begin{equation}
\Gamma_{\qv}=B\sqrt{2\lambda}\sqrt{\lambda^2q_x^4+q_z^2}
 \sqrt{\sqrt{\lambda^2q_x^4+q_z^2}+\lambda q_x^2}.
 \label{Gamma_q}
\end{equation}
The resulting effective surface Hamiltonian is convenient for studying the surface properties of the system, with
the bulk properties obtained from Eq. (\ref{uqy}).
Equivalently, this result can be obtained by solving the Euler-Lagrange
equation and specifying the boundary conditions, as in the nematic
liquid crystal cell with surface disorder \cite{usFRGPRE}.

By comparing the induced surface elastic energy $\Gamma_{\qv}\sim Kq_x^3$ ($q_z \ll\lambda q_x^2$) with surface pinning due to
rubbing on the substrate, $W_Qq_x^2$, we obtain an important length scale
\begin{equation}
l_W=\frac{K}{W_Q}.
\label{l_W}
\end{equation}
On longer scales, the surface rubbing dominates, and $\Gamma_{\qv}$ can be approximated by $\Gamma_{\qv}\sim B\sqrt{2\lambda}q_z^{3/2}$.
In this regime the effective kernel is given by
\begin{equation}
\Gamma_{\qv}+W_Qq_x^2\approx B\sqrt{2\lambda}q_z^{3/2}+W_Qq_x^2,
\end{equation}
with an anisotropic ``elasticity'' for small $q_x$, which differs markedly from the bulk smectic elasticity and $\Gamma_{\qv}$ surface smectic
elasticity \textit{without} a rubbed substrate.

However, for a weak pinning $W_Q$ this scale can be much larger than the sample region and thus $W_Q$ can be neglected. For the remainder of the paper we will focus on this more interesting nonrubbed substrate limit.

\subsection{Replicated model}
In treating random heterogeneous systems it is often
convenient to work with an effective translationally invariant field
theory. This is possible via the \mbox{standard} replica ``trick'' \cite{Anderson}, at
the expense of introducing $n$ replica fields (with the $n\rightarrow
0$ limit taken at the end of the calculation). The disorder-averaged
free energy is given by ${\overline F}=-T\overline{\ln Z}=-T\lim_{n\rightarrow 0}{\overline{Z^n}-1\over n}$, with
\begin{equation}
\overline{Z^n}=\int\left[du_0^\alpha\right]e^{-H_{surface}^{(r)}[u_0^\alpha]/T},
\end{equation}
where the effective translationally-invariant replicated Hamiltonian
$H_{surface}^{(r)}[u_0^\alpha]$ is given by
\begin{eqnarray}
&&\hspace{-1 cm}H_{surface}^{(r)}\nonumber\\
&=&\sum_{\alpha}^n\int\frac{d^{d-2}q_x dq_z}{(2\pi)^{d-1}}
\frac{\Gamma_{\qv}}{2}
\left|u_0^{\alpha}(q_x,q_z)\right|^2\nonumber\\
&&+\frac{1}{4T}\sum_{\alpha,\beta}\int d^{d-2}xdz\Delta_f
\left|\partial_x\left[u_0^{\alpha}(x,z)-u_0^{\beta}(x,z)\right]\right|^2\nonumber\\
&&-\frac{1}{2T}\sum_{\alpha,\beta}\int d^{d-2}x dz R_v\left[u_0^{\alpha}(x,z)-u_0^{\beta}(x,z)\right].
\label{Hsr}
\end{eqnarray}

The advantage of the dimensional reduction in Sec.~\ref{sec:smecticDimensionalReduction} is that formally the
problem becomes quite similar to the extensively studied bulk
random pinning model \cite{DSFisherFRG,Nattermann,GiamarchiLedoussalFluxLattice_PRL,
GiamarchiLedoussalFluxLattice_PRB} in one lower
dimension but at the expense of a modified long-range elasticity encoded in $\Gamma_{\qv}$,
Eq.~\rf{Gamma_q}.
As we observe from the coarse-graining procedure, in general the random orientational pinning may also be field dependent,
with $\Delta_f[u_0^{\alpha}(x,z)-u_0^{\beta}(x,z)]$ a generic periodic function.
We therefore generalized to the form in (\ref{Hsr}).

\section{Estimate of the finite smectic domain scales}
\label{sec:Imary-Ma}
The qualitative features of the response of the smectic cell to a heterogeneous substrate can be understood through
a generalization of the Imry-Ma argument \cite{ImryMa} to the surface-pinning
problems \cite{FeldmanVinokurPRL,usFRGPRL,usFRGPRE}. In this section, we estimate the characteristic length scales
beyond which pinning energies become significant in distorting smectic order.

\subsection{Bulk Imry-Ma analysis}
\label{sec:bulk_Imry_ma}
For an ordered region of $L_x\times L_z$ on the heterogeneous
substrate the distortions decay into the bulk within a depth of $L_y\sim L_x$.
Consequently, the elastic energy cost of such region scales as
\begin{equation}
E_e\sim L_x^{d-1}L_z \left[K(a/L_x^2)^2+B(a/L_z)^2\right].
\end{equation}
The two elastic contributions balance for
$L_x^2\sim \lambda L_z$ (where $\lambda=\sqrt{K/B}$),
as an expression of a Virial theorem, justified in detail in
in Sec.~\ref{sec:larkinLength_smectic}.
With this anisotropic scaling the elastic energy reduces to
\begin{equation}
E_e\sim Ba^2\lambda^2L_x^{d-3}
\sim Ba^2\lambda^{(d+1)/2}L_z^{(d-3)/2}.
\label{Imry_ma_E_e_bulk}
\end{equation}

In such region, the interaction of the layers with the random positional pinning can lower the energy
by a typical value
\begin{eqnarray}
E_v&\sim& V_p\sqrt{N_p}\sim \Delta_v^{1/2}\sqrt{L_x^{d-2}L_z}\frac{\xi_0^{(d-1)/2}}{a^{(d-1)/2}}\nonumber\\
&&\sim \Delta_v^{1/2}L_x^{d/2}/\sqrt{\lambda}\nonumber\\
&&\sim \Delta_v^{1/2}\lambda^{(d-2)/4}L_z^{(d/4)},
\label{Imry_ma_E_v}
\end{eqnarray}
where $V_p$ is the typical random pinning strength with zero mean
and variance $\Delta_v\approx V_p^2/\xi_0^{d-1}$ ($\xi_0$ is the pinning correlation length) and $N_p$ is the number
of surface pinning sites. Comparing between $E_v$ and $E_e$ shows that for dimension $d<d^v_{lc}=6$,
arbitrary weak random positional pinning
always dominates over the elastic energy on sufficiently long scales,
\begin{equation}
L_x>\xi_x\sim(B^2/\Delta_v)^{1/(6-d)},
\end{equation}
and
\begin{equation}
L_z>\xi_z\sim (B^2/\Delta_v)^{2/(6-d)}.
\end{equation}
Thus, long-range smectic order is always destabilized on these long scales by an arbitrarily weak random positional pinning.

Similarly, the random orientational pinning can lower the energy through interaction with the layers
by a typical amount,
\begin{equation}
E_f\sim V_h\sqrt{N_h}a/L_x\sim \Delta_f^{1/2} L_x^{d/2-1}\sim\Delta_f^{1/2}L_z^{(d-2)/4}.
\label{Imry_ma_E_f}
\end{equation}
Comparing $E_f$ and
$E_e$, we find that arbitrarily weak random orientational pinning destroys long range smectic order
for $d<d^f_{lc}=4$ on sufficiently long scales
\begin{equation}
L_x>\xi_x\sim(B^2/\Delta_f)^{1/(4-d)},
\end{equation}
and
\begin{equation}
L_z>\xi_z\sim(B^2/\Delta_f)^{2/(4-d)}.
\end{equation}

Reassuringly, as we will see, these length scales, identified through the Imry-Ma argument, are consistent with the domain sizes
given in Eqs.~(\ref{LLrandomTilt}, \ref{LLrandomPosition}), obtained through a detailed field theoretical calculation.

\subsection{Surface Imry-Ma analysis}
\label{sec:surface_Imry_ma}
A complementary (but equivalent to bulk) way of estimating the surface-pinning  and bulk ordering competition is by using the substrate model derived in Sec.~\ref{sec:smecticDimensionalReduction}.
From the effective surface theory given in Eq.~(\ref{effectiveH0}) (with the pinning
of $W_Q$ neglected), for an ordered region of $L_x\times L_z$ on the heterogeneous surface,
the elastic kernel $\Gamma_{\qv}$ scales according to
\begin{equation}
\Gamma_{\qv}\sim B\lambda^2L_x^{-3}\sim B\lambda^{1/2}L_z^{-3/2}.
\end{equation}
Therefore the elastic energy cost is given by
\begin{equation}
E_e\sim L_x^{d-2}L_zB\lambda^2a^2L_x^{-3}\sim B\lambda^2a^2L_x^{d-3}
\end{equation}
or in terms of $L_z$
\begin{equation}
E_e\sim L_x^{d-2}L_z B\lambda^{1/2}a^2L_z^{-3/2}\sim B\lambda^{1/2}a^2L_z^{(d-3)/2},
\end{equation}
with the same scaling as Eq.~(\ref{Imry_ma_E_e_bulk}).
Comparing with the scaling of the random pinning energies given in Eqs.~(\ref{Imry_ma_E_v})
and (\ref{Imry_ma_E_f}), the Imry-Ma argument of
effective surface Hamiltonian leads to the same critical dimension and characteristic lengths as
in Sec.~\ref{sec:bulk_Imry_ma}.

\section{Short scale ``Larkin'' analysis: random force approximation}
\label{sec:smectic_IntermediateResults}
\subsection{Random force (linear) approximation}
\label{sec:randomTorque_smectic}
The importance of surface pinning can be assessed by computing smectic layer
distortions $\overline{\langle u_0^2\rangle}$ at $y=0$ surface (dominated by the
zero-temperature distortions) within the Larkin
approximation \cite{Larkin}, which amounts to a linear random force
approximation,
$F(\xv)=\partial_{u_0}V[u_0(\xv),\xv]\bigg|_{u_0=0}$
to the random potential $V[u_0(\xv),\xv]$, with inherited Gaussian
statistics and variance
\begin{equation}
\overline{F(\xv)F(\xv')}
\equiv\Delta_v\delta^{d-1}(\xv-\xv')=-R_v''(0)\delta^{d-1}(\xv-\xv').
\end{equation}
In momentum space the pinning energy becomes
\begin{equation}
H_{pin}=\int \frac{d^{d-1}q_x}{(2\pi)^{d-1}}
\Big[iq_x h(q_x,q_z)
-F(q_x,q_z)\Big]u_0(-q_x,-q_z).
\label{randomTorque_pinning_smectic}
\end{equation}

Within this approximation, the
correlation of the layer distortion on the random substrate ($y=0$) is
\begin{equation}
C_{Larkin}(\qv)=\frac{T}{\Gamma_{\qv}}+\frac{\Delta_fq_x^2+\Delta_v}
  {\Gamma_{\qv}^2},\label{C_Larkin}
\end{equation}
including the thermal contribution
\begin{equation}
C_{T,Larkin}(q)=\frac{T}{\Gamma_{\qv}},\label{C_T_smectic}
\end{equation}
and the contributions from both types of surface disorders,
\begin{equation}
C_{\Delta,Larkin}(q)=\frac{\Delta_fq_x^2+\Delta_v}
  {\Gamma_{\qv}^2}. \label{C_D_smectic}
\end{equation}

At long scales of interest, the contribution from
thermal fluctuation is
clearly subdominant to the zero temperature random pinning driven distortions.
We therefore focus on this latter disorder dominant contribution.

\subsubsection{Power counting}
For various ranges of $q_x$ and $q_z$, the elastic kernel $\Gamma_{\qv}$
exhibits the following asymptotics
\begin{equation}
\Gamma_{\qv}\approx\left\{\begin{array}{ll}
B\sqrt{2\lambda}q_z^{3/2}, &\lambda q_x^2\ll q_z,\\
B\sqrt{2}\lambda^2q_x^3, &\lambda q_x^2\gg q_z.
\end{array}\right.
\end{equation}
We note that there is a $2:1$ ratio in the powers of $q_x$ and $q_z$
that reflects the anisotropy of the underlying smectic state.
This anisotropic scaling is crucial in our estimates below the
critical dimensions for the importance of surface pinning.

The correlation function contribution from random \textit{orientational} pinning is given by
\begin{equation}
C_{\Delta,Larkin}(q)=\frac{\Delta_f q_x^2}{\Gamma_{\qv}^2}
\approx \frac{\Delta_fq_x^2}{\left(B\sqrt{2\lambda}q_z^{3/2}+B\sqrt{2}\lambda^2q_x^3\right)^2}.
\end{equation}
Thus, the scaling of smectic zero-temperature distortions is approximately given by
\begin{subequations}
\begin{eqnarray}
\overline{\langle u^2\rangle}&\sim& \Delta_f \int_{1/L_x} \frac{q_x^2d^{d-2}q_x dq_z}{q_x^6}\sim L_x^{4-d},\\
\overline{\langle u^2\rangle}&\sim& \Delta_f\int_{1/L_z} \frac{q_x^2d^{d-2}q_x dq_z}{q_z^3}\sim L_z^{(4-d)/2},
\end{eqnarray}
\label{d_lc_orientational}
\end{subequations}
and leads to the lower critical dimension $d^f_{lc}=4$ for the random
\textit{orientational} pinning.

For the random \textit{positional} pinning, we have
\begin{equation}
C_{\Delta,Larkin}(q)=\frac{\Delta_v}{\Gamma_{\qv}^2}
\approx \frac{\Delta_v}{\left(B\sqrt{2\lambda}q_z^{3/2}+B\sqrt{2}\lambda^2q_x^3\right)^2},
\end{equation}
which gives
\begin{subequations}
\begin{align}
\overline{\langle u^2\rangle} \sim \Delta_v\int_{1/L_x} \frac{d^{d-2}q_xdq_z}{q_x^6}\sim L_x^{6-d},\\
\overline{\langle u^2\rangle} \sim \Delta_v\int_{1/L_z} \frac{d^{d-2}q_xdq_z}{q_z^3} \sim L_z^{(6-d)/2},
\end{align}\label{d_lc_positional}
\end{subequations}
leading to $d^v_{lc}=6$ as the lower critical dimension for the positional pinning.

This agrees with our Imry-Ma analysis of the previous section and is to be contrasted with the bulk predictions of Ref.[\onlinecite{RTaerogelPRB}],
which finds
\begin{subequations}
\begin{align}
\overline{\langle u^2\rangle}\sim \Delta_f\int \frac{q_x^2d^{d-1}q_xdq_z}{(q_x^4+q_z^2)^2}\sim \int_{1/L_x} \frac{q_x^{d+2}dq_x}{q_x^8}\sim L_x^{5-d},\\
\overline{\langle u^2\rangle}\sim \Delta_v\int \frac{d^{d-1}q_xdq_z}{(q_x^4+q_z^2)^2}\sim \int_{1/L_x} \frac{q_x^{d}dq_x}{q_x^8}\sim L_x^{7-d},
\end{align}
\end{subequations}
with $d^{bulk}_{lc} = 5$ and $7$ for for bulk orientational and positional random pinnings,
respectively.
Thus, similarly to the surface disordered nematic cell discussed in Ref.~[\onlinecite{usFRGPRL}] and [\onlinecite{usFRGPRE}],
here, too, the restriction of the pinning to a surface reduces its lower critical
dimensions down by one.

\subsection{Domain size}
\label{sec:larkinLength_smectic}
The utility of the Larkin approximation is that it predicts the range of
its own validity, limited to length scales where the distortion of $u_0$ is small.
From Eqs. (\ref{d_lc_orientational}) and (\ref{d_lc_positional}) it is clear that
below the respective $d_{lc}$, the distortions diverge at long scales,
signaling an instability of the smectic state and associated breakdown of Larkin approximation.

We identify the substrate extent $L_x$ and $L_z$ at which
these smectic distortions grow large to the order of the layer spacing $a$
as (the so-called) Larkin domain lengths $\xi_x$ and $\xi_z$ \cite{ChargeDensityWave,VortexGlass}.
More specifically, Larkin lengths are defined as
\begin{equation}
\overline{\langle u^2_0(x,z)\rangle}=a^2,
\end{equation}
which clearly depends on
both surface orientational and positional disorder \cite{DSFisherFRG,GiamarchiLedoussalFluxLattice_PRB,GiamarchiLedoussalFluxLattice_PRL,LedoussalWiese,BalentsFisher, commentRM}. Thus, the values of
$\xi_{x,z}$ depend on the relative strengths of these two types of disorder.
For $L_x \ll \sqrt{\lambda L_z}$ in 3D, a simple analysis (with details relegated to Appendix \ref{app:LakinLengths_smectic}) gives
\begin{equation}
\overline{\left\langle u_0^2(\xv)\right\rangle}
\approx\int_\qv \frac{\Delta_fq_x^2+\Delta_v}{\Gamma_{\qv}^2}
\approx \frac{1}{3c}\frac{\Delta_v}{ B^2\lambda^3}L_x^3+\frac{1}{c}\frac{\Delta_f}{ B^2\lambda^3}L_x,
\end{equation}
with $c=\frac{4\pi^2}{\pi-2}\approx 34.6$, leading to the equation for $\xi_x$
\begin{equation}
\frac{1}{3c}\frac{\Delta_v}{ B^2\lambda^3}\xi_x^3+\frac{1}{c}\frac{\Delta_f}{ B^2\lambda^3}\xi_x\equiv a^2.
\label{equation_of_xi_x}
\end{equation}
In the opposite limit of $L_z < L_x^2/\lambda$ we instead get
\begin{equation}
\overline{\langle u_0^2(\xv)\rangle}
\approx\int_\qv \frac{\Delta_fq_x^2+\Delta_v}{\Gamma_{\qv}^2}
\approx \frac{1}{3c}\frac{\Delta_v}{B^2\lambda^{3/2}}L_z^{3/2}+\frac{1}{c}\frac{\Delta_f}{B^2\lambda^{5/2}}L_z^{1/2},
\end{equation}
which leads to the equation for a $\xi_z$,
\begin{equation}
\frac{1}{3c}\frac{\Delta_v}{B^2\lambda^{3/2}}\xi_z^{3/2}+\frac{1}{c}\frac{\Delta_f}{B^2\lambda^{5/2}}\xi_z^{1/2}\equiv a^2.
\label{equation_of_xi_z}
\end{equation}
As discussed in Appendix \ref{app:LakinLengths_smectic}, above equations are
to be understood in the scaling sense and not as precise quantitative conditions.

The Larkin lengths $\xi_{x,z}$ in Eqs.~(\ref{equation_of_xi_x}) and (\ref{equation_of_xi_z})
give the size of the smectic
domains on the random substrate
within which the elastic energy dominates over the pinning,
which is effectively weak and Larkin approximation is valid.

More generally, in the presence of both types of pinning, the smectic domain
size is determined by the minimum of $\xi_{x,z}^f$ and $\xi_{x,z}^v$, calculated in Appendix \ref{app:LakinLengths_smectic}
\begin{equation}
\xi_{x,z}=\xi_{x,z}(\xi_{x,z}^f,\xi_{x,z}^v)\approx \mbox{min}\{\xi_{x,z}^f,\xi_{x,z}^v\},
\label{domain_size}
\end{equation}
in which
\begin{eqnarray}
\left\{ \begin{array}{lll}
\xi_z^f&\approx&\Big[c\frac{B^2\lambda^{5/2}a^2}{\Delta_f}\Big]^{2},\\
\xi_x^f&=&c\frac{B^2\lambda^3a^2}{\Delta_f}\approx \sqrt{\lambda \xi_z^f},
\end{array}\right.
\label{LLrandomTilt}
\end{eqnarray}
and
\begin{eqnarray}
\left\{ \begin{array}{lll}
\xi_z^v&\approx&\Big[3c\frac{B^2\lambda^{3/2}a^2}{\Delta_v}\Big]^{2/3},\\
\xi_x^v&=&\Big[3c\frac{B^2\lambda^3a^2}{\Delta_v}\Big]^{1/3}\approx\sqrt{\lambda\xi_z^v}.
\end{array}\right.
\label{LLrandomPosition}
\end{eqnarray}

The highly anisotropic domain constructed by $\xi_x$ and $\xi_z$ is illustrated in Fig.~\ref{fig:smecticCartoon}.

\subsection{ Surface correlation at short scales}
\label{Sec:LarkinCorrelation}

\begin{figure}[htbp]
\centering
\includegraphics[height=5 cm]{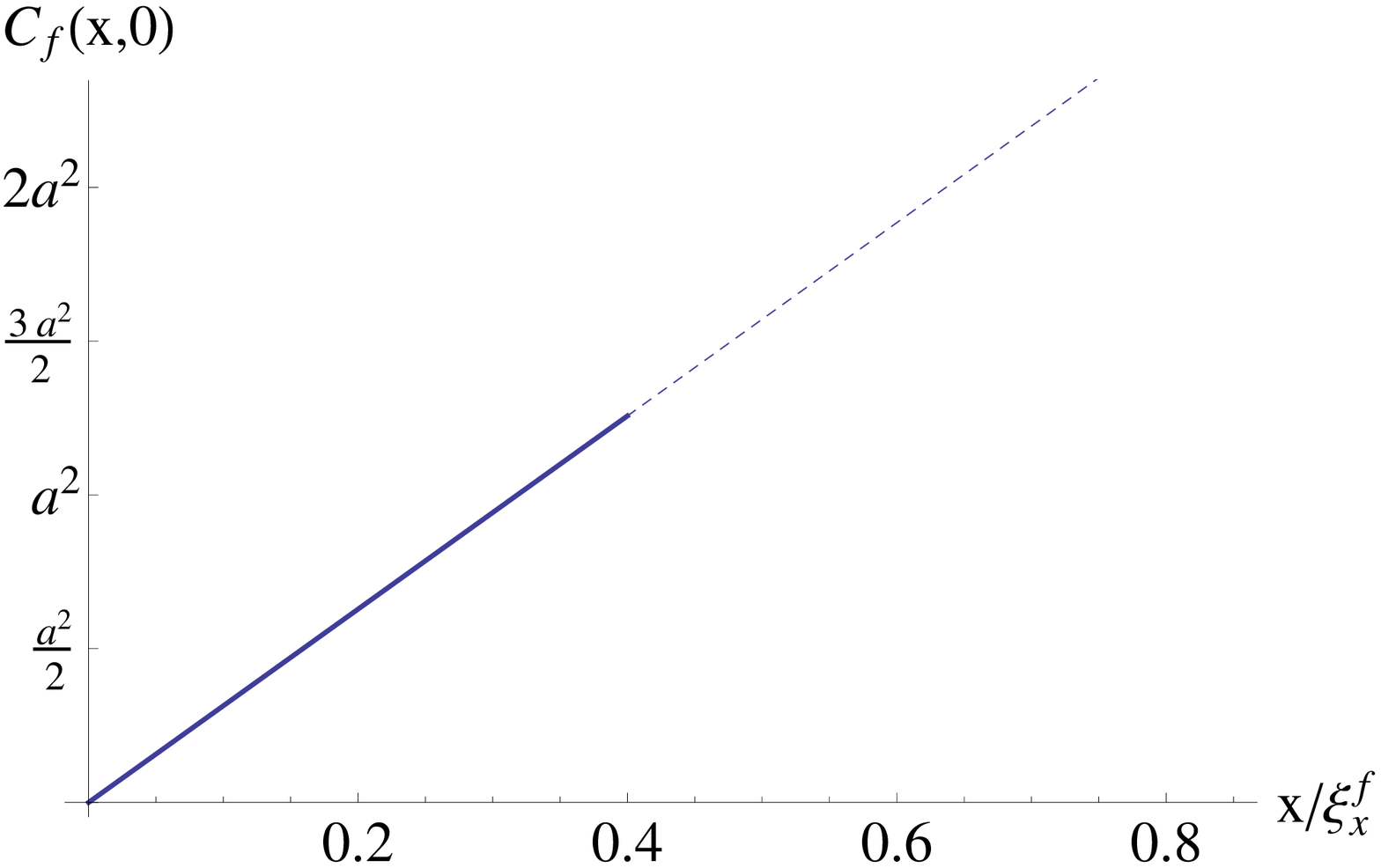}\\
\includegraphics[height=5 cm]{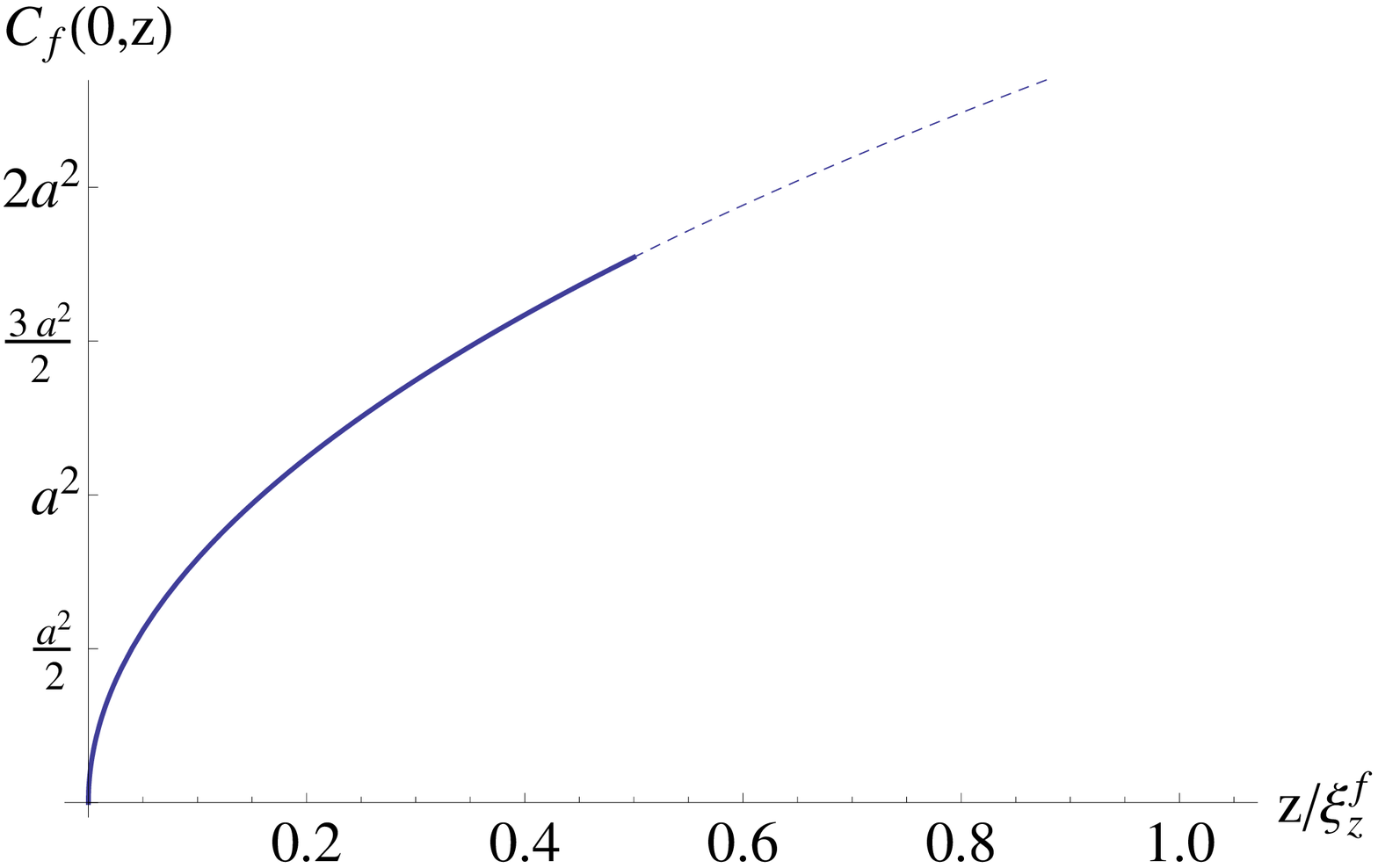}
\caption{(Color online) The correlation functions on the heterogeneous substrate along and perpendicular to the layers,
with dominant orientational disorder valid for $(x,z)\ll\xi_{x,z}$.
On longer scales nonlinear nature of pinning needs to be taken into account, which we do via RG and matching methods in Sec.~\ref{sec:smectic_FRG}.
}
\label{fig:tiltCorrelation}
\end{figure}

\begin{figure}[h]
\centering
\includegraphics[height=4.5 cm]{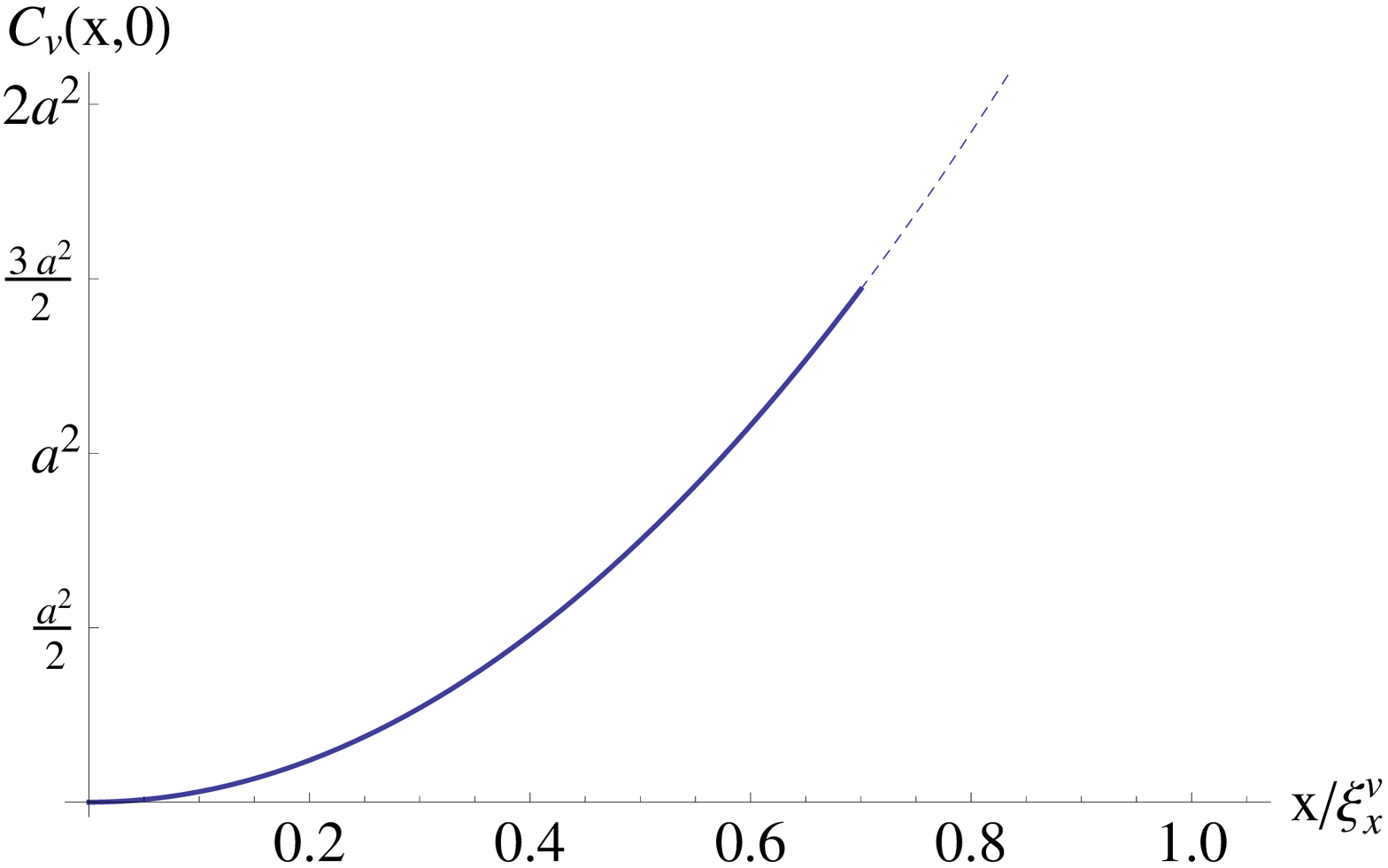}\\
\includegraphics[height=4.5 cm]{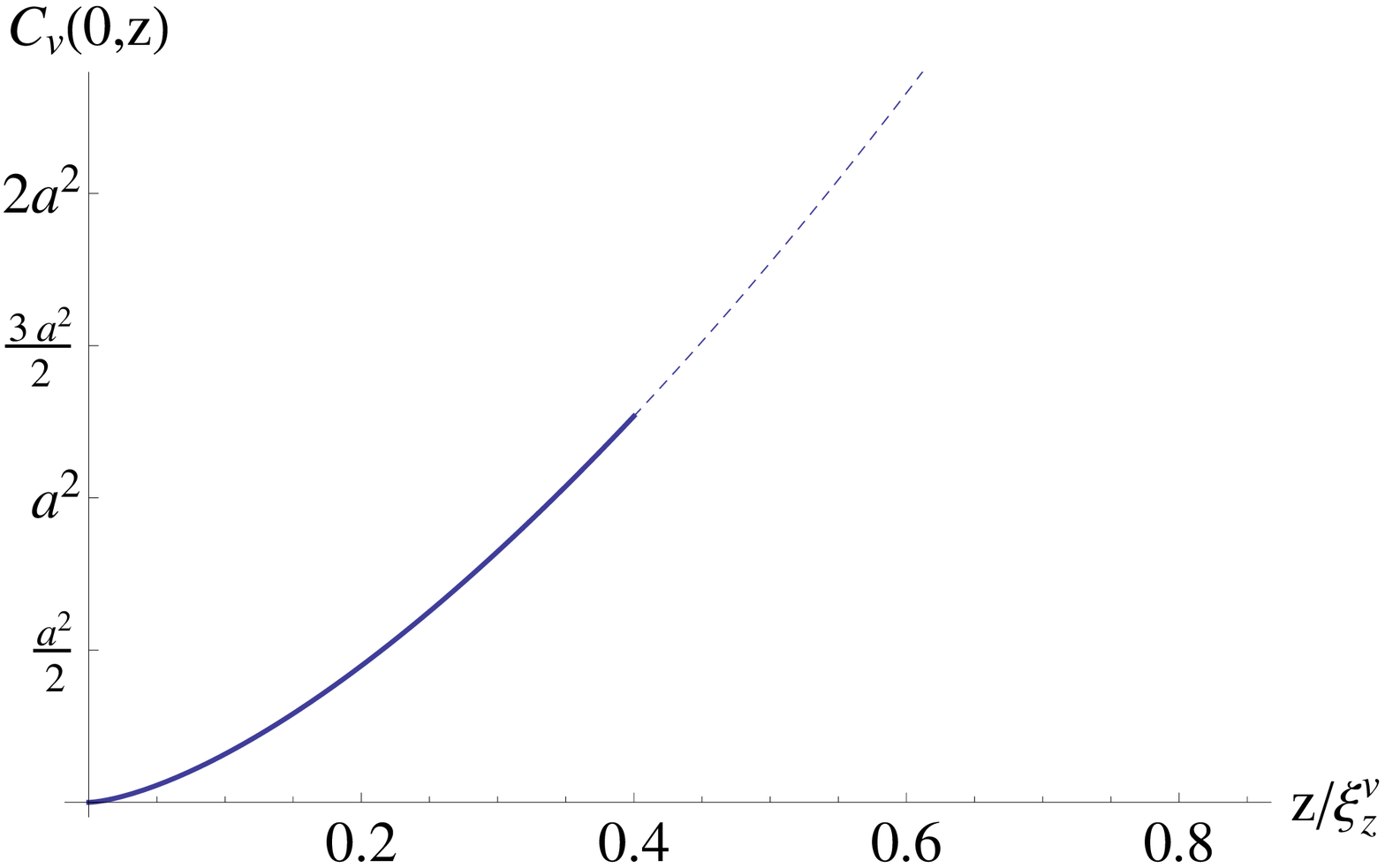}
\caption{(Color online) The correlation function on the heterogeneous substrate along and perpendicular to smectic layers, with dominant positional pinning, valid only inside Larkin domains, $(x,z)\ll\xi_{x,z}$. On longer scales an RG analysis is necessary.}
\label{fig:positionCorrelation}
\end{figure}

As shown in Appendix \ref{app:correlationfunctions}, with Eq.~(\ref{C_Larkin})
{\bf t}he $C(x,z)=\overline{\langle [u_0(x,z)-u_0(0,0)]^2 \rangle}$ can be easily calculated within the
Larkin regime ($x\ll\xi_x$ and $z\ll\xi_z$), and along the layer ($x$)
and layer normal ($z$) direction, the orientational pinning contribution is given by
\begin{equation}
\left\{\begin{array}{lll}
C_f(x,0)&\approx& \pi a^2x/\xi_x^f,\\
C_f(0,z)&\approx& \sqrt{2\pi}a^2\sqrt{z/\xi_z^f}\approx \sqrt{2\pi} a^2 \sqrt{\lambda z}/\xi_x^f,
\end{array}\right.
\label{correlation_orientational_shortScale}
\end{equation}
where $\xi^f_{x,z}$ are given in Eq.~(\ref{LLrandomTilt}).
At short scales $x\ll\xi_x$ and $z\ll\xi_z$
the behaviors of these correlation functions are plotted in Fig.~\ref{fig:tiltCorrelation}.
The contribution
from random positional pinning is
\begin{eqnarray}
\left\{\begin{array}{lll}
C_v(x,0)&\approx& 3a^2\frac{x^2}{(\xi_x^v)^2},\\
C_v(0,z)&\approx& \sqrt{8\pi}a^2\left(\frac{z}{\xi_z^v}\right)^{3/2},
\end{array}\right.
\label{correlation_positional_shortScale}
\end{eqnarray}
in which $\xi^v_{x,z}$ are given in Eq.~(\ref{LLrandomPosition}).
At short scales, $x\ll\xi_x$ and $z\ll\xi_z$,
these correlation functions are plotted in Fig.~\ref{fig:positionCorrelation}.

Thus, within finite smectic domains $(x,z)\ll\xi_{x,z}$,
on the random substrate we predict anisotropic power law correlations given above.

\subsection{Estimate of distortions into the bulk}
The nature of the distortions in the bulk of the cell, $y> 0$, can be deduced from the $u(\qv,y)$ in (\ref{uqy}) and the distortions $u_0(\qv)$ on the heterogeneous surface computed above.  The main feature of the $u(\qv,y)$ is that the amplitude of distortions at wavevector $\qv=(q_x,q_z)$ exponentially decays into the bulk with length
\begin{equation}
\xi_{\qv}=1/D(\qv)=\left(\frac{2\lambda}{\sqrt{\lambda^2q_x^4+q_z^2}+\lambda q_x^2}\right)^{1/2},
\end{equation}
where $D(\qv)=\frac{1}{\sqrt{2\lambda}}\sqrt{\sqrt{\lambda^2q_x^4+q_z^2}+\lambda q_x^2}$.

For surface smectic order limited by domain size $(\xi_x,\xi_z)$ the characteristic substrate surface wavevector $\qv_L$ is set by $(\xi_x^{-1},\xi_z^{-1})$ and leads to
\begin{eqnarray}
\xi_y&\simeq & \left[D(\xi_x^{-1},\xi_z^{-1})\right]^{-1}\nonumber\\
&=&\left(\frac{2\lambda}{\sqrt{\lambda^2\xi_x^{-4}+\xi_z^{-2}}+\lambda \xi_x^{-2}}\right)^{1/2}\approx \xi_x,
\end{eqnarray}
in which $\xi_x\approx\sqrt{\lambda \xi_z}$ is used. Thus we predict that for $y>\xi_y$ (ignoring the more subtle distortions on scales beyond the Larkin scale, see Sec.~\ref{sec:smectic_FRG}) the smectic surface distortion will anneal away into the bulk.

This allows us to define a $y$-dependent smectic domain size via a Larkin criterion on
\begin{eqnarray}
\overline{\langle u^2(y)\rangle}&=&\int_{\xi_x^{-1}(y),\xi_z^{-1}(y)}
 \frac{\Delta_fq_x^2}{\Gamma_{\qv}^2} \left|\phi(\qv,y)\right|^2 \frac{dq_xdq_z}{(2\pi)^2} \nonumber\\
 &\sim& \int_{\xi_x^{-1}(y),\xi_z^{-1}(y)}\overline{\left\langle \left|u_0(\qv)\right|^2\right\rangle}e^{-2 D(\qv)y}\frac{dq_xdq_z}{(2\pi)^2} \nonumber\\
 &\sim& a^2\frac{\xi_x(y)}{\xi_x}e^{-2D\left[\xi_x^{-1}(y),\xi_z^{-1}(y)\right]y},
\end{eqnarray}
where we assume dominant orientational pinning (positional pinning can be similarly analyzed).

As usual, $\xi_x(y)$ is set by $\overline{\langle u^2(y)\rangle}\simeq a^2$; thus, we have
\begin{equation}
\xi_x(y)\simeq \xi_x e^{2y/\xi_x(y)}.
\end{equation}
A self-consistent solution for this equation is simply
\begin{equation}
\xi_x(y)\sim\max{(\xi_x,2y)},
\end{equation}
showing near healing of smectic order (but see Sec.~\ref{sec:smectic_FRG}) at a distance $\xi_x$ into the bulk.

\section{Physics  on scales beyond smectic domains}
\label{sec:smectic_FRG}
On length scales longer than the crossover scales $\xi_{x,z}$,
the distortions of $u_0$ grow into a nonlinear regime, where the random-force (Larkin) approximation is inadequate, and the effects of the surface
disorders must be treated nonperturbatively. As
with bulk disorder problems, this can be done systematically using an
FRG analysis \cite{DSFisherFRG,GiamarchiLedoussalFluxLattice_PRB,GiamarchiLedoussalFluxLattice_PRL,LedoussalWiese,BalentsFisher}.

\subsection{Renormalization group analysis}
\label{sec:FRG_calculation_smectic}
We employ the standard
momentum-shell RG transformation \cite{Wilson} by separating the layer distortion
field into long- and short-scale contributions according to
$u_0^\alpha(x,z)=u_{0<}^\alpha(x,z) +u_{0>}^\alpha(x,z)$ and
perturbatively in the surface disorder integrate out the high
wave-vector fields $u_{0>}^\alpha(x,z)$ that take support in an
infinitesimal shell $\Lambda/b < q_x < \Lambda\equiv 1/a$, with
$b=e^{\delta\ell}$. We follow this with a rescaling of lengths and of the
long wavelength part of the field in real space,
\begin{eqnarray}
x&=&b\, x',\label{xb}\\
z&=&b^w z',\\
u_{0<}^{\alpha}(bx', b^w z')&=&b^{\phi}u_0^{\alpha}(x',z'),\label{phib}
\end{eqnarray}
and in momentum space,
\begin{eqnarray}
q_x&=&b^{-1}q_x',\label{qb}\\
q_z&=&b^{-w}q_z',\\
u_{0<}^{\alpha}(b^{-1}q_x',b^{-w}q_z')&=&b^{d-2+w+\phi}u_0^{\alpha}(q_x',q_z'),\label{phiqb}
\end{eqnarray}
to restore the UV cutoff back to $\Lambda=1/a$. Because a smectic liquid crystal
is periodic under translations by a multiple of the layer spacing, it is convenient to choose
the arbitrary field dimension $\phi=0$ \cite{RTaerogelPRB} and take $w=2$
so the associated smectic period and length $\lambda=\sqrt{K/B}$ are not rescaled under
the RG transformation.

For simplicity and clarity of presentation we focus on an infinitely thick cell with the Hamiltonian given in Eq.~(\ref{Hsr}),
where both disorder strengths are general functions of $u_0(x,z)$, and
employ the functional renormalization group method.
The above rescaling leads to zeroth order RG flows of the
effective couplings, which for a thick cell are
given by
\begin{eqnarray}
\Gamma_{\qv}'(b)&=&b^d\Gamma_{\qv},\\
\label{flow_Gamma_q}
B'(b)&=&b^{d-3}B,\\
\Delta_f'(u,b)&=&b^{d-2}\Delta_f(u),\\
R_v'(u,b)&=&b^{d}R_v(u),
\label{flow_Delta_v}
\end{eqnarray}
where $T$ is kept fixed and $\Delta_f(u)$ is a periodic function of the field $u$.

The Hamiltonian can be separated into three parts,
\begin{equation}
H_{surface}^{(r)}=H_0+H_{\Delta_f}+H_{\Delta_v},
\end{equation}
where $H_{\Delta_f}$ and $H_{\Delta_v}$ are given by
\begin{eqnarray}
H_{\Delta_f}&=&\frac{1}{4T}\sum_{\alpha,\beta}\int d^{d-2}xdz\Delta_f\left[u_0^{\alpha}(x,z)-u_0^{\beta}(x,z)\right]\nonumber\\
&&\times \left|\partial_x\left[u_0^{\alpha}(x,z)-u_0^{\beta}(x,z)\right]\right|^2,\\
H_{\Delta_v}&=&-\frac{1}{2T}\sum_{\alpha,\beta}\int d^{d-2}x dz R_v\left[u_0^{\alpha}(x,z)-u_0^{\beta}(x,z)\right],\nonumber\\
\end{eqnarray}
and $H_0$ has already been given by the elastic term in Eq.~(\ref{Hsr}).

We limit the FRG analysis to one-loop order, performing the momentum shell
integration over the high-wave-vector components $u_{0>}^{\alpha}$ perturbatively
in the nonlinearities $\Delta_f[u_0^{\alpha}(x,z)-u_0^{\beta}(x,z)]$ and
$\Delta_v[u_0^{\alpha}(x,z)-u_0^{\beta}(x,z)]$. The change in the Hamiltonian
due to this coarse-graining is given by
\begin{eqnarray}
\delta H_s^{(r)}&=& \left\langle H_{\Delta_f}\right\rangle_>+\left\langle H_{\Delta_v}\right\rangle_>
 -\frac{1}{2T}\left\langle H_{\Delta_f}^2\right\rangle_>^c \nonumber\\
&& -\frac{1}{2T}\left\langle H_{\Delta_v}^2\right\rangle_>^c
 -\frac{1}{T}\left\langle H_{\Delta_f}H_{\Delta_v}\right\rangle_>^c+\dots .
\label{HorderExpansion}
\end{eqnarray}

Relegating the detailed calculations to Appendix \ref{app:FRG_2ndOrder_smectic},
this coarse-graining procedure leads to functional RG flow equations for $\Delta_f(u)$ and $R_v(u)$
\begin{widetext}
\begin{eqnarray}
\partial_{\ell}\Delta_f(u,\ell)&=&(d-2)\Delta_f(u)+\eta\Delta_f''(u)-\frac{A}{2\Lambda^2} R_v''(u)R_v^{''''}(u)-A_4\Delta_f(u)\Delta_f''(u)\nonumber\\
 &&-A_5\Big[\Delta_f(u)\Delta_f''(u)
-\Delta_f(u)\Delta_f''(0)-\Delta_f(0)\Delta_f''(u)\Big]
 -A_6\left[\frac{1}{2}\Delta_f'(u)\Delta_f'(u)-\Delta_f'(u)\Delta_f'(0)\right]\nonumber\\
 &&+A\left[\Delta_f''(u) R_v''(u)-\Delta_f''(0)R_v''(u)-\Delta_f''(u)R_v''(0)\right],
\label{Delta_fFullFlow}\\
\partial_{\ell} R_v(u,\ell)&=&d R_v(u)-\zeta\Delta_f(u)+\eta R_v''(u)
 +A\left[\frac{1}{2} R_v''(u)R_v''(u) -R_v''(u)R_v''(0)\right]
 +A_3\Big[\frac{1}{2}\Delta_f(u)\Delta_f(u)\nonumber\\
 &&-\Delta_f(u)\Delta_f(0)\Big]
 -A_5\left[\Delta_f(u)R_v''(u)-\Delta_f(0)R_v''(u)-\Delta_f(u)R_v''(0)\right],
\label{Delta_vFullflow}
\end{eqnarray}
\end{widetext}
in which $\eta=\frac{T}{2\pi B\lambda}$, $A=\frac{\pi-2}{8\pi^2B^2\lambda^3\Lambda^3}$ and other coefficients are nonuniversal constants given in Appendix \ref{app:FRG_2ndOrder_smectic}.

\subsection{Single harmonic form of positional pinning}
\label{sec:simple_Rf_smectic}
 In general, the fully nonlinear functions $\Delta_f(u)$ and $R_v(u)$ in
(\ref{Hsr}) need to be treated under the functional renormalization group coarse graining
\cite{DSFisherFRG,GiamarchiLedoussalFluxLattice_PRL,GiamarchiLedoussalFluxLattice_PRB,RTaerogelPRB}.
However, at finite temperature and in 3D, it is clear from Eqs.~(\ref{Delta_fFullFlow}) and (\ref{Delta_vFullflow}) that, at long scales, the pinning is dominated by the
field-independent $\Delta_f$ and the lowest harmonic of the positional disorder \cite{RTaerogelPRB},
\begin{equation}
R_v(u)=\Delta_v\cos{(q_0 u)}/q_0^2,
\end{equation}
where $q_0=2\pi/a$ is the smectic wave number.
Because, at finite $T$, $\eta$ is nonzero, all higher, $n>1$ harmonics of $R_v(u)$ in Eq.~(\ref{R_v_full_form}) are less relevant
with eigenvalues $\lambda_n=3-\eta n^2 q_0^2<\lambda_1$ and, thus, can be neglected.

\begin{figure}[htbp]
\centering
  \includegraphics[width=8 cm]{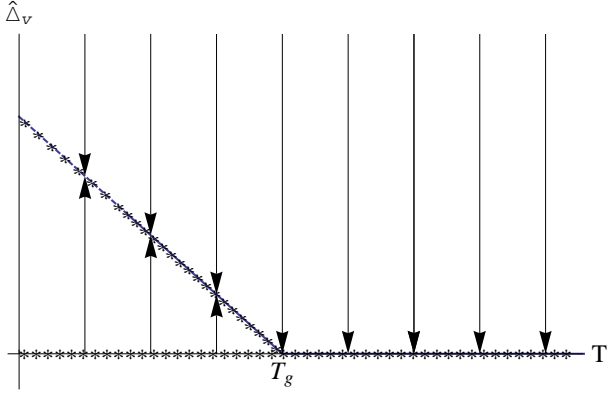}\\
  \caption{(Color online) RG flow of $\hat{\Delta}_v$ in $d=3$: for $T<T_g=6\pi B\lambda/q_0^2$, it flows to
  a fixed line $\hat{\Delta}_v^*=(3-\frac{Tq_0^2}{2\pi B\lambda})$ and for $T>T_g$
  it flows to disorder-free fixed line $\hat{\Delta}_v^*=0$. This corresponds to a glass transition at $T_g$
  from a high-temperature thermal smectic phase to a low-temperature randomly pinned smectic glass.}
  \label{fig:DeltavFlowmap3D_2}
\end{figure}

With these considerable simplifications,
the flow of the surface disorder strengths in Eqs.~(\ref{Delta_fFullFlow}) and (\ref{Delta_vFullflow}) reduce to
\begin{eqnarray}
\partial_{\ell}\Delta_f&=&\Delta_f+\frac{A}{4\Lambda^2}q_0^2 \Delta_v^2,\label{Delta_fFlowSimplified}\\
\partial_{\ell}\Delta_v&=&\left(3-\eta q_0^2\right)\Delta_v-Aq_0^2\Delta_v^2.\label{Delta_vFlowSimplified}
\end{eqnarray}
The dimensionless coupling of the random positional pinning, $\hat{\Delta}_v=Aq_0^2\Delta_v=6\pi^2(a/\xi^v_x)^3$, which is dimensionless measure of the random positional pinning,
flows according to
\begin{equation}
\partial_{\ell}\hat{\Delta}_v=\left(3-\frac{Tq_0^2}{2\pi B\lambda}\right)\hat{\Delta}_v-\hat{\Delta}_v^2.
\label{Delta_v_flow}
\end{equation}
Taking an ansatz $\hat{\Delta}_v(\ell)=f(\ell)e^{(3-\frac{Tq_0^2}{2\pi B\lambda})\ell}$, the solution to this flow equation can be readily obtained:
\begin{equation}
\hat{\Delta}_v(\ell)=\frac{\left(3-\frac{Tq_0^2}{2\pi B\lambda}\right)\hat{\Delta}_v(0)e^{(3-\frac{Tq_0^2}{2\pi B\lambda})\ell}}
 {\left(3-\frac{Tq_0^2}{2\pi B\lambda}\right)+\hat{\Delta}_v(0)\left[e^{(3-\frac{Tq_0^2}{2\pi B\lambda})\ell}-1\right]}.
\label{flow_solution_Delta_v}
\end{equation}
Thus we find that $\hat{\Delta}_v$ decays to the
$\hat{\Delta}_v^*=0$ fixed line for $T>6\pi B\lambda/q_0^2\equiv T_{g}$, and flows to fixed line
\begin{equation}
\hat{\Delta}_v^*=\left(3-\frac{Tq_0^2}{2\pi B\lambda}\right)
\label{Delta_v_fixed}
\end{equation}
for $T<T_{g}$, as illustrated in Fig.~\ref{fig:DeltavFlowmap3D_2}.

We therefore predict that a 3D surface disordered smectic cell
exhibits a Cardy-Ostlund-like \cite{CardyOstlund} phase transition at $T_{g}=6\pi B\lambda/q_0^2$,
between a high temperature thermal smectic phase, in which, at long scales, the surface positional pinning
is averaged away by thermal fluctuations and a low-temperature random-pinned smectic glass phase controlled by the nontrivial $\hat{\Delta}_v^*(T)$ fixed line.  This
surface pinned state (and the associated transition) is quite similar
to the super-rough phase of a crystal surface grown on a random
substrate \cite{Toner_DiVincenzo_sr}, the 1+1 vortex glass phase of
flux-line vortices (confined to a plane) in type II
superconductors \cite{DSFisherBG,VortexGlass}, and 3D smectic
liquid crystal pinned by a random porous environment of, e.g.,
aerogel \cite{RTaerogelPRB}.

The dimensionless coupling of the surface random orientational pinning,
$\hat{\Delta}_f=A q_0^4\Delta_f=8\pi^4a/\xi^f_x$, flows according to
\begin{equation}
\partial_{\ell}\hat{\Delta}_f=\hat{\Delta}_f+ \pi^2 \hat{\Delta}_v^2.
\label{dimensionlessDelta_f}
\end{equation}
From this equation, we learn that even if the bare value of $\hat{\Delta}_f$ is zero, under
coarse graining, at long scales the random-field pinning, $\Delta_v$, generates a nonzero
random orientational pinning $\hat{\Delta}_f(\ell)$.
Since for $T>T_{g}$, $\hat{\Delta}_v\rightarrow 0$, the flow of $\hat{\Delta}_f$ reduces to $\partial_{\ell}\hat{\Delta}_f=\hat{\Delta}_f$, with a trivial solution
$\hat{\Delta}_f(\ell)=\hat{\Delta}_fe^{\ell} $.
It is clear that the random orientational pinning is strongly relevant
even in the high temperature phase, $T>T_g$, with the latter distinguished by the irrelevance of the
positional pinning, $\hat{\Delta}_v\rightarrow 0$.

In contrast, in the low-temperature phase, $T<T_{g}$, the dimensionless strength of surface positional disorder approaches a
nonzero value, $\hat{\Delta}_v\rightarrow \hat{\Delta}_v^*>0$.
Replacing $\hat{\Delta}_v$ in Eq.~(\ref{dimensionlessDelta_f})
by its fixed point value, we obtain
\begin{equation}
\hat{\Delta}_f(\ell)=\Big[\hat{\Delta}_f+\pi^2(\hat{\Delta}_v^*)^2\Big]e^{\ell}
 -\pi^2(\hat{\Delta}_v^*)^2.
\label{Delta_f_lowT_solution}
\end{equation}
Similarly to the bulk disorder result \cite{RTaerogelPRB}, as $\ell\rightarrow\infty$,
this has the same asymptotic behavior as in the high-temperature phase, $\hat{\Delta}_{f0}e^{\ell}$,
but with the enhanced nonuniversal amplitude $\hat{\Delta}_{f0}$ acquiring an additional contribution that scales with  $(T_{g}-T)^2$ for $T<T_{g}$.

\subsection{Matching analysis}
\label{sec:matching}
As discussed earlier, correlation function on scales shorter than the Larkin domain size
can be computed within the random-force approximation in Sec.~\ref{sec:smectic_IntermediateResults}.
However, at scales longer than the domain size, effective pinning grows large
(compared to the elastic energy) and can, therefore, no longer be treated perturbatively
within the random force model.
Nevertheless, we can utilize the RG and matching method to effectively overcome this difficulty.
It allows us to establish a relation between a correlation function at
long scales (beyond the domain size),
which is impossible to calculate directly due to the aforementioned infrared divergences,
to this correlation function at short scales (below the domain size), which can be easily calculated
in a controlled perturbation theory \cite{NelsonRudnick,RudnickNelson,GiamarchiLedoussalFluxLattice_PRL,GiamarchiLedoussalFluxLattice_PRB,
Toner_DiVincenzo_sr,RTaerogelPRB,usFRGPRE} of Sec.~\ref{sec:smectic_IntermediateResults}.

For $T<T_{g}$, the surface positional disorder is relevant.
However, near the Cardy-Ostlund transition $T_{g}$,
$(3-\eta q_0^2)\ll 1$, and the fixed point value of dimensionless
positional pinning is small ($\hat{\Delta}_v^*\ll 1$).
Making use of the matching method, we establish a relation between
replicated correlation functions at short and long scales:
\begin{eqnarray}
&&\hspace{-0.6 cm}C_{\alpha\beta}\left[q_x,q_z;\Gamma_{\qv}(0),\Delta_f(0),\Delta_v(0)\right]\nonumber\\
&\equiv& \frac{\left\langle u_{0}^{\alpha}(q_x,q_z)u_{0}^{\beta}(q_x',q_z')\right\rangle}{\delta^2(\qv+\qv')}\nonumber\\
&=&e^{(1+\omega)\ell}
 C_{\alpha\beta}\left[e^{\ell}q_x,e^{\omega\ell}q_z;\Gamma_{\qv}(\ell),\Delta_f(\ell),\Delta_v(\ell)\right].
\end{eqnarray}
We then choose the value of $\ell=\ln{(\Lambda/q_x)}$ so the rescaled momentum
$q_x$ is at the cutoff $\Lambda$ and therefore the correlation function on the right-hand
side of the above equation can be safely evaluated perturbatively. This gives
\begin{eqnarray}
&&\hspace{-0.6 cm}C_{\alpha\beta}\left[q_x,q_z;\Gamma_{\qv}(0),\Delta_f(0),\Delta_v(0)\right]\nonumber\\
&=&\left(\frac{\Lambda}{q_x}\right)^{1+\omega}
 C_{\alpha\beta}\left[\Lambda,(\Lambda/q_x)^{\omega}q_z;\Gamma_{\qv}^*(\ell^*),\Delta_f(\ell^*),\Delta_v(\ell^*)\right],\nonumber\\
 \label{matching_formula}
\end{eqnarray}
and for $\ell\rightarrow \infty$ corresponds to long scales with $q_x\rightarrow 0$, where $\ell^*$ is big enough so the
disorder strengths on the right-hand side can be replaced by their fixed values.

Safely calculating the correlation function at the cutoff scale on the right-hand side using the Larkin analysis of Sec.~\ref{sec:randomTorque_smectic} we obtain
\begin{eqnarray}
&&\hspace{-0.6 cm} C_{\alpha\beta}\left[\Lambda,(\Lambda/q_x)^{\omega}q_z;\Gamma_{\qv}^*(\ell^*),\Delta_f(\ell^*),\Delta_v(\ell^*)\right]\nonumber\\
&=&\frac{T\delta_{\alpha\beta}}{\Gamma_{\qv}(\ell^*)}+\frac{\Delta_f(\ell^*)\Lambda^2+\Delta_v(\ell^*)}{\Gamma_{\qv}^2(\ell^*)},
\end{eqnarray}
where the $\ell^*$ dependent parameters the recursion relations Eq.~(\ref{flow_Gamma_q})-(\ref{flow_Delta_v}) at
$\ell^*=\ln{(\Lambda/q_x)}$:
\begin{eqnarray}
\Gamma_{\qv}(\ell^*)&=&\left(\frac{\Lambda}{q_x}\right)^3\Gamma_{\qv},\\
\Delta_f(\ell^*)&=&\frac{\hat{\Delta}_f(\ell^*)}{A(\ell^*)q_0^4}
 =\frac{\Lambda}{q_x}\left[\Delta_f+\frac{\pi^2}{Aq_0^4}(\hat{\Delta}_v^*)^2\right],\\
\Delta_v(\ell^*)&=&\frac{\hat{\Delta}_v(\ell^*)}{A(\ell^*)q_0^2}=\frac{\hat{\Delta}_v^*}{Aq_0^2}.
\end{eqnarray}
Here $A(\ell)=A$ is a constant given by Eq.~(\ref{A_1}). In the last two equations
we have taken $q_x$ to be small enough and, therefore, $\ell^*$ large enough so the $\hat{\Delta}_v(\ell^*)$ has flowed to its fixed point value
given in Sec.~\ref{sec:simple_Rf_smectic}, and the low-temperature solution of $\hat{\Delta}_f(\ell)$
is used, with $\ell^*$ large enough that the second term in Eq.~(\ref{Delta_f_lowT_solution}) can be ignored.

Combining these ingredients, we therefore find
\begin{eqnarray}
\hspace{-1 cm}&&\hspace{-0.6 cm}C_{\alpha\beta}\left[q_x,q_z;\Gamma_{\qv},\Delta_f,\Delta_v\right]\nonumber\\
\hspace{-1 cm}&=&\frac{T\delta_{\alpha\beta}}{\Gamma_{\qv}}
+\frac{\Delta_f\left[1+\frac{\xi^f_x}{8\pi^2a}\left(\hat{\Delta}_v^*\right)^2\right]q_x^2
+\frac{q_x^3}{\Lambda^3}\frac{1}{Aq_0^2}\hat{\Delta}_v^*}{\Gamma_{\qv}^2},
\label{matched_correlator}
\end{eqnarray}
in which $\Gamma_{\qv}$ is the bare long-range surface elasticity kernel given in (\ref{Gamma_q}), $\Delta_f$ is the bare strength of
orientational disorder strength, and $\hat{\Delta}_v^*$ is the fixed value given in (\ref{Delta_v_fixed}).
It is easy to see that the $\frac{q_x^3}{\Lambda^3}\hat{\Delta}_v^*$ term arising from the
random positional pinning is negligible as $q_x\rightarrow 0$
compared with the contribution of the surface orientational pinning.
Thus, the smectic phonon correlation functions in the high- and low-temperature phases are
only subtly distinguished by the enhanced amplitude of surface orientational pinning
from $\Delta_f$ to
\begin{equation}
\Delta_f(T)=\Delta_f\left[1+\frac{\xi^f_x}{8\pi^2a}\left(\hat{\Delta}^*_v\right)^2\right].
\end{equation}

\begin{figure}[h]
  \centering
  \includegraphics[width=8 cm]{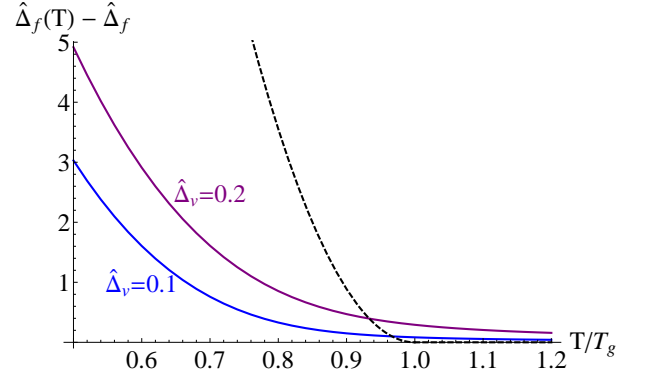}
  \caption{(Color online) The exact correction to the long-scale dimensionless strength of the orientational pinning as a function of temperature, obtained by numerically integrating Eqs.~(\ref{flow_solution_Delta_v}) and (\ref{dimensionlessDelta_f}) for different ``bare'' values of the positional disorder: $\hat{\Delta}_v=0.1$  and $\hat{\Delta}_v=0.2$. As a comparison, the dashed curve is the approximation based on the fixed point value of $\hat{\Delta}_v$ used in Eqs.~(\ref{delta_f_T_definition}), (\ref{delta_f_T}) and (\ref{Delta_f_lowT_solution}).}
  \label{fig:TrueCorrection}
\end{figure}

The above matching calculation gives the long-scale momentum-space surface correlation function
\begin{equation}
C(\qv)=\frac{\Delta_f\delta_f(T)q_x^2}{\Gamma_{\qv}^2},
\end{equation}
in which $\delta_f(T)=\Delta_f(T)/\Delta_f$, as given in Eq.~(\ref{delta_f_T}),
distinguishes the high- ($T>T_g$) and low- ($T<T_g$) temperature phases.

Moreover, the correction to the orientational disorder at long scales, $\Delta_f(T)-\Delta_f$, that we approximated in Eq.~(\ref{dimensionlessDelta_f}) by the fixed point value of the positional disorder which is zero for $T>T_g$ and quadratic in $(T_g-T)$ for $T<T_g$, is quantitatively inaccurate. The exact solution of the flow equation for $\hat{\Delta}_f$ given in Eq.~(\ref{dimensionlessDelta_f}) with $\hat{\Delta}_v(\ell)$ in (\ref{flow_solution_Delta_v}) leads to a long scale correction of $\hat{\Delta}_f$ in the correlation function that depends on the ``bare'' value of $\hat{\Delta}_v$ and is  non-zero for $T>T_g$, as shown in Fig.~\ref{fig:TrueCorrection}.
This difference does not change the qualitative behavior of the system, e.g., the experimental feature of the x-ray scattering peaks described in Sec.~\ref{sec:X-ray}. However, it may be important for quantitative comparison with experiments.

\subsection{Surface correlation function at long scales}
\label{sec:long_scale_correlation}
\begin{figure}[htbp]
\centering
  \includegraphics[width=7 cm]{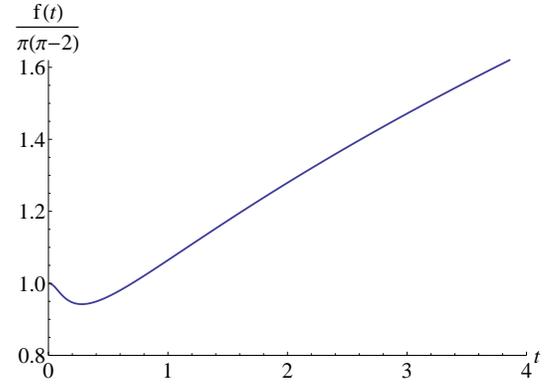}\\
  \caption{
 (Color online) The value of dimensionless function $f(t)$ with $t=\lambda z/x^2$, which has limiting behaviors given in Eq.~(\ref{f_t_limits}).
}
  \label{fig:f_t}
\end{figure}
As discussed in Appendix \ref{app:full_long_scale_correlation}, the full behavior of the real-space long scale correlations can be summarized as
\begin{equation}
C(x,z)=\frac{\Delta_f(T)}{4\pi^2B^2\lambda^3}xf(t),
\label{Cxz_long_full}
\end{equation}
in which $t=\frac{\lambda z}{x^2}$ and $f(t)$ (plotted in Fig.~\ref{fig:f_t}) is given by
\begin{eqnarray}
f(t)&=&\pi(\pi-2)+\pi\left[2+2 e^{-\frac{1}{4t}}\sqrt{\pi t}-\pi\mathrm{erfc}\left(\frac{1}{2\sqrt{t}}\right)\right]\nonumber\\
&&+\frac{\sqrt{2}}{3\sqrt{t}}\bigg[3t\Gamma\left(-\frac{1}{4}\right)\Gamma\left(\frac{3}{4}\right) \mathrm{_1F_2}\left(-\frac{1}{4};\frac{1}{2},\frac{5}{4};\frac{1}{64t^2}\right)\nonumber\\
&&-\Gamma\left(\frac{1}{4}\right)\Gamma\left(\frac{5}{4}\right) \mathrm{_1F_2}\left(\frac{1}{4};\frac{3}{2},\frac{7}{4};\frac{1}{64t^2}\right)\bigg],
\label{f_t}
\end{eqnarray}
where $\mathrm{_1F_2}(a_1;b_1,b_2;x)$ is the generalized hypergeometric function. The limiting behavior of $f(t)$ is given by
\begin{equation}
f(t)\approx\left\{\begin{array}{ll}
\pi(\pi-2)-4\pi t^2,&\mbox{ for } t\ll 1,\\
2.64 \sqrt{t},&\mbox{ for } t\gg 1.
\end{array}\right.
\label{f_t_limits}
\end{equation}

\begin{figure}[htbp]
\centering
  \includegraphics[width=8.5 cm]{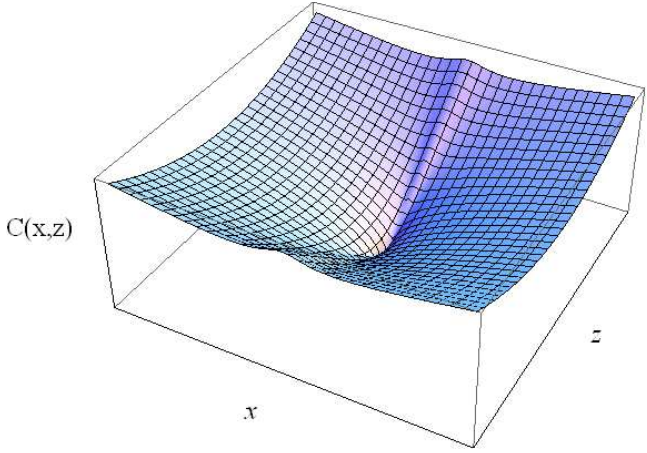}\\
  \caption{(Color online) The long-scale surface phonon correlation function $C(x,z)$,
  exhibiting limiting behavior $C(x,z)\sim x$ and $\sqrt{z}$ for small and
  large values of $t=\lambda z/x^2$, respectively.
  }
  \label{fig:Cxz_regions}
\end{figure}

\begin{figure}[htbp]
\centering
  \includegraphics[width=7 cm]{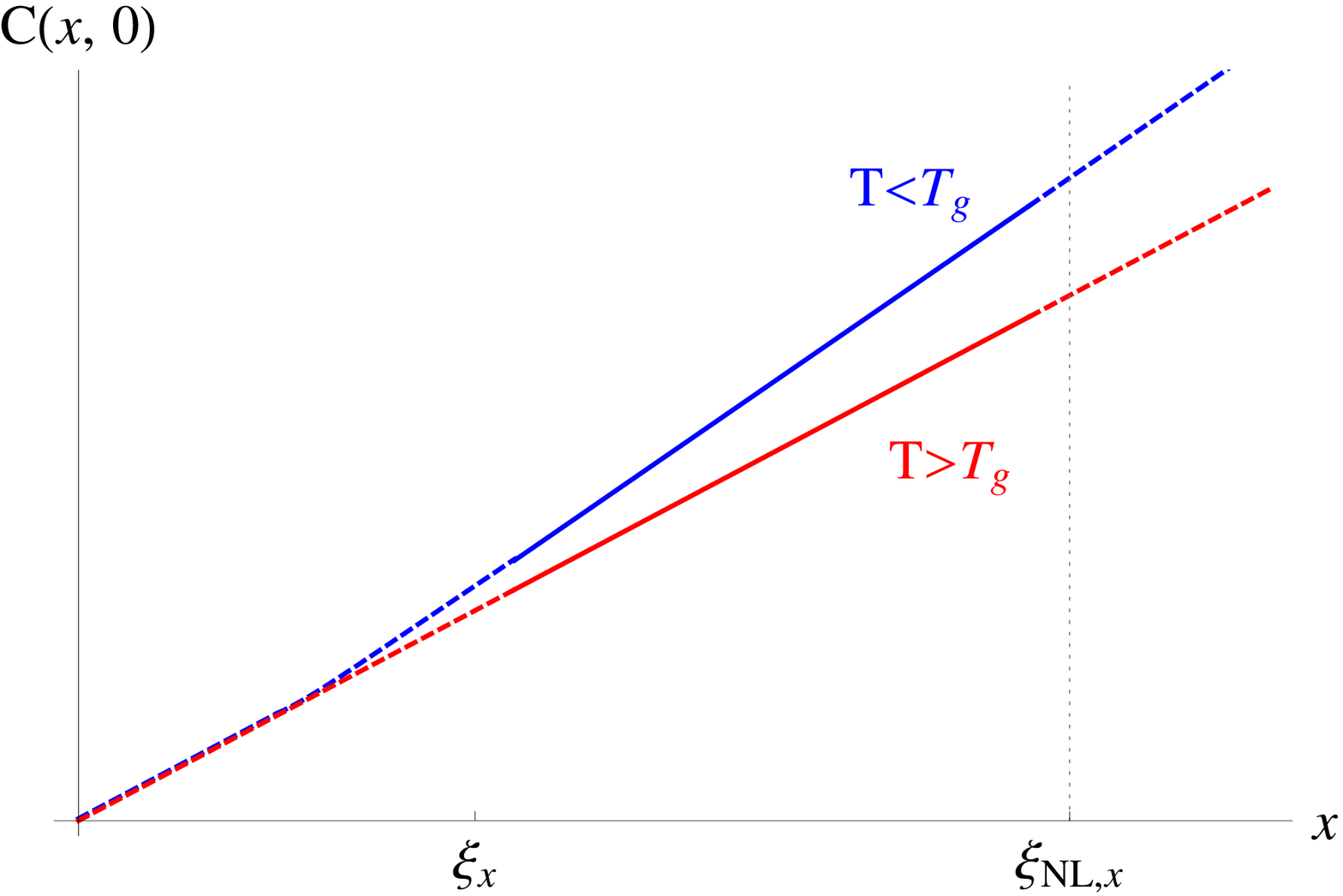}\\
  \includegraphics[width=7 cm]{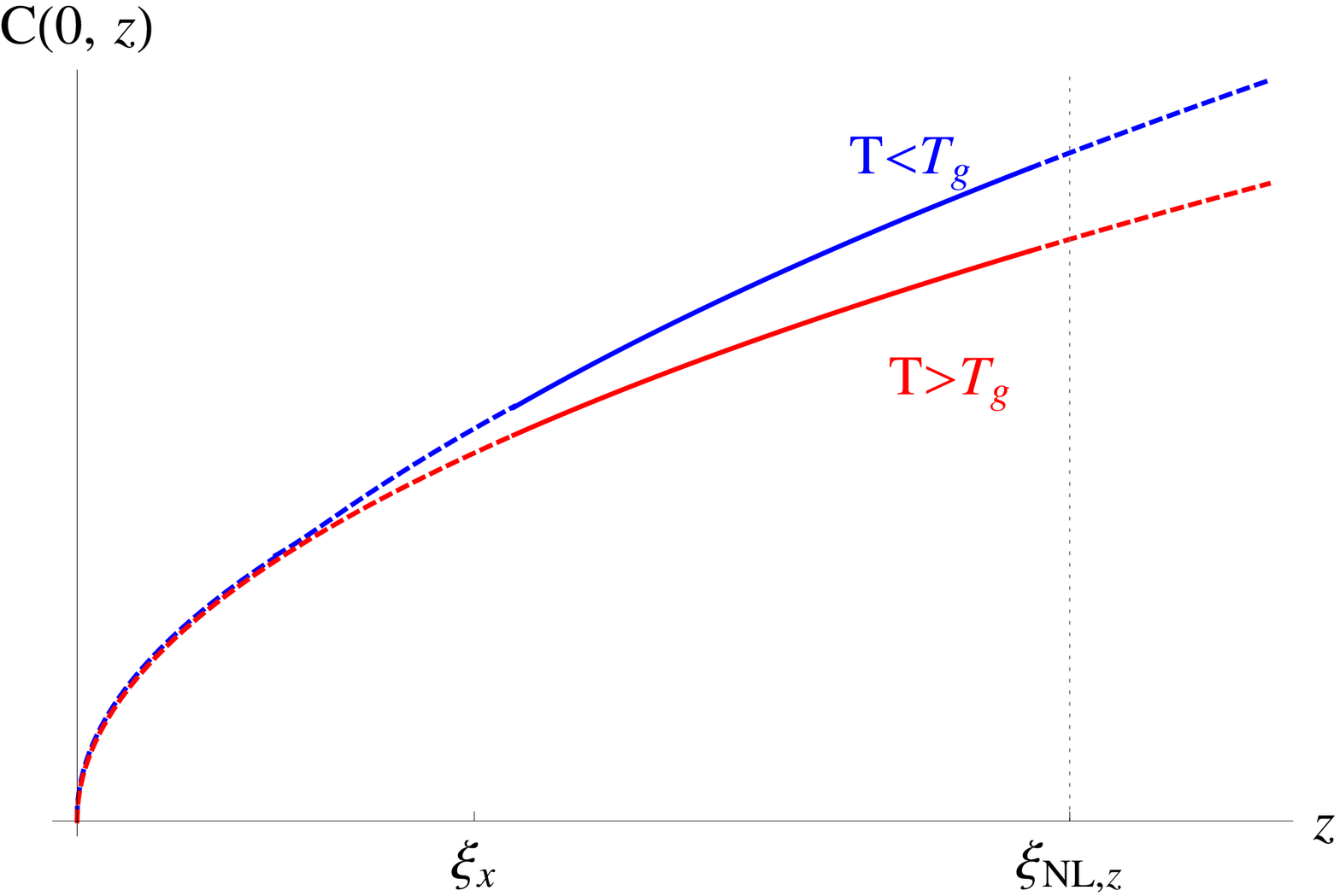}\\
  \caption{(Color online) The two-point smectic phonon correlation function  along $x$ and $z$ axes on the random substrate at $y=0$.
   For temperature $T>T_g$ the correlation is the same as the short-scale result,
   (\ref{correlation_orientational_shortScale}), while below $T_g$ it is modified
   by the enhanced amplitude $\delta_f(T)$, [(\ref{delta_f_T_definition}) and (\ref{delta_f_T})].
  }
  \label{fig:Cf_full}
\end{figure}

In the $(x,z)$ plane, the above limits correspond to regions near the $x$ and $z$ axes, where simple asymptotic form of the correlation function $C(x,z)$ is available, as shown in Fig.~\ref{fig:Cxz_regions}.
In real space along $x$, it is given by
\begin{eqnarray}
C(x,0)&\approx& 2\frac{(\pi-2)\Delta_f(T)}{4\pi^2B^2\lambda^3}\frac{\pi x}{2} \nonumber\\
&=& \pi a^2\delta_f(T)x/\xi_x^f=\pi a^2x/\xi_x(T),
\label{Cx0_long}
\end{eqnarray}
where $\xi_x(T)=\xi_x^f/\delta_f(T)$. Along $z$ we instead find
\begin{eqnarray}
C(0,z)&\approx& 2a^2\delta_f(T)\sqrt{\frac{\pi}{2}}\sqrt{\frac{z}{\xi_z^f}} \nonumber\\
&=&\sqrt{2\pi}a^2\sqrt{z/\xi_z(T)},
\label{C0z_long}
\end{eqnarray}
where $\xi_z(T)=\xi_z^f/\delta_f^2(T)$. Thus at long scales the correlation function
asymptotically scales identically to that in the Larkin's perturbative regime, (\ref{correlation_orientational_shortScale}), but with an amplitude enhanced below $T_g$, as illustrated in Fig.~\ref{fig:Cf_full}.

Within the harmonic elastic theory (i.e., neglecting elastic nonlinearities and dislocations) we expect the above predictions to be asymptotically exact. We explore the role of elastic nonlinearities in Sec.~\ref{sec:nonlinearity}, where we show that they indeed are expected to modify our predictions on scales longer than the nonlinear length scale $\xi_{NL}$ but leave the challenging problem of their study to future research.

\section{Experimental predictions}
\label{sec:Experiments}
We next explore the experimental signatures of our predictions for the surface pinned smectic cell, focusing on polarized light microscopy and x-ray scattering.

\subsection{Cross-polarized light microscopy}
\label{sec:microscopy_smectic}
In smectic cells with weak random substrate pinning, the molecular orientation
is expected to vary slowly in the bulk, and under standard conditions we expect the
Mauguin limit be valid with polarization of light following local optic axis.
Under such conditions, a convenient geometry to probe the substrate-driven smectic distortion is
that of a crossed polarizer-analyzer pair, with the polarizer aligned with a well rubbed back
substrate. In this geometry, the transmission vanishes in the absence of smectic distortions. Thus,
transmitted light directly images the optic axis distortions on the front heterogeneous substrate \cite{usFRGPRL,usFRGPRE}.
This provides a sensitive way to investigate statistics of the surface smectic layer distortions $\partial_x u_0(x,z)$
in a smectic cell weakly pinned by a dirty substrate.

Here, we only consider smectic order distortions on the short scales (within the Larkin regime) of thick smectic cells
where the effects are most pronounced. The subtle long scale asymptotic behavior predicted here is unlikely to be accessible in current experiments, as it requires currently unavailable detailed quantitative analysis. Short-scale treatment below can also be readily generalized to long scales.
We take the back substrate to be rubbed and, therefore, with the optic (nematic) axis well aligned along the $z$ axis,
while near the front $y=0$ substrate the smectic order is subject to random orientational and positional pinning of the dirty substrate.

As alluded to above, in this weakly pinned Mauguin limit optical transmission
is determined by the azimuthal distortions of the optic axis (layer normal) orientation $\phi_0(x,z)=\phi(x,y,z)|_{y=0}$ on the random substrate.
In this geometry with the polarizer at the back substrate along the $z$
axis, and analyzer at the front substrate along the $x$ axis, the transmitted optical field is given by
\begin{equation}
E_{out}(x,z)=E_{in}\sin{\phi_0(x,z)}\approx -E_{in}\frac{\partial u_0(x,z)}{\partial x}.
\end{equation}
This gives the polarized light microscopy intensity,
\begin{equation}
I_{out}(x,z)\approx I_{in} |\partial_x u_0(x,z)|^2.
\end{equation}
Thus, the average transmitted intensity is proportional to the variance
of the smectic layer normal fluctuations on the heterogeneous substrate,
\begin{equation}
\frac{\overline{I}_{out}}{I_{in}}\approx\overline{ |\partial_x u_0(x,z)|^2}.
\end{equation}
From the local intensity we can also obtain a correlation function of the transmitted intensity, related to spatial smectic phonon correlations:
\begin{eqnarray}
&&\hspace{-0.6 cm} C^{I}(x-x',z-z')\nonumber\\
&=&\overline{I_{out}(x,z)I_{out}(x',z')}\nonumber\\
&\approx&I_{in}^2\overline{ |\partial_x u_0(x,z)|^2 |\partial_{x'} u_0(x',z')|^2} \nonumber\\
&\approx& \overline{I}_{out}^2\left\{1+2\left[\frac{\overline{\partial_x u_0(x,z)\partial_{x'} u_0(x',z')}}{\overline{ |\partial_x u_0(x,z)|^2}}\right]^2\right\},
\label{I_I_correlation_definition}
\end{eqnarray}
where we utilized Gaussian approximation valid at short scales.

\subsubsection{Surface random orientational pinning}

For a dirty substrate where the orientational pinning dominates over positional disorder, the transmitted intensity averaged over the Larkin domain can be straightforwardly calculated and is given by
\begin{eqnarray}
\frac{\overline{I}_{out}}{I_{in}}&=&\overline{|\partial_x u_0(x,z)|^2} \nonumber\\
&=&\Delta_f\int\frac{dq_x dq_z}{(2\pi)^2}q_x^4 \frac{1}{\Gamma_{\qv}^2} \nonumber\\
&=&\frac{(\pi-2)\Delta_f}{4\pi^2 B^2\lambda^3}\int_{1/\xi_x^L}^{1/a} dq_x \nonumber\\
&\approx&\frac{\Delta_f}{c B^2\lambda^3a}.
\end{eqnarray}
The smectic bulk modulus dependence can, in principle, be probed through its well-understood temperature dependence
\cite{BelliniScience,CriticalBehaviorNSmA}.

In the regime of dominant orientational disorder, the transmission intensity-intensity correlation function along the $x$ axis, $C^I(x,0)$ can be straightforwardly computed by noticing that
\begin{eqnarray}
\hspace{-0.6 cm}\frac{\overline{ \partial_x u_0(x+x',z)\partial_{x'} u_0(x',z)}}{\overline{ |\partial_x u_0(x,z)|^2}}
&=&a\int_{0}^{1/a}\cos{(q_xx)} dq_x \nonumber\\
&=&\frac{\sin{(x/a)}}{x/a}.
\end{eqnarray}
and thus we find
\begin{equation}
C^{I}(x,0)=\overline{I}_{out}^2\left\{1+2\left[\frac{\sin{(x/a)}}{x/a}\right]^2\right\}.
\label{Icorr_x_tilt}
\end{equation}

Similarly, the light intensity-intensity correlation along the $z$ axis is given by
\begin{equation}
C^I(0,z)=\overline{I}_{out}^2\left\{1+2 \left[\sqrt{\frac{\pi a}{2z}}C_F\left(\sqrt{\frac{2z}{\pi a}}\right)\right]^2\right\},
\label{Icorr_z_tilt}
\end{equation}
where $C_F(z)=\int_0^z \cos{(\pi t^2/2)}dt$ is the Fresnel integral.

These correlation functions probe surface orientational order, that for weak pinning is long range ordered, with structure
only on the scale of the layer spacing $a$.

\subsubsection{Surface random positional pinning}
For cells dominated  by positional pinning the average transmission intensity is given by
\begin{eqnarray}
\frac{\overline{I}_{out}}{I_{in}}&=&\overline{ |\partial_xu_0(x,z)|^2} \nonumber\\
&=&\frac{(\pi-2)\Delta_v}{4\pi^2 B^2\lambda^3}\int_{1/\xi_x}^{1/a}\frac{ dq_x}{q_x^2} \nonumber\\
&\approx&\frac{(\pi-2)\Delta_v}{4\pi^2 B^2\lambda^3}\xi_x,
\end{eqnarray}
where we cutoff the average by the smectic domain size, $\xi_x\simeq\xi_x^v$, given in Eq.~(\ref{LLrandomPosition}).

The intensity-intensity correlation function along the $x$ axis calculated from (\ref{I_I_correlation_definition}) is given by:
\begin{eqnarray}
C^{I}(x,0)&\approx&\overline{I}_{out}^2\left\{1+2\left[\frac{1}{\xi_x}\int_{1/\xi_x}^{1/a}
  \frac{\cos{(q_xx)}}{q_x^2}dq_x \right]^2\right\} \nonumber\\
  &\approx&\overline{I}_{out}^2\left[1+2\left(1-\frac{\pi x}{2\xi_x}\right)^2\right],
\label{Icorr_x_positional}
\end{eqnarray}
where we have extended the upper bound of integral to infinity,
\begin{eqnarray}
&&\hspace{-0.6 cm}\frac{1}{\xi_x}\int_{1/\xi_x}^{1/a}\frac{\cos{(q_xx)}}{q_x^2}dq_x \nonumber\\
&\approx& -\frac{\pi x}{2\xi_x}+\cos{(\frac{x}{\xi_x})}+\frac{x}{\xi_x}\mathrm{Si}{(\frac{x}{\xi_x})} \nonumber\\
&\approx& 1-\frac{\pi x}{2\xi_x},
\end{eqnarray}
valid for $0<a \ll x\ll \xi_x$ and where $\mathrm{Si}(z)=\int_0^z \frac{\sin{(t)}}{t}dt$
is the sine integral function.

Similarly, at constant $x$ along the layer normal, the intensity correlation function is given by
\begin{eqnarray}
C^I(0,z)&=&\overline{I}_{out}^2\left\{1+2 \left[\frac{1}{2\sqrt{\xi_z}}\int_{1/
\xi_z}^{1/a} \frac{\cos{(q_z z)}}{q_z^{3/2}}dq_z \right]^2\right\} \nonumber\\
&\approx& \overline{I}_{out}^2\left[1+2 \left(
1-\sqrt{\frac{\pi z}{2\xi_z}}  \right)^2\right] \nonumber\\
&=&\overline{I}_{out}^2\left(3-4\sqrt{\frac{\pi z}{2\xi_z}}+\frac{\pi z}{\xi_z} \right),
\label{Icorr_z_positional}
\end{eqnarray}
valid for  $0<a \ll z\ll \xi_z$. The intensity-intensity correlations decay to their asymptotic (transmission) value on the scale of smectic domains $\xi_{x,z}$.

In fact, we expect the intensity-intensity correlations predicted by Eqs.~(\ref{Icorr_x_tilt}), (\ref{Icorr_z_tilt}), (\ref{Icorr_x_positional}) and (\ref{Icorr_z_positional}) to be observable in crossed polarized light microscopy on a thick smectic cell with a single dirty substrate.

\subsection{X-ray scattering and characteristic lengths}
\label{sec:X-ray}
Experimental studies of a smectic cell \cite{ClarkSmC,CDJonesThesis} observed that the smectic x-ray scattering peak broadens significantly as temperature lowers towards the A-C transition. In this section, we analyze the x-ray scattering utilizing the smectic cell model and its analysis from earlier sections.

Following standard treatment of x-ray scattering of a smectic liquid crystal \cite{Caille_french,ClarkBellocq_Xray_Sm,ChenToner_biaxial},
for a wave vector $k_z$ near $k_z=\pm q_0=\pm 2\pi/a$, the scattering intensity is proportional to the structure function
$S(\vec{k})$, which is the Fourier transform of the density-density correlation function. The smectic density $\rho(\vec{r})$ periodic along $z$ can be represented by
\begin{eqnarray}
\rho(\vec{r}) = \sum_{n} \rho_n e^{i q_n [z + u(\vec{r})]},
\end{eqnarray}
where $q_n = nq_0 = 2n\pi/a$.

From this we obtain:
\begin{equation}
S(\vec {k}) \propto \sum_{n}\rho_n^2 \int d\vec{r} e^{-i\vec{k}\cdot\vec{r}}e^{i q_n z}
\langle e^{i q_n [u(\vec{r} ) - u(0)]}\rangle
\end{equation}
Focusing on the lowest order $n=1$ peak at $q_0$, and using a Gaussian approximation for the statistics of $u$ distortions (expected to be valid on short scales), we find
\begin{equation}
S(\vec{k})\propto \int d\vec{r} e^{-i (k_z-q_0) z - i\vec{k}_{\perp}\cdot \vec{r}_{\perp}}
e^{-\frac{q_0^2}{2} \overline{\langle[u(\vec{r} ) - u(0)]^2\rangle}}.
\end{equation}

For smectic distortions confined to the vicinity of the random substrate \cite{commentXray},
the structure function reduces to
\begin{equation}
S(k_x, \tilde{k}_z) =  \int dx dz e^{-i \tilde{k}_z z - ik_x\cdot x}
e^{-\frac{q_0^2}{2} C(x,z)},
\end{equation}
in which $\tilde{k}_z=k_z-q_0$.

A detailed analysis of the peak requires the full form of the phonon correlation function $C(x,z)$,
which is only available at long scale but is somewhat complicated to analyze. However, as discussed by Chen and Toner
\cite{ChenToner_biaxial}, the behavior near and away from the Bragg peak is respectively controlled by long- and short-length-scale correlations.

For $k_x<\sqrt{\lambda^{-1}\tilde{k}_z}$,
the main contribution to the
structure function comes from $x>\sqrt{\lambda z}>\xi_x$; thus, approximation $C(x,z)\approx \pi a^2 x/\xi_x(T)$ can be used.
At finite $k_z$ we thereby have
\begin{eqnarray}
S(k_x,\tilde{k}_z)\sim S_x(k_x)&=&\int dx e^{-ik_xx} e^{-2\pi^3 x/\xi_x(T)} \nonumber\\
&=&\frac{4\pi^3/\xi_x(T)}{k_x^2+4\pi^6[\xi_x(T)]^{-2}},
\label{Sx_kx}
\end{eqnarray}
where the peak width is given by $\Delta k_x=2\pi^3 [\xi_x(T)]^{-1}\propto \delta_f(T)$.

For $\sqrt{\lambda^{-1}\tilde{k}_z}<k_x$,
similarly, a long-scale phonon correlation can be approximated as
$C(x,z)\approx \sqrt{2\pi} a^2 \sqrt{z/\xi_z(T)}$, and at finite $k_x$ the structure function
\begin{widetext}
\begin{eqnarray}
S(k_x,\tilde{k}_z)&\sim& S_z(\tilde{k}_z)\nonumber\\
&=&\int dz e^{-i\tilde{k}_zz}e^{-2\pi^2\sqrt{2\pi}\sqrt{\lambda z}/\xi_x(T)} \nonumber\\
&=&\frac{2\pi^3\xi_z(T)}{\left[\tilde{k}_z\xi_z(T)\right]^{3/2}}
\left\{\cos{\left[\frac{2\pi^5}{\tilde{k}_z\xi_z(T)}\right]}
\left[1-2\mathrm{F_C}{\left(\frac{2\pi^2}{\sqrt{\tilde{k}_z\xi_z(T)}}\right)}\right]+
\sin{\left[\frac{2\pi^5}{\tilde{k}_z\xi_z(T)}\right]}\left[1-2\mathrm{F_S}{\left(\frac{2\pi^2}{\sqrt{\tilde{k}_z\xi_z(T)}}\right)}\right]
\right\}\nonumber\\
&\approx& \frac{\xi_z(T)}{2\pi^5}\left\{1-\frac{15\left[\tilde{k}_z\xi_z(T)\right]^2}{16\pi^{10}}\right\}, \mbox{ for } \tilde{k}_z\xi_z(T)\ll 1,
\label{sz_kz}
\end{eqnarray}
\end{widetext}
in which $\mathrm{F_C}(x)$ and $\mathrm{F_S}(x)$ are Fresnel integrals.
Thus, we find that the $S_z(\tilde{k}_z)$ peak width is given by $\Delta\tilde{k}_z\propto[\xi_z(T)]^{-1}\propto \delta_f^2(T)$.

\begin{figure}
  \centering
  \includegraphics[width=8.5 cm]{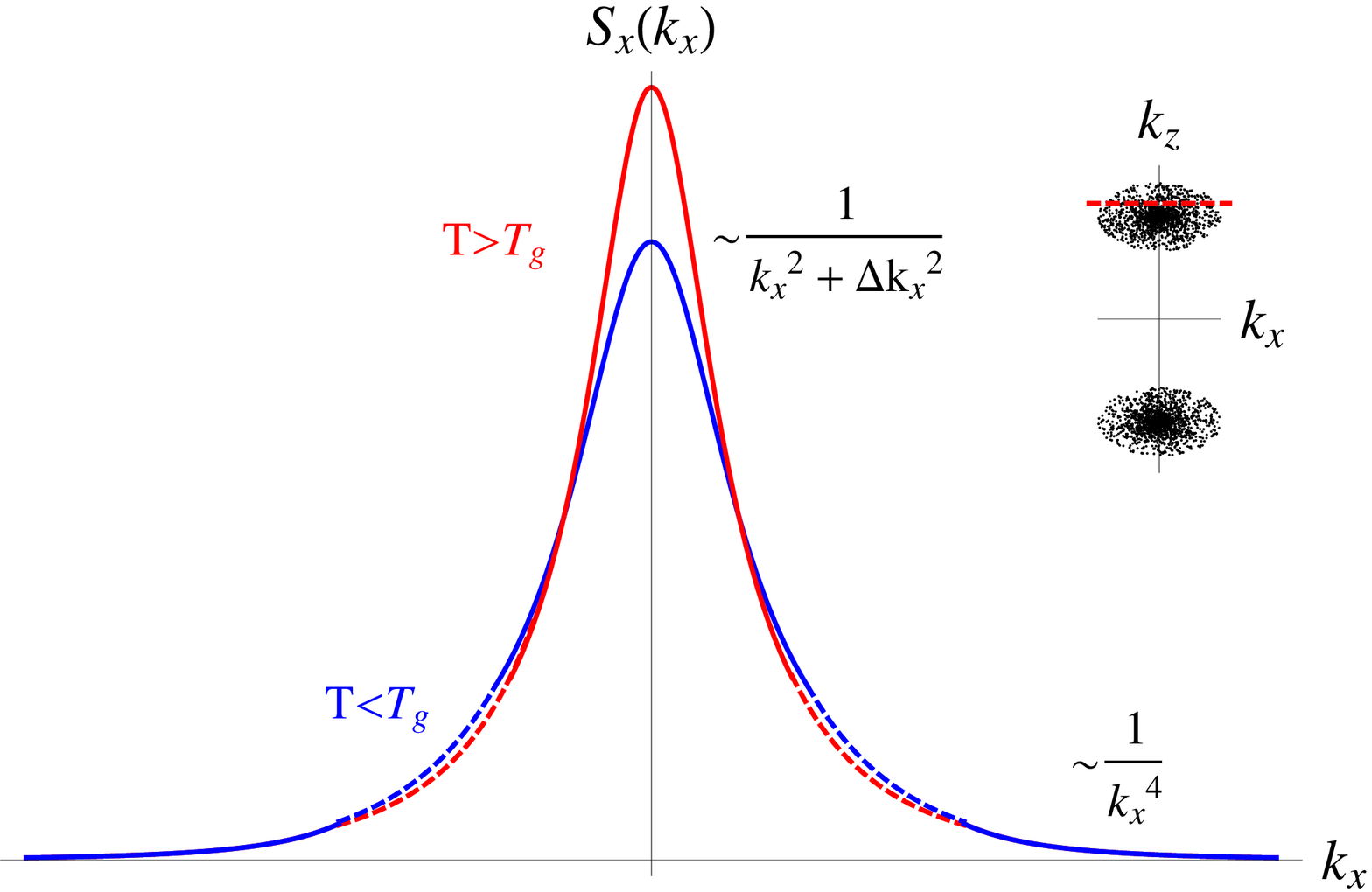}\\
  \includegraphics[width=8.5 cm]{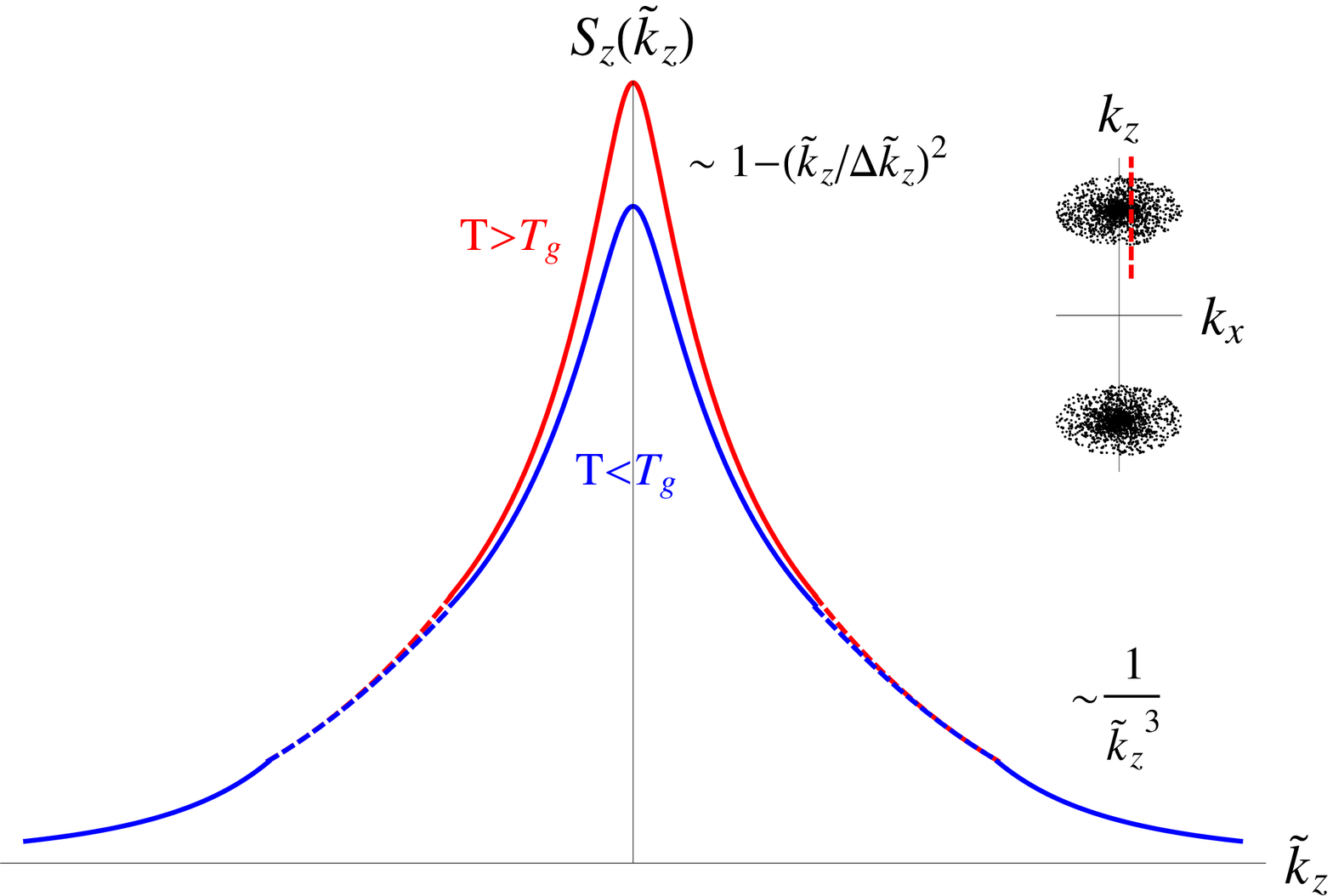}
  \caption{(Color online) Structure function from Eqs.~(\ref{Sx_kx}), (\ref{sz_kz}), and (\ref{Sxz_large_k}), showing the x-ray scattering peak profile at finite $k_x$ and $\tilde{k}_z$ for $T>T_g$ and $T<T_g$,
 in which the position of the plotted profile function is shown on the Bragg peak by a dashed line in the inset of each figure.
 The peaks broaden at temperatures below $T_g$. }\label{fig:BraggPeak}
\end{figure}

Above details of the peak profile are summarized in Fig.~\ref{fig:BraggPeak}.
We thus predict that the x-ray peak widths along $k_x$ and $\tilde{k}_z$ will broaden with reduced temperature below the CO phase transition temperature $T_g$.
This prediction for the finite smectic order characterized by $\xi_{x,z}(T)$ may
have already been observed as the aforementioned precipitous x-ray peak broadening in
cooled smectic liquid crystal cells with a random substrate~\cite{CDJonesThesis,ClarkSmC}. More detailed experimental studies are necessary to verify this conjecture.

Away from the Bragg peak, the structure function is dominated by a short-scale correlation,
well approximated by $e^{-\frac{q_0^2}{2}C(x,z)}\approx 1-\frac{q_0^2}{2}C(x,z)$. At large $k_x$ and $\tilde{k}_z$ we thus obtain
\begin{equation}
S(k_x,\tilde{k}_z)\approx \frac{q_0^2}{2}C_{\Delta}(k_x,\tilde{k}_z),
\end{equation}
with limits
\begin{eqnarray}
&&\hspace{-0.6 cm}S(k_x,\tilde{k}_z)\nonumber\\
&\approx& \frac{q_0^2}{2}
\frac{\Delta_fk_x^2+\Delta_v}{2B^2\lambda}
\frac{\sqrt{\lambda^2k_x^4+\tilde{k}_z^2}-\lambda k_x^2}{\tilde{k}_z^2(\lambda^2k_x^4+\tilde{k}_z^2)}\nonumber\\
&\approx&\left\{\begin{array}{ll}
\frac{\pi^2\Delta_f}{2B^2\lambda^4a^2}\frac{1}{k_x^4}&\mbox{ for } k_x\gg \sqrt{\lambda^{-1}\tilde{k}_z},\\
\frac{\pi^2(\Delta_fk_x^2+\Delta_v)}{B^2\lambda a^2}\frac{1}{\tilde{k}_z^3}&\mbox{ for } k_x\ll\sqrt{\lambda^{-1}\tilde{k}_z},\\
\end{array}\right.
\label{Sxz_large_k}
\end{eqnarray}
as illustrated in Fig.~\ref{fig:BraggPeak}.

\section{Limits of validity}
\label{sec:validity_of_theory}

\subsection{Analysis of elastic nonlinearity}
\label{sec:nonlinearity}
So far we have limited our analysis (see Sec.~\ref{sec:smecticmodel}) of a smectic cell with a random substrate to
a \textit{harmonic} elastic description neglecting the
nonlinear elasticity. However, these are known to be important
in smectic systems in the presence of thermal \cite{ClarkMeyer,GrinsteinPelcovits}
and quenched random \cite{RTaerogelPRB,RTAnomalousPRL} fluctuations.
In this section, we formulate the treatment of these nonlinearities and estimate their effects in surface disordered smectic cells.

A complete description that includes elastic nonlinearities and random surface pinning (the latter limited to a Larkin harmonic description) is captured by the replicated energy functional
\begin{widetext}
\begin{eqnarray}
H^{(r)}&=&H_{bulk}+H_{pin}\nonumber\\
&=&\sum_{\alpha}\int d^{d-2}xdz\int_0^{\infty} dy
\left\{\frac{K}{2}\left(\nabla^2_\perp u^{\alpha}\right)^2 +\frac{B}{2}\left[\partial_z  u^{\alpha}-\frac{1}{2}\left(\nabla_{\perp}u^{\alpha}\right)^2\right]^2\right\}\nonumber\\
&&-\frac{1}{2T}\sum_{\alpha,\beta}^n \int d^{d-2}xdzdy\delta(y)
\left[\Delta_f \partial_x u_0^{\alpha}(x,z)\partial_x u_0^{\beta}(x,z)
+ \Delta_v u_0^{\alpha}(x,z)u_0^{\beta}(x,z)\right] \nonumber\\
&=&\sum_{\alpha}\int d^{d-2}xdz\int_0^{\infty} dy
\left[\frac{K}{2}\left(\nabla^2_\perp u^{\alpha}\right)^2 +\frac{B}{2}\left(\partial_z  u^{\alpha}\right)^2\right]\nonumber\\
&&-\frac{1}{2T}\sum_{\alpha,\beta}^n \int d^{d-2}xdzdy\delta(y)
\left[\Delta_f \partial_x u_0^{\alpha}(x,z)\partial_x u_0^{\beta}(x,z)
+ \Delta_v u_0^{\alpha}(x,z)u_0^{\beta}(x,z)\right]\nonumber\\
&&-\sum_{\alpha}\int d^{d-2}xdzdy\frac{B}{2}\left(\partial_zu^{\alpha}\right)\left(\nabla_{\perp} u^{\alpha}\right)^2
+\sum_{\alpha}\int d^{d-2}xdzdy\frac{B}{8}\left(\nabla_{\perp}u^{\alpha}\right)^4 \\
&\equiv& H_0+H_1.
\end{eqnarray}
\end{widetext}

To assess the influence of $H_1$ nonlinearities we perform $(d-1)$ dimensional coarse-graining at fixed $y$,  perturbatively in $H_1$, which allows us to utilize
the harmonic correlation function derived in Sec.~\ref{sec:smectic_IntermediateResults},
\begin{eqnarray}
&&\hspace{-0.6 cm}C(q_x,q_z,y) \nonumber\\
&=&\overline{\langle |u(q_x,q_z,y)|^2 \rangle} \nonumber\\
&=&\Big(\frac{T}{\Gamma_{\qv}}\delta_{\alpha,\beta}+\frac{\Delta_f q_x^2+\Delta_v}{\Gamma_{\qv}^2}\Big)\phi^2(\bfq,y),
\label{C_qx_qz_y}
\end{eqnarray}
where $\phi(\bfq,y)$ is given in Eq.~(\ref{uqy}).
The lowest nontrivial correction is given by
\begin{widetext}
\begin{eqnarray}
\delta H&=& -\frac{1}{2T}\left\langle H_1^2\right\rangle_>^c
=-\frac{B^2}{4T}\sum_{\alpha,\beta}\int d^{d-2}xdydzd^{d-2}x'dy'dz' \left[\partial_zu^{\alpha}(x,y,z)\right]\left[\partial_zu^{\beta}(x',y',z')\right]\nonumber\\
&&\times \left(\overline{\left\langle \left[\nabla_{\perp}u^{\alpha}(x,y,z)\right]\left[\nabla_{\perp}'u^{\beta}(x',y',z')\right]\right\rangle_>}\right)^2,
\label{dudududu}
\end{eqnarray}
\end{widetext}
where the correlator of short-scale modes is given by [see (\ref{C_qx_qz_y})]
\begin{eqnarray}
&&\hspace{-0.7cm}\overline{\left\langle \left[\partial_xu^{\alpha}(x,y,z)\right]\left[\partial_{x'}u^{\beta}(x',y',z')\right]\right\rangle_>}\nonumber\\
&=&\int \frac{d^{d-2}k_x dk_z}{(2\pi)^{d-1}} \Big(\frac{T\delta_{\alpha\beta}}{\Gamma_{\bf k}}+\frac{\Delta_fk_x^2+\Delta_v}{\Gamma_{\bf k}^2}\Big)k_x^2
e^{ik_x(x-x')}\nonumber\\
&&\times e^{ik_z(z-z')}\phi({\bf k},y)\phi({\bf k},y').
\end{eqnarray}
Redefining variables $X=(x+x')/2$, $\delta x=x-x'$, $Z=(z+z')/2$, and $\delta z=z-z'$,
the dominant contribution is given by
\begin{widetext}
\begin{equation}
\delta H  = \frac{1}{2} \sum_{\alpha}\int d^{d-2}XdZ \int dy dy' \delta B(y,y') \left[\partial_z u^{\alpha}(X,y,Z)\right] \left[\partial_z u^{\alpha}(X,y',Z)\right],
\label{nonlinear_correction_formula}
\end{equation}
where
\begin{eqnarray}
\delta B(y,y')&=&-\frac{B^2}{2T}\int d^{d-2}\delta x d\delta z \left(\overline{\langle\left[\nabla_{\perp}u^{\alpha}(x,y,z)\right]\left[\nabla_{\perp}'u^{\beta}(x',y',z')\right]\rangle_> }\right)^2 \nonumber\\
&=&-\frac{B^2}{2T}\int \frac{d^{d-2}k_x}{(2\pi)^{d-2}}\frac{dk_z}{2\pi}
\Big(\frac{T}{\Gamma_{\bf k}}+\frac{\Delta_f k_x^2+\Delta_v}{\Gamma_{\bf k}^2}\Big)^2k_x^4
\phi^2({\bf k},y)\phi^2({\bf k},y') \\
&\approx& -B^2\int \frac{d^{d-2}k_x}{(2\pi)^{d-2}}\frac{dk_z}{2\pi}\frac{\Delta_f k_x^2+\Delta_v}{\Gamma_{\bf k}^3}
k_x^4 \phi^2({\bf k},y)\phi^2({\bf k},y').
\end{eqnarray}
\end{widetext}

Fourier transform with respect to $X$ and $Z$, (\ref{nonlinear_correction_formula}) gives
\begin{eqnarray}
\delta H  &=& \frac{1}{2} \sum_{\alpha}\int \frac{d^{d-2}q_xdq_z}{(2\pi)^{d-1}} \int dy dy' \delta B(y,y') q_z^2 |u_0^{\alpha}(\bfq)|^2 \nonumber\\ &&\phi(\bfq,y)\phi(\bfq,y') \nonumber\\
&\equiv& \frac{1}{2} \sum_{\alpha}\int \frac{d^{d-2}q_xdq_z}{(2\pi)^{d-1}} \delta B(\qv) q_z^2 |u_0^{\alpha}(\bfq)|^2 ,
\end{eqnarray}
in which the momentum-dependent correction to the smectic compressional modulus in the effective surface theory, $\delta B(\qv)$, is given by
\begin{eqnarray}
\delta B(\qv)&=&\int dy dy' \delta B(y,y')\phi(\bfq,y)\phi(\bfq,y') \nonumber\\
&=&-B^2\int \frac{d^{d-2}k_x}{(2\pi)^{d-2}}\frac{dk_z}{2\pi} \frac{\Delta_f k_x^2+\Delta_v}{\Gamma_{\bf k}^3} k_x^4  b^2(\qv,{\bf k}),\nonumber\\
\end{eqnarray}
with the function $b(\qv,{\bf k})$ given by
\begin{equation}
b(\qv,{\bf k})=\int_0^{\infty} dy \phi^2({\bf k},y)\phi(\qv,y).
\end{equation}

We thus obtain the effective randomly surface pinned smectic model that includes the effect of smectic nonlinear elasticity. Although formally well defined, so far we have not succeeded in systematically treating this correction analytically. We leave such an analysis to future studies.

Focusing on the leading exponential contribution allows us to carry out $y$ and $y'$ integrals, giving
\begin{equation}
\delta B(\qv)\approx -B^2\int \frac{d^{d-2}k_x d k_z}{(2\pi)^{d-1}} \frac{\Delta_f k_x^2+\Delta_v}{\Gamma_{\bf k}^3\left[2D({\bf k})+D(\qv)\right]^2}k_x^4,
\end{equation}
where $D(\qv)=\frac{1}{\sqrt{2\lambda}}\sqrt{\sqrt{\lambda^2q_x^4+q_z^2}+\lambda q_x^2}$, and, thus, in the $\qv\rightarrow 0$ limit it gives
\begin{eqnarray}
\delta B(\qv\rightarrow 0)&\approx&\delta B_0
\approx -\frac{B^2}{4}\int \frac{d^{d-2}k_xdk_z}{(2\pi)^{d-1}}
\frac{\Delta_f k_x^2+\Delta_v}{\Gamma_{\bf k}^3D^2({\bf k})}k_x^4\nonumber\\
&\sim& \frac{\Delta_f}{B} L_x^{5-d}+\frac{\Delta_v}{B} L_x^{7-d}
\end{eqnarray}

Equating $\delta B_0 q_z^2$ to the harmonic kernel in Eq.~(\ref{Gamma_q}) allows us to define nonlinear crossover length $\xi_{NL,x}$ beyond which nonlinear elastic effects must be taken into account.
Focusing on orientational pinning (that we showed is dominant at long scales, see Sec.~\ref{sec:smectic_FRG}), we find
\begin{eqnarray}
\delta B_0 q_z^2|_{q_z=(\xi_{NL,z})^{-1}}&\sim& \Gamma_{\qv},\nonumber\\
\frac{\Delta_f}{B\lambda^{2}}(\xi_{NL,x})^{5-d-4} &\sim& K (\xi_{NL,x})^{-3},
\end{eqnarray}
which gives
\begin{equation}
\xi_{NL,x} \sim (K^2/\Delta_f)^{1/(4-d)},
\end{equation}
as the length above which the nonlinear elasticity dominates \cite{comment_positional_nonlinear}.

Based on this we expect that although nonlinear elasticity is relevant below $d_f^{NL}=4$ for the orientational pinning, it is only relevant at length scales much larger than the (Larkin) domain size $\xi_x$.
Thus, to study system properties at length scales smaller than the nonlinear length $\xi_{NL,x}$, it is safe to ignore the nonlinear elasticity. Here we employed this approximation and left the full analysis of nonlinear elasticity for future
study \cite{unpublishedUS}.

\subsection{Stability of orientational order}
\label{sec:Orientational_stability}
All of the above analysis of smectic (positional) order was predicated on the assumption that the long-range nematic (orientational) order is stable. Below we examine this assumption.

In the limit of dominant orientational pinning, the mean-squared distortion of orientational order on the heterogeneous substrate at $y=0$ is given by
\begin{eqnarray}
\overline{\langle|\delta{\bf n}(\xv)|^2\rangle}&\approx &\overline{\langle |\partial_x u(\xv)|^2\rangle}\nonumber\\
&=&\Delta_f\int\frac{dq_xdq_z}{(2\pi)^2}q_x^4 \frac{1}{\Gamma_{\qv}^2} \nonumber\\
&=&\frac{(\pi-2)\Delta_f}{4\pi^2 B^2\lambda^3}\int_{1/L}^{1/a} dq_x \nonumber\\
&\approx&\frac{\Delta_f}{c B^2\lambda^3a}\nonumber\\
&=&a/\xi_x^f\ll 1.
\label{orientational_fluctuation}
\end{eqnarray}
Evidently, orientational order is indeed long ranged for weak orientational pinning.

In contrast, we find that for dominant positional pinning, the distortion of nematic director diverges at long scales according to
\begin{eqnarray}
\overline{\langle|\delta{\bf n}(\xv)|^2\rangle}&\approx &\overline{ \langle|\partial_xu_0(x,z)|^2\rangle} \nonumber\\
&=&\frac{(\pi-2)\Delta_v}{4\pi^2 B^2\lambda^3}\int_{1/L}^{1/a}\frac{ dq_x}{q_x^2} \nonumber\\
&\approx&\frac{(\pi-2)\Delta_v}{4\pi^2 B^2\lambda^3}L.
\end{eqnarray}
This implies that the orientational order is destroyed at scales above
\begin{equation}
\xi_{O,x}=\frac{cB^2\lambda^3}{\Delta_v}=\frac{1}{3} (\xi_x^v)^3/a^2\gg \xi_x^v,
\label{Orientational_length_x}
\end{equation}
set by the condition on $L$ such that $\delta n \approx 1$.
Similarly, along $z$, the length scale beyond which the mean squared orientational
distortion $\overline{\langle|\delta{\bf n}(\xv)|^2\rangle}$ grows to order $1$ is
\begin{equation}
\xi_{O,z}\approx\left(\frac{cB^2\lambda^{5/2}}{\Delta_v}\right)^2\sim \xi_{O,x}^2/\lambda\gg \xi_z^v.
\label{Orientational_length_z}
\end{equation}

Naively this purely harmonic calculation implies that the analysis and prediction of previous sections are only valid up to this finite nematic domain scale, $\xi_{O,x}$ and $\xi_{O,z}$.
However, based on Eq.~(\ref{matched_correlator}) and the supporting RG and matching analysis, we note that at long scales the positional pinning becomes subdominant to purely orientational pinning, whether it is relevant for $T<T_g$ or irrelevant for $T>T_g$ (in the former case just modifying the strength of the effective orientational pinning). Thus, we conclude based on this and Eq.~(\ref{orientational_fluctuation}) that the orientational order is indeed stable for weak smectic surface pinning.

\section{Smectic cell with homeotropic alignment}
\label{sec:parallel_cell_smectic}
So far our focus has been on a smectic cell in the bookshelf geometry,
corresponding to molecular alignment parallel to the substrate, with the smectic layer normal lying in the plane of the substrate. As we demonstrated in previous
sections, random surface pinning in this geometry has significant qualitative effects on the smectic order, even in the bulk of the cell, destabilizing it on a random substrate.

In this section, for completeness, we also analyze a cell with a homeotropic alignment, with smectic layers running parallel to the substrate. As we show below, in this geometry the problem reduces to that of a surface-pinned $xy$ model. It thereby maps onto the previously studied planar nematic order with surface random pinning \cite{usFRGPRL,usFRGPRE}.

In this homeotropic geometry the cell is described by the Hamiltonian
\begin{equation}
 H_{sm}=\int d^{d-1}x\int_0^w dz
\left[\frac{K}{2}(\nabla^2_\perp u)^2 +\frac{B}{2}(\partial_z
  u)^2\right]
 + H_{pin},
\label{Hsmectic_parallel_cell}
\end{equation}
with the layer normal $\hat{z}$ perpendicular to the substrate.
Substrate randomness introduces distortions in the preferred layer normal
\begin{equation}
\hat{n}=(\delta n_x,\delta n_y, \hat{z})/\sqrt{1+\delta n_x^2+\delta n_y^2},
\end{equation}
leading to pinning energy
\begin{eqnarray}
&&\hspace{-0.6 cm}H_{pin}\nonumber\\
&=&\int
d^{d-1}x dz \delta(z)\left\{-W(\nh\cdot\hat{z})^2
-[\nh\cdot \gv(\rv)]^2-V(u,\rv)\right\} \nonumber\\
&\approx&\int
d^{d-1}x dz \delta(z)\bigg[-W+\frac{W}{2}(\delta n_x^2+\delta n_y^2)
- {\bf h}(\xv)\cdot\delta {\bf n}\nonumber\\
&&-V(u,\xv)\bigg] \nonumber\\
&\approx&\int
d^{d-1}x dz \delta(z)\left[\frac{W}{2}(\nabla_{\perp} u)^2
-{\bf h}(\xv)\cdot(\nabla_{\perp}u)-V(u,\xv)\right].\nonumber\\
\label{Hsm_pin_parrallel_cell}
\end{eqnarray}
Above we made use of the approximate relation between gradient of the smectic layers and the nematic director deviation
$\delta {\bf n}=(\delta n_x,\delta n_y,0)\approx \nabla_{\perp}u$, and in the last
equation ignored irrelevant constants. The first pinning term encodes the average substrate homeotropic alignment, $\left[\nh\cdot\gv(\rv)\right]^2\sim {\bf h}(\xv)\cdot(\nabla_{\perp} u)$ captures the random orientational pinning on the substrate, and the last term is the random positional pinning at $z=0$.

As for the bookshelf geometry, it is convenient to carry out the dimensional reduction to the surface
field $u(x,y,z)|_{z=0}\equiv u_0(x,y)$ at $z=0$ in a half-infinite cell.
To this end we solve the corresponding Euler-Lagrange equation subject to the substrate boundary condition,
\begin{equation}
K\nabla_{\perp}^4u(x,y,z)-B\partial_z^2u(x,y,z)=0,\nonumber
\end{equation}
whose solution for a thick cell is given by
\begin{equation}
u(\qv,z)=u_0(\qv)e^{-\lambda q^2 z}.
\end{equation}
Plugging this back into the bulk Hamiltonian and integrating over the $z$ degrees of freedom for $z>0$,
we obtain the effective surface Hamiltonian
\begin{eqnarray}
H_{sm}^{surf}&=&\int\frac{d^2q}{(2\pi)^2}\frac{\Gamma_0}{2}q^2|u_0(\qv)|^2\nonumber\\
&&-\int d^{d-1}x dz \delta(z)\left[ {\bf h}(\xv)\cdot(\nabla_{\perp}u)+V(u,\xv)\right],\nonumber\\
\end{eqnarray}
where
\begin{equation}
\Gamma_0 = \sqrt{KB} +W.
\end{equation}
In the momentum space the random pinning is given by
\begin{eqnarray}
H_{pin}=-\int\frac{d^2q}{(2\pi)^2}\left[
i\qv\cdot{\bf h}(-\qv) u_0(\qv)
+f(-\qv)u_0(\qv)\right].\nonumber\\
\end{eqnarray}

Thus, as anticipated in the Introduction, in this homeotropic geometry the effects of the random substrate in a smectic cell reduce to that of a well-studied random-field 2D $xy$-model \cite{DSFisherFRG,Nattermann,CardyOstlund,Toner_DiVincenzo_sr,DoussalGiamarchi_2D_RFxy} for the displacement of the smectic layer in contact with the random substrate.
In a weak pinning limit (where effects of dislocations and of nonlinear elasticity can be safely neglected) it is predicted to lead to weak breakdown of long-ranged smectic order with log-squared phonon correlations \cite{Toner_DiVincenzo_sr,VortexGlass,villainVG}. The random substrate pinning is significantly less disruptive in this geometry and can be used to test numerous theoretical predictions for the 2D random-field $xy$-model \cite{usFRGPRL,usFRGPRE}.

\section{Conclusions}
\label{sec:conclusion}
\subsection{Summary}

Motivated by a number of liquid crystal experiments on smectic
cells \cite{CDJonesThesis,ClarkSmC}, in this paper we studied the smectic liquid crystal order perturbed
by a randomly heterogeneous substrate. Demonstrating that the
resulting random surface pinning comes in two, orientational and
positional, forms, we introduced a general model for a heterogeneous
smectic cell and analyzed the response and stability of its smectic
order to a random substrate. We thereby demonstrated rigorously that
smectic long-range order in a thick cell (without a rubbed substrate)
at long scales is unstable to arbitrarily weak surface disorder,
replaced by random smectic domains whose size and correlations on the
heterogeneous substrate and in the bulk we have computed. We showed that
for weak random pinning the smectic order is thereby replaced by a
three-dimensional smectic-glass-like state, that undergoes a
temperature and pinning-driven Cardy-Ostlund-like phase transition
between a high-temperature weakly-pinned smectic and a low-temperature
smectic glass, where smectic layers in bookshelf geometry are strongly
pinned by the dirty substrate.

We computed the statistics of the induced positional and orientational
distortions within and on scales beyond the smectic domains.  We also
analyzed the transmission signal in the polarized light microscopy and
the x-ray scattering structure function in such a smectic cell. We
showed that these can be used to experimentally probe our predictions
of the finite smectic correlation length and the statistics of the
substrate induced smectic textures. Based on this, we suggested that
the precipitous broadening of the x-ray peak with reduced temperature
observed by Jones and Clark \cite{ClarkSmC,CDJonesThesis} may be associated with the
predicted smectic glass phase transition. More systematic experimental
and theoretical studies are necessary to explore this
further. Finally, we also demonstrated that in the geometry where
layers run parallel to the random substrate, the distortions on the
substrate are governed by the well-studied 2D random-field $xy$-model.

\subsection{Future directions}

While our current study makes significant progress toward
understanding of randomly surface-pinned smectics,
it also stimulates a number of interesting open questions
that we leave to
future research. Probably the most challenging of these is the nature
of the locally smectic state in the presence of strong random surface
pinning. This regime will exhibit distortions beyond the elastic
approximation, resulting in the proliferation of dislocations that are
notoriously difficult to treat analytically, requiring sophisticated
numerical analysis.

A more systematic (e.g., RG) treatment of nonlinear elasticity, which
we only briefly touched on via a self-consistent scaling analysis, is
another challenging problem requiring a simultaneous treatment of
pinning and elastic anharmonicities.

Throughout this manuscript we have also assumed that the average
strain of the pinned smectic state vanishes. While this is true in
equilibrium, experimental observations \cite{ClarkSmC,CDJonesThesis,discussionClark}
suggest that the system may have difficult time equilibrating as the
bulk layer spacing changes significantly with temperature, while the
layers remain pinned at the random substrate. Thus, as the system is
cooled into the smectic state, we expect a significant build up of
strain near the random substrate, with the relaxation rate limited by
nucleation and motion of edge dislocations. It is clear that
accounting for this background strain is crucial for a detailed
understanding of current experiments \cite{ClarkSmC,CDJonesThesis}. We are currently
exploring this interesting effect \cite{unpublishedUS}.

Finally, our earlier work on nematic cells \cite{usFRGPRL,usFRGPRE} and current
study of smectic cells \cite{usSmecticEPL} lend themselves to generalizations
to a broad array of other liquid-crystal phases, such as, for example
the cholesteric and smectic-$C$ states. The effect of the random
substrate depends qualitatively on the nature of the phases and we leave their analysis to future studies.

We hope that the present study will stimulate further
theoretical and experimental work on these and many other open
questions.

\acknowledgments
We thank Noel Clark, Matthew Glaser, Christopher Jones, Joe Maclennan, Robert Meyer, Ivan Smalyukh and David Walba for discussions and acknowledge support by the NSF through DMR-1001240 and MRSEC DMR-0820579.

\appendix

\section{Solution of the Euler-Lagrange equation}
\label{app:solution_complexIntegral}
In this appendix, we
study the solution to the Euler-Lagrange equation of the smectic liquid crystal cells subject to a boundary condition of a random surface texture $u(x,y=0,z)=u_0(x,z)$ or equivalently random force $f(x,z)$ as in (\ref{smectic_eom_bulk}),
\begin{equation}
K\nabla_{\perp}^4u-B\partial_z^2 u=f(x,z)\delta(y),\nonumber
\end{equation}
where we used a surface pinning force
\begin{equation}
H_{pin}=-\int dx dy dz \delta(y)f(x,z)u(x,y,z),
\end{equation}
to encode the boundary condition $u(x,y=0,z)=u_0(x,z)$.

\begin{figure}[b]
\centering
\includegraphics[height=5 cm]{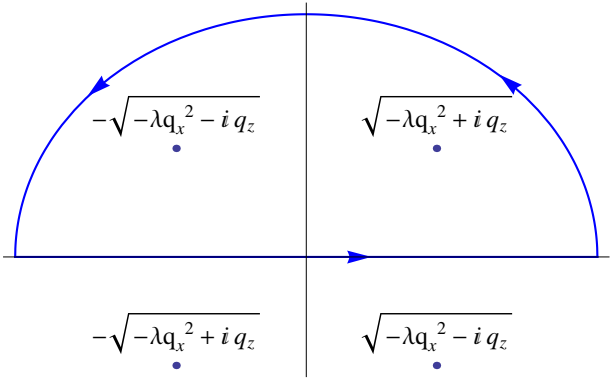}\\
\caption{(Color online) The integral (\ref{ComplexIntegral}) represented by
 a contour integral closed in the upper-half of the complex plane with
 poles shown.}
\label{fig:CounterIntegral}
\end{figure}

In Fourier space Eq.~(\ref{smectic_eom_bulk}) reduces to
\begin{equation}
\left[K(q_x^2+q_y^2)^2+Bq_z^2\right]u(q_x,q_z,q_y)=f(q_x,q_z),
\end{equation}
giving
\begin{equation}
u(q_x,q_z,q_y)=\frac{f(q_x,q_z)}{K(q_x^2+q_y^2)^2+Bq_z^2}.
\end{equation}
Then, for $y>0$ we have
\begin{eqnarray}
\hspace{-0.6 cm}u(q_x,q_z,y)
&=&\int_{-\infty}^{\infty} \frac{dq_y}{2\pi}
  \frac{f(q_x,q_z)e^{iq_yy}}{K(q_x^2+q_y^2)^2+Bq_z^2} \nonumber\\
&=&\frac{f(q_x,q_z)}{2\pi B}\int_{-\infty}^{\infty} dq_y
  \frac{e^{iq_yy}}{\lambda^2(q_x^2+q_y^2)^2+q_z^2} \nonumber\\
&=&\frac{f(q_x,q_z)}{2\pi B\sqrt{\lambda}}\int_{-\infty}^{\infty} d\overline{q}_y
  \frac{e^{i\overline{q}_y \overline{y}}}{(\lambda q_x^2+\overline{q}_y^2)^2+q_z^2},
\label{ComplexIntegral}
\end{eqnarray}
with the associated length $\lambda=\sqrt{K/B}$,
$\overline{q}_y=\sqrt{\lambda}q_y$, and $\overline{y}=y/\sqrt{\lambda}$.
As shown in Fig.~\ref{fig:CounterIntegral}, (\ref{ComplexIntegral})
can be readily evaluated by use of a contour integration, closing the contour in the upper half of the complex plane.
Defining $q_p=\sqrt{-\lambda q_x^2+iq_z}$, the four poles
are given by $\pm q_p$ and $\pm q_p^*$.
Closing the contour in the upper half-plane, we find
\begin{widetext}
\begin{eqnarray}
u(q_x,q_z,y)&=&\frac{f(q_x,q_z)}{2\pi B\sqrt{\lambda}}\int_{-\infty}^{\infty} d\overline{q}_y
  \frac{e^{i\overline{q}_y\overline{y}}}{(\overline{q}_y-q_p)(\overline{q}_y-q_p^*)
(\overline{q}_y+q_p)(\overline{q}_y+q_p^*)} \nonumber\\
&=&\frac{f(q_x,q_z)}{2\pi B\sqrt{\lambda}}2\pi i\left[\frac{e^{iq_p \overline{y}}}{(q_p-q_p^*)
(q_p+q_p)(q_p+q_p^*)}+\frac{e^{-iq_p^* \overline{y}}}{(-q_p^*-q_p)(-q_p^*-q_p^*)
(-q_p^*+q_p)}\right]\nonumber\\
&=&\frac{f(q_x,q_z)}{2B\sqrt{\lambda}} i\left[\frac{e^{iq_p \overline{y}}}{4iq_p \mbox{Re}(q_p)\mbox{Im}(q_p)}
+\frac{e^{-iq_p^*\overline{y}}}{4iq_p^*\mbox{Re}(q_p)\mbox{Im}(q_p)}\right] \nonumber\\
&=&\frac{f(q_x,q_z)}{2B\sqrt{\lambda}}\frac{1}{2\mbox{Re}(q_p)\mbox{Im}(q_p)}\mbox{Re}\left[\frac{e^{iq_p \overline{y}}}{q_p}\right],
\end{eqnarray}
\end{widetext}
and
\begin{equation}
q_p^2=-\lambda q_x^2+iq_z=|q_p^2|e^
{i\left[\pi-\arctan{\left(\frac{q_z}{\lambda q_x^2}\right)}\right]},
\end{equation}
with $\left|q_p^2\right|=\sqrt{\lambda^2 q_x^4+q_z^2}$. Noting that
\begin{eqnarray}
q_p&=&\left(\lambda^2 q_x^4+q_z^2\right)^{1/4}e^{i\left[\frac{\pi}{2}
 -\half\arctan{\left(\frac{q_z}{\lambda q_x^2}\right)}\right]},\\
\mbox{Re}(q_p)&=&\left(\lambda^2 q_x^4+q_z^2\right)^{1/4}
 \sin{\left[\half\arctan{\left(\frac{q_z}{\lambda q_x^2}\right)}\right]} \nonumber\\
&=&\frac{1}{\sqrt{2}}\sqrt{\sqrt{\lambda^2q_x^4+q_z^2}-\lambda q_x^2},\\
\mbox{Im}(q_p)&=&\left(\lambda^2 q_x^4+q_z^2\right)^{1/4}
 \cos{\left[\half\arctan{\left(\frac{q_z}{\lambda q_x^2}\right)}\right]} \nonumber\\
&=&\frac{1}{\sqrt{2}}\sqrt{\sqrt{\lambda^2q_x^4+q_z^2}+\lambda q_x^2}.
\end{eqnarray}

We find
\begin{widetext}
\begin{eqnarray}
u(q_x,q_z,y)&=&\frac{f(q_x,q_z)}{2B\sqrt{2\lambda}}
 \frac{1}{q_z\sqrt{\lambda^2q_x^4+q_z^2}}
 e^{-\frac{y}{\sqrt{2\lambda}}\sqrt{\sqrt{\lambda^2q_x^4+q_z^2}+\lambda q_x^2}}\nonumber\\
&& \times\Big[\sqrt{\sqrt{\lambda^2q_x^4+q_z^2}-\lambda q_x^2}
 \cos{\Big(\frac{y}{\sqrt{2\lambda}}\sqrt{\sqrt{\lambda^2q_x^4+q_z^2}-\lambda q_x^2}\Big)}\nonumber\\
&& +\sqrt{\sqrt{\lambda^2q_x^4+q_z^2}+\lambda q_x^2}
 \sin{\Big(\frac{y}{\sqrt{2\lambda}}\sqrt{\sqrt{\lambda^2q_x^4+q_z^2}-\lambda q_x^2}\Big)}\Big].
 \label{uqxqzy_app}
\end{eqnarray}
\end{widetext}
On the $y=0$ substrate,
\begin{eqnarray}
u_0(q_x,q_z)&\equiv&u(q_x,q_z,y=0)\nonumber\\
&=&\frac{f(q_x,q_z)}{2B\sqrt{2\lambda}}
\frac{\sqrt{\sqrt{\lambda^2q_x^4+q_z^2}-\lambda q_x^2}}{q_z\sqrt{\lambda^2q_x^4+q_z^2}}.
\end{eqnarray}
Combining this with Eq.~(\ref{uqxqzy_app}) allows us to express $u(q_x,q_y,y)$ in terms of this substrate deformation
\begin{eqnarray}
&&\hspace{-0.6 cm}u(q_x,q_z,y)
=u_0(q_x,q_z)
 e^{-\frac{y}{\sqrt{2\lambda}}\sqrt{\sqrt{\lambda^2q_x^4+q_z^2}+\lambda q_x^2}}\nonumber\\
 &&\times\Bigg[\frac{\sqrt{\sqrt{\lambda^2q_x^4+q_z^2}+\lambda q_x^2}}{\sqrt{\sqrt{\lambda^2q_x^4+q_z^2}-\lambda q_x^2}}
 \sin{\Big(\frac{y}{\sqrt{2\lambda}}\sqrt{\sqrt{\lambda^2q_x^4+q_z^2}-\lambda
q_x^2}\Big)} \nonumber\\
&&
+\cos{\Big(\frac{y}{\sqrt{2\lambda}}\sqrt{\sqrt{\lambda^2q_x^4+q_z^2}-\lambda q_x^2}\Big)}\Bigg], \nonumber
\end{eqnarray}
given as Eq.~(\ref{uqy}) in the main context.

\section{Domain size in the surface disordered smectic cell in bookshelf geometry}
\label{app:LakinLengths_smectic}
In this appendix we present the details of the analysis of the Larkin lengths
setting the domain size and shape within which smectic order is stable and the weak random pinning is treated perturbatively.
To make analytical progress we focus on extreme limits of dominant positional pinning $\Delta_v$ or dominant orientational pinning $\Delta_f$, but not both.
The calculation here is only precise in the scaling sense and is not quantitatively accurate.

\subsection{Larkin lengths for dominant surface orientational pinning}
Focusing on the case of dominant random orientational disorder and generalizing
the analysis to $d$ dimensions with $(d-2)$-dimensional $x$, we find that the smectic phonon
variance is given by
\begin{equation}
\overline{\langle u^2_0(x,z)\rangle}=\frac{\Delta_f}{2B^2\lambda}
\int \frac{d^{d-2}q_xdq_z}{(2\pi)^{d-1}}
q_x^2\frac{\sqrt{\lambda^2q_x^4+q_z^2}-\lambda q_x^2}{q_z^2(\lambda^2q_x^4+q_z^2)}.
\end{equation}

This integral is divergent at small $q$ and needs cutoffs set by the domain size $L_x$ and $L_z$. For $L_x\ll \sqrt{\lambda L_z}$, one can integrate $q_z$ out, cut off the IR integration by $q_x\sim 1/L_x$, and find
\begin{eqnarray}
&&\frac{\Delta_f}{4\pi B^2\lambda} \int\frac{d^{d-2}q_x}{(2\pi)^{d-2}}
 \frac{\pi-2}{\lambda^2 q_x^2} \nonumber\\
&=&\frac{(\pi-2)\Delta_f}{4\pi B^2\lambda^3}\frac{S_{d-2}}{(2\pi)^{d-2}}\int_{1/L_x}^{1/a}\frac{q_x^{d-3}dq_x}{q_x^2} \nonumber\\
&=&\frac{(\pi-2)\Delta_f}{4\pi B^2\lambda^3}C_{d-2}\frac{1}{4-d}L_x^{4-d},
\end{eqnarray}
diverging for $d < d_{lc} =4$, where $S_d$ is the surface area of a $d$-dimensional unit sphere,
and $C_d=S_d/(2\pi)^d$.
Defining the Larkin domain size $\xi^f_x$ as the length scale over which $u_{rms}$ is on the order of smectic layer spacing, we find
\begin{equation}
\xi_x^f=\Big[\frac{4(4-d)\pi }{(\pi-2)C_{d-2}}\frac{B^2\lambda^3a^2}{\Delta_f}\Big]^{1/(4-d)},
\end{equation}
which for the physically relevant case of $d=3$ is given by
\begin{equation}
\xi_x^f=c\frac{B^2\lambda^3a^2}{\Delta_f},
\end{equation}
in which $c=\frac{4\pi^2 }{\pi-2}\approx 34.6$.

At the lower critical dimension of $d=4$, $u_{rms}$ diverges logarithmically leading to
\begin{equation}
\xi_x^f=a \exp{\left[\frac{4\pi}{\pi-2}\frac{(2\pi)^2}{S_2}\frac{B^2\lambda^3a^2}{\Delta_f}\right]}.
\end{equation}

Similarly for the Larkin length along $z$, integrating over $q_x$ first, we find
\begin{eqnarray}
&&\hspace{-0.6 cm}\overline{\langle u^2_0(x,z)\rangle}\nonumber\\
&=&\frac{\Delta_f}{2B^2\lambda}\int \frac{dq_z}{2\pi}\int\frac{d^{d-2}q_x}{(2\pi)^{d-2}}
q_x^2\frac{\sqrt{\lambda^2q_x^4+q_z^2}-\lambda q_x^2}{q_z^2(\lambda^2q_x^4+q_z^2)} \nonumber\\
&=&\frac{\Delta_f}{2B^2\lambda}\frac{S_{d-2}}{(2\pi)^{d-2}}
\int \frac{dq_z}{2\pi}\int_0^{\infty}
\frac{\sqrt{\lambda^2q_x^4+q_z^2}-\lambda q_x^2}{q_z^2(\lambda^2q_x^4+q_z^2)}q_x^{d-1}dq_x \nonumber\\
&=&\frac{\Delta_f}{2\pi B^2\lambda^{d/2+1}}\frac{S_{d-2}}{(2\pi)^{d-2}}
 \frac{\Gamma(\half-\frac{d}{4})\Gamma(\frac{d}{4})-\pi^{3/2}\sec(\frac{d\pi}{4})}{4\sqrt{\pi}}\nonumber\\
&&\times \int_{1/L_z}^{1/a} dq_z q_z^{\frac{d}{2}-3} \nonumber\\
&=&\frac{1}{2B(d)}\frac{\Delta_f}{B^2\lambda^{d/2+1}} \frac{L_z^{2-d/2}}{2-d/2},
\end{eqnarray}
in which
\begin{equation}
\frac{1}{B(d)}=\frac{1}{\pi}\frac{S_{d-2}}{(2\pi)^{d-2}}\frac{\Gamma(\half-\frac{d}{4})
\Gamma(\frac{d}{4})-\pi^{3/2}\sec(\frac{d\pi}{4})}{4\sqrt{\pi}}.
\end{equation}
This gives
\begin{equation}
\xi_z^f=\Big[(4-d)B(d)\frac{B^2\lambda^{d/2+1}a^2}{\Delta_f}\Big]^{2/(4-d)},
\end{equation}
which in the physically relevant three dimensions reduces to
\begin{equation}
\xi_z^f=\Big[B(3)\frac{B^2\lambda^{5/2}a^2}{\Delta_f}\Big]^{2},
\end{equation}
where $B(3)\approx 37.5$ is roughly the same as $c$.
At the lower critical dimension we instead find
\begin{equation}
\xi_z^f=a \exp{\left[\frac{2}{B(4)}\frac{B^2\lambda^3a^2}{\Delta_f}\right]}.
\end{equation}

\subsection{Larkin lengths for dominant surface positional pinning}
For a smectic cell where positional disorder dominates, the variance of smectic distortions is given by
\begin{eqnarray}
\overline{\langle u^2_0(x,z)\rangle}&=&\int\frac{dq_x}{2\pi}\int_{-\infty}^{\infty}
 \frac{dq_z}{2\pi}  \frac{\Delta_v}{2B^2\lambda}
\frac{\sqrt{\lambda^2q_x^4+q_z^2}-\lambda q_x^2}{q_z^2(\lambda^2q_x^4+q_z^2)} \nonumber\\
&=& \frac{\Delta_v}{4\pi B^2\lambda} \int\frac{dq_x}{2\pi}
 \frac{\pi-2}{\lambda^2 q_x^4} \nonumber\\
&=&\frac{\Delta_v}{4\pi B^2\lambda}\frac{2(\pi-2)}{2\pi\lambda^2}\int_{1/L_x}^{1/a}\frac{dq_x}{q_x^4} \nonumber\\
&=&\frac{(\pi-2)\Delta_v}{12\pi^2 B^2\lambda^3} L_x^3,
\end{eqnarray}
which diverges as $L_x^3$ in 3D.

Consequently, the Larkin length along $x$ is given by
\begin{equation}
\xi_x^v=\Big[3c\frac{B^2\lambda^3a^2}{\Delta_v}\Big]^{1/3}.
\end{equation}
Similarly, the Larkin length along $z$ is instead given by
\begin{equation}
\xi_z^v=\Big[A(3)\frac{B^2\lambda^{3/2}a^2}{\Delta_v}\Big]^{2/3},
\end{equation}
where $A(3)=\frac{12\pi^{5/2}}{16\Gamma^2(\frac{5}{4})-\sqrt{2\pi^3}}\approx 39.83$, roughly the same as $c$.

More generally in $d$ dimensions the variance is given by…
\begin{eqnarray}
\overline{\langle u^2_0(x,z)\rangle}&=&\frac{\Delta_v}{4\pi B^2\lambda} \int\frac{d^{d-2}q_x}{(2\pi)^{d-2}}
 \frac{\pi-2}{\lambda^2 q_x^4} \nonumber\\
&=&\frac{(\pi-2)\Delta_v}{4\pi B^2\lambda^3}\frac{S_{d-2}}{(2\pi)^{d-2}}\int_{1/L_x}^{1/a}\frac{q_x^{d-3}dq_x}{q_x^4} \nonumber\\
&=&\frac{(\pi-2)\Delta_v}{4\pi B^2\lambda^3}C_{d-2}\frac{1}{6-d}L_x^{6-d},
\end{eqnarray}
which is diverging for $d<d_{lc}^v=6$, leading to the Larkin length along $x$
given by:
\begin{equation}
\xi_x^v=\Big[\frac{4(6-d)\pi }{(\pi-2)C_{d-2}}\frac{B^2\lambda^3a^2}{\Delta_v}\Big]^{1/(6-d)}.
\end{equation}
And similarly along $z$
\begin{equation}
\xi_z^v=\Big[\frac{(6-d)A(d)}{3}\frac{B^2\lambda^{d/2}a^2}{\Delta_v}\Big]^{2/(6-d)},
\end{equation}
in which
\begin{equation}
\frac{1}{A(d)}=\frac{1}{\pi}\frac{S_{d-2}}{(2\pi)^{d-2}}\frac{\Gamma(1-\frac{d}{4})\Gamma(\frac{d-2}{4})
 -\pi^{3/2}\csc(\frac{d\pi}{4})}{12\sqrt{\pi}}.
\end{equation}

At the lower critical dimension $d=6$, $u_{rms}$ diverges logarithmically, leading to
\begin{eqnarray}
\xi_x^v&=&a \exp{\left[\frac{4\pi}{\pi-2}\frac{(2\pi)^4}{S_4}\frac{B^2\lambda^3a^2}{\Delta_v}\right]},\\
\xi_z^v&=&a \exp{\left[\frac{2A(6)}{3}\frac{B^2\lambda^3a^2}{\Delta_v}\right]}.
\end{eqnarray}

\section{Calculation of correlation functions at short scales (Larkin regime)}
\label{app:correlationfunctions}
Here we present the details of the correlation function of smectic phonons $u_0(x,z)$ on the dirty substrate, $y=0$, focusing on the short-scale Larkin regime. In 3D it is given by a Fourier transform of Eq.~(\ref{C_D_smectic}),
\begin{eqnarray}
&&\hspace{-0.6  cm}C(x,z)\nonumber\\
&=&\overline{\langle [u_0(x,z)-u_0(0,0)]^2\rangle}\nonumber\\
&=&2\int\frac{dq_z}{2\pi}\frac{d q_x}{2\pi}
\frac{\Delta_fq_x^2+\Delta_v}{(\Gamma_{\qv})^2}(1-e^{i q_x x+iq_zz}),
\end{eqnarray}
where within the Larkin regime the integrations are bounded by the intermediate
length scales $\xi_{x,z}$ given in Sec.~\ref{sec:larkinLength_smectic}.
We treat the two forms of orientational and positional pinnings individually, for simplicity assuming one or the other (but not both) dominates.

\subsection{Dominant surface orientational disorder}
With dominant surface orientational disorder, the correlation function is
given by
\begin{eqnarray}
&&\hspace{-0.6  cm}C_f(x,z)\nonumber\\
&=&\overline{\langle [u_0(x,z)-u_0(0,0)]^2\rangle}\nonumber\\
&=&\int\frac{dq_z}{2\pi}\frac{d q_x}{2\pi}
\frac{\Delta_fq_x^2}{2B^2\lambda}\frac{\sqrt{\lambda^2q_x^4+q_z^2}-\lambda q_x^2}{q_z^2(\lambda^2q_x^4+q_z^2)}
(1-e^{iq_x x+iq_zz}).\nonumber\\
\end{eqnarray}
For the dependence along $x$, we first integrate out $q_z$,
obtaining (for $x\ll\xi_x$)
\begin{eqnarray}
C_f(x,0)&\approx& 2\frac{(\pi-2)\Delta_f}{4\pi^2 B^2\lambda^3}
\int_{1/\xi_x^f}^{\infty} q_x^{-2}[1-\cos{(q_xx)}]dq_x \nonumber\\
&\approx& 2\frac{(\pi-2)\Delta_f}{4\pi^2 B^2\lambda^3} \frac{\pi x}{2} \nonumber\\
&\approx& \pi a^2 x/\xi_x^f.
\end{eqnarray}
For the dependence along $z$, we instead have
\begin{eqnarray}
C_f(0,z)
&\approx&
2\frac{a^2}{2\sqrt{\xi_z^f}}
 \int_{1/\xi_z^f}^{\infty}dq_z \frac{1-\cos{(q_z z)}}{q_z^{3/2}} \nonumber\\
&\approx& \sqrt{2\pi}a^2 \sqrt{z/\xi_z^f},
\end{eqnarray}
where the definition of $\xi_{x,z}^f$ is given in Sec.~\ref{sec:larkinLength_smectic}. These correlation functions with dominant orientational disorder
are plotted in Fig.~\ref{fig:tiltCorrelation}.

\subsection{Dominant surface positional disorder}
With dominant random positional pinning, the correlation function in 3D is given by
\begin{eqnarray}
&&\hspace{-0.6  cm}C_v(x,z)\nonumber\\
&=&\overline{\langle [u_0(x,z)-u_0(0,0)]^2\rangle }\nonumber\\
&=&2\int\frac{dq_z}{2\pi}\frac{dq_x}{2\pi}
\frac{\Delta_v}{2B^2\lambda}\frac{\sqrt{\lambda^2q_x^4+q_z^2}-\lambda q_x^2}{q_z^2(\lambda^2q_x^4+q_z^2)}
 (1-e^{iq_{x}\cdot x+iq_zz}).\nonumber\\
\end{eqnarray}
Focusing first on the $x$ dependence at $z=0$, we integrate out $q_z$ and find
\begin{eqnarray}
&&\hspace{-0.6  cm}C_v(x,0)\nonumber\\
&\approx& 2
\frac{(\pi-2)\Delta_v}{4\pi^2 B^2\lambda^3}
\int_{1/\xi_x^v}^{\infty} q_x^{-4}\left[1-\cos{(q_xx)}\right]dq_x\nonumber\\
&\approx& 2 \frac{(\pi-2)\Delta_v}{4\pi^2 B^2\lambda^3}
\left[\int_{1/\xi_x^v}^{1/x} q_x^{-4}\frac{(q_xx)^2}{2}dq_x+\int_{1/x}^{\infty}q_x^{-4}dq_x \right]\nonumber\\
&=&  \frac{(\pi-2)\Delta_v}{4\pi^2 B^2\lambda^3}
 \xi_x^v x^2 ,
\end{eqnarray}
where we simplified the integral by separating it into $q_x<1/x$
and $q_x>1/x$ parts with corresponding approximations. Utilizing the condition from Sec.~\ref{sec:larkinLength_smectic}
and Appendix \ref{app:LakinLengths_smectic} that $\frac{(\pi-2)\Delta_v}{4\pi^2 B^2\lambda^3}=3a^2/(\xi_x^v)^3$,
the correlation function reduces to a simple form,
\begin{equation}
C_v(x,0)\approx 3 a^2 \frac{ x^2}{(\xi_x^v)^2}.
\end{equation}

Equivalently, this integral can be evaluated exactly in terms of special functions and then a small $x \ll \xi_x^v$ limit is taken. Consistent with the above estimate,  we find
\begin{eqnarray}
&&\hspace{-0.6 cm}C_v(x,0)\nonumber\\
&\approx& 2\frac{(\pi-2)\Delta_v}{4\pi^2 B^2\lambda^3}
\int_{1/\xi_x^v}^{\infty} q_x^{d-7}[1-\cos{(q_xx)}]dq_x \nonumber\\
&=& 2\frac{(\pi-2)\Delta_v}{4\pi^2 B^2\lambda^3}
\bigg\{\frac{1}{3}(\xi_x^v)^3+\frac{1}{6}\xi_x^v\left[x^2-2(\xi_x^v)^2\right]\cos{(\frac{x}{\xi_x^v})}\nonumber\\
&& +\frac{1}{6}x(\xi_x^v)^2\sin{(\frac{x}{\xi_x^v})}-\frac{x^3}{12}\left[\pi -2\mbox{ }\mathrm{Si}{(\frac{x}{\xi_x^v})}\right]\bigg\} \nonumber\\
&\approx& 3a^2 \frac{x^2}{(\xi_x^v)^2},
\end{eqnarray}
where $\mathrm{Si}(z)=\int_0^z \sin{(t)}\frac{dt}{t}$ is the sine integral function.

To obtain the short-scale $z$ dependence of the correlation function, we instead first integrate over $q_x$, obtaining
\begin{eqnarray}
C_v(0,z)
&\approx& 2 \frac{\Delta_v}{4\pi^2 B^2\lambda} \frac{16\Gamma^2(\frac{5}{4})-\sqrt{2\pi^3}}{2\sqrt{\pi\lambda}}
 \int_{1/\xi_z^v}^{\infty}dq_z \frac{1-e^{iq_zz}}{q_z^{5/2}} \nonumber\\
&\approx& 2 \frac{\Delta_v}{4\pi^2 B^2\lambda} \frac{16\Gamma^2(\frac{5}{4})-\sqrt{2\pi^3}}{2\sqrt{\pi\lambda}}
\frac{2\sqrt{2\pi}}{3}z^{3/2} \nonumber\\
&\approx& \sqrt{8\pi}a^2\left(\frac{z}{\xi_z^v}\right)^{3/2},
\end{eqnarray}
where the definition of $\xi_z^v$ from Sec.~\ref{sec:larkinLength_smectic} was used. The correlation functions with dominant positional disorder
are plotted in Fig.~\ref{fig:positionCorrelation}.

As discussed in the main body of the paper, the validity of this perturbative
analysis is limited to short scales of the Larkin regime.
To understand the behavior at longer distance, where the Larkin treatment
breaks down, we utilize the renormalization group analysis.

\section{FRG analysis to the second order}
\label{app:FRG_2ndOrder_smectic}
In this appendix we present the details of the FRG analysis by evaluating each term in Eq.~(\ref{HorderExpansion}).
Averaging over the high-wave-vector fields in
the surface orientational disorder term in Eq.~(\ref{HorderExpansion}) we find
\begin{widetext}
\begin{eqnarray}
\langle H_{\Delta_f}\rangle_>&=&\frac{1}{4T}\sum_{\alpha,\beta}\int d^{d-2}xdz\int_{\kappa}\tilde{\Delta}_f(\kappa)\left\langle e^{i\kappa \left[u_0^{\alpha}(x,z)-u_0^{\beta}(x,z)\right]}
\left|\partial_x\left[u_0^{\alpha}(x,z)-u_0^{\beta}(x,z)\right]\right|^2\right\rangle_> \nonumber\\
&=&\frac{1}{4T}\sum_{\alpha,\beta}\int d^{d-2}xdz\int_{\kappa}\tilde{\Delta}_f(\kappa)
e^{i\kappa \left[u_{0<}^{\alpha}(x,z)-u_{0<}^{\beta}(x,z)\right]} \nonumber\\
&&\times\left\langle e^{i\kappa \left[u_{0>}^{\alpha}(x,z)-u_{0>}^{\beta}(x,z)\right]}
\left|\partial_x\left[u_{0<}^{\alpha}(x,z)-u_{0<}^{\beta}(x,z)
  +u_{0>}^{\alpha}(x,z)-u_{0>}^{\beta}(x,z)\right]\right|^2\right\rangle_> \nonumber\\
&=&\frac{1}{4T}\sum_{\alpha,\beta}\int d^{d-2}xdz\int_{\kappa}\tilde{\Delta}_f(\kappa)
e^{i\kappa \left[u_{0<}^{\alpha}(x,z)-u_{0<}^{\beta}(x,z)\right]}
\bigg\langle e^{i\kappa \left[u_{0>}^{\alpha}(x,z)-u_{0>}^{\beta}(x,z)\right]}\nonumber\\
&&\times\left\{\left|\partial_x\left[u_{0<}^{\alpha}(x,z)-u_{0<}^{\beta}(x,z)\right]\right|^2+
\left|\partial_x\left[u_{0>}^{\alpha}(x,z)-u_{0>}^{\beta}(x,z)\right]\right|^2\right\}\bigg\rangle_>,
\label{Delta_f_>}
\end{eqnarray}
\end{widetext}
in which the cross term is linear in $u_>$ and, thus, vanishes in momentum-shell RG. The key part of the first term
in the above equation is
\begin{eqnarray}
&&\hspace{-0.6 cm}\langle e^{i\kappa \left[u_{0>}^{\alpha}(x,z)-u_{0>}^{\beta}(x,z)\right]}\rangle_>\nonumber\\
&=&\frac{1}{Z_0^>}\int [du_{0>}^{\alpha}]e^{i\kappa \left[u_{0>}^{\alpha}(x,z)-u_{0>}^{\beta}(x,z)\right]}e^{-H_0^>/T} \nonumber\\
&=&e^{-\kappa^2f_{\alpha \beta}},
\end{eqnarray}
in which $Z_0^>$ is the partition function of the high-wave vector components of the system
\begin{equation}
Z_0^>=\int [du_{0>}^{\alpha}]e^{-H_0^>/T},
\end{equation}
and $f_{\alpha \beta}$ is given by
\begin{eqnarray}
f_{\alpha\beta}&=&C_{T,\alpha\alpha}^>(0)-G_{T,\alpha,\beta}^>(0) \nonumber\\
&=&\int_{\Lambda e^{-\delta\ell}}^{\Lambda}\frac{d^{d-2}q_x}{(2\pi)^{d-2}}\int_{-\infty}^{
\infty}\frac{dq_z}{2\pi}\frac{T}{\Gamma_{\qv}}(1-\delta_{\alpha\beta}) \nonumber\\
&\equiv&\eta(1-\delta_{\alpha\beta})\delta\ell.\label{eta_definition}
\end{eqnarray}
For $d=3$ it is easy to show that $\eta=\frac{T}{2\pi B\lambda}$, which is
a constant under the renormalization flow. We can see that with the expansion
$e^{-\kappa^2f_{\alpha\beta}}=1-\kappa^2f_{\alpha\beta}+\cdots$, the zeroth-order term contributes
through rescaling and the first-order term leads to the random
orientational pinning nonlinearity
\begin{equation}
\delta\Delta_f^{(1)}(u)=\eta\Delta_f''(u)\delta\ell,
\end{equation}
with $'$ indicating derivation respect to $u$.

Clearly, the average in the last term of Eq.~(\ref{Delta_f_>})
contributes to the surface positional pinning nonlinearity $R_v(u)$. To lowest
order, we calculate the average
\begin{eqnarray}
&&\hspace{-0.6 cm}\left\langle e^{i\kappa \left[u_{0>}^{\alpha}(x,z)-u_{0>}^{\beta}(x,z)\right]}
\left|\partial_x\left[u_{0>}^{\alpha}(x,z)-u_{0>}^{\beta}(x,z)\right]\right|^2\right\rangle_>\nonumber\\
&\approx&\left\langle \left|\partial_x\left[u_{0>}^{\alpha}(x,z)-u_{0>}^{\beta}(x,z)\right]\right|^2\right\rangle_> \nonumber\\
&=&2\int_{\Lambda e^{-\delta\ell}}^{\Lambda}\frac{d^{d-2}q_x}{(2\pi)^{d-2}}\int_{-\infty}^{
\infty}\frac{dq_z}{2\pi}\frac{T q_x^2}{\Gamma_{\qv}}(1-\delta_{\alpha\beta}) \nonumber\\
&\equiv& 2\zeta(1-\delta_{\alpha\beta})\delta\ell,
\end{eqnarray}
while for $d=3$ we have $\zeta=\frac{\Lambda^2}{2\pi B\lambda}$, which contributes
\begin{equation}
\delta R_v^{(1a)}(u)=-\zeta\Delta_f(u)\delta\ell.
\label{deltaDelta_v1}
\end{equation}

We now consider the surface positional disorder in Eq.~(\ref{HorderExpansion}),
\begin{widetext}
\begin{eqnarray}
&&\hspace{-0.6 cm}\langle H_{\Delta_v}\rangle_> \nonumber\\
&=&-\sum_{\alpha,\beta}\frac{1}{2T}\int d^{d-2}xdz\left\langle R_v\left[u_0^{\alpha}(x,z)-u_0^{\beta}(x,z)\right]\right\rangle_> \nonumber\\
&\approx&-\sum_{\alpha,\beta}\frac{1}{2T}\int d^{d-2}xdz
 \left\langle R_v\left[u_{0<}^{\alpha}(x,z)-u_{0<}^{\beta}(x,z)\right]
 +\frac{1}{2} R_v''\left[u_{0<}^{\alpha}(x,z)-u_{0<}^{\beta}(x,z)\right]
\left[u_{0>}^{\alpha}(x,z)-u_{0>}^{\beta}(x,z)\right]^2 \right\rangle_> \nonumber\\
&=&-\sum_{\alpha,\beta}\frac{1}{2T}\int d^{d-2}xdz
 \left\{ R_v\left[u_{0<}^{\alpha}(x,z)-u_{0<}^{\beta}(x,z)\right]
   +\frac{1}{2}R_v''\left[u_{0<}^{\alpha}(x,z)-u_{0<}^{\beta}(x,z)\right]
\left\langle\left[u_{0>}^{\alpha}(x,z)-u_{0>}^{\beta}(x,z)\right]^2\right\rangle_>\right\},\nonumber\\
\end{eqnarray}
\end{widetext}
in which the first term gives a correction through rescaling, the linear order term in $u_{0>}^{\alpha}$
makes a vanishing contribution, and the last term
\begin{equation}
\delta R_v^{(1b)}(u)=\eta R_v''(u)\delta\ell,
\end{equation}
[together with Eq. (\ref{deltaDelta_v1})] gives a first-order contribution to the surface positional disorder.

To study the second-order contribution in the RG, we start with the $(R_v)^2$ term in (\ref{HorderExpansion}) and obtain
\begin{widetext}
\begin{eqnarray}
\hspace{-0.8 cm} -\frac{1}{2T}\left\langle H_{\Delta_v}^2\right\rangle_>^c
&=&-\frac{1}{(2T)^3}\sum_{\alpha_1,\beta_1,\alpha_2,\beta_2}\int d^{d-2}xdz\int d^{d-2}x'dz'\left\langle R_v\left[u_0^{\alpha_1}(x,z)-u_0^{\beta_1}(x,z)\right]
 R_v\left[u_0^{\alpha_2}(x',z')-u_0^{\beta_2}(x',z')\right] \right\rangle_>^c \nonumber\\\
&=&-\frac{1}{8T^3}\sum_{\alpha_1,\beta_1,\alpha_2,\beta_2}\int_{x,z}\int_{x',z'}
R_v''\left[u_{0<}^{\alpha_1}(x,z)-u_{0<}^{\beta_1}(x,z)\right]
R_v''\left[u_{0<}^{\alpha_2}(x',z')-u_{0<}^{\beta_2}(x',z')\right] \nonumber\\
&&\times\frac{1}{4}\left\langle\left[u_{0>}^{\alpha_1}(x,z)-u_{0>}^{\beta_1}(x,z)\right]^2
\left[u_{0>}^{\alpha_2}(x',z')-u_{0>}^{\beta_2}(x',z')\right]^2 \right\rangle_>^c,
\end{eqnarray}
\end{widetext}
in which the average is given by
\begin{eqnarray}
\hspace{-0.2 cm}&&\hspace{-0.6 cm}\frac{1}{4}\left\langle\left[u_{0>}^{\alpha_1}(x,z)-u_{0>}^{\beta_1}(x,z)\right]^2
\left[u_{0>}^{\alpha_2}(x',z')-u_{0>}^{\beta_2}(x',z')\right]^2\right\rangle_>^c\nonumber\\
\hspace{-0.2 cm}&=&\left[C_T^>(\delta x,\delta z)\right]^2(\delta_{\alpha_1\alpha_2}+\delta_{\alpha_1\alpha_2}\delta_{\beta_1\beta_2}\nonumber\\
&& -\delta_{\alpha_1\beta_2}\delta_{\beta_1\beta_2}
 -\delta_{\alpha_2\beta_1}\delta_{\beta_1\beta_2}),
\end{eqnarray}
where $\delta x=x-x'$ and $\delta z=z-z'$, and the three replica correction contributed by the $\delta_{\alpha_1\alpha_2}$ term is irrelevant
relative to the two-replica terms and is, thus, neglected. The second-order contribution from
the random positional pinning then becomes
\begin{widetext}
\begin{eqnarray}
\hspace{-0.6 cm} -\frac{1}{2T}\left\langle H_{\Delta_v}^2\right\rangle_>^c
&=&-\frac{1}{8T^3}\sum_{\alpha,\beta}\int_{x,z}\int_{x',z'}\left[C_T^>(\delta x,\delta z)\right]^2
\bigg\{R_v''\left[u_{0<}^{\alpha}(x,z)-u_{0<}^{\beta}(x,z)\right]
R_v''\left[u_{0<}^{\alpha}(x',z')-u_{0<}^{\beta}(x',z')\right]\nonumber\\
&&-2R_v''\left[u_{0<}^{\alpha}(x,z)-u_{0<}^{\beta}(x,z)\right]
 R_v''(0)\bigg\} \nonumber\\
&\approx&-\frac{1}{8T^3}\sum_{\alpha,\beta}\int_{x,z}\int_{\delta x,\delta z}\left[C_T^>(\delta x,\delta z)\right]^2
\bigg\{R_v''\left[u_{0<}^{\alpha}(x,z)-u_{0<}^{\beta}(x,z)\right]
R_v''\left[u_{0<}^{\alpha}(x,z)-u_{0<}^{\beta}(x,z)\right]\nonumber\\
&&+\frac{1}{2} R_v''\left[u_{0<}^{\alpha}(x,z)-u_{0<}^{\beta}(x,z)\right]
R_v^{''''}\left[u_{0<}^{\alpha}(x,z)-u_{0<}^{\beta}(x,z)\right]
 \left|\partial_x\left[u_{0<}^{\alpha}(x,z)-u_{0<}^{\beta}(x,z)\right]\delta_x\right|^2\nonumber\\
&&-2 R_v''\left[u_{0<}^{\alpha}(x,z)-u_{0<}^{\beta}(x,z)\right]
R_v''(0)\bigg\},
\label{Delta_vDelta_vcontribution}
\end{eqnarray}
\end{widetext}
with
\begin{eqnarray}
A\delta\ell &=&\frac{2}{4T^2}\int_{\delta x,\delta z}\left[C_T^>(\delta x,\delta z)\right]^2 \nonumber\\
&=&\frac{1}{2T^2}\int_{\Lambda e^{-\delta\ell}}^{\Lambda}\frac{d^{d-2}q_x}{(2\pi)^{d-2}}\int_{-\infty}^{
\infty}\frac{dq_z}{2\pi}\frac{T^2}{\Gamma_{\qv}^2},
\label{A_1}
\end{eqnarray}
and
\begin{eqnarray}
A_2\delta\ell&=&\frac{1}{4T^2}\int_{\delta x,\delta z}(\delta_x)^2\left[C_T^>(\delta x,\delta z)\right]^2 \nonumber\\
&\approx&\frac{1}{4T^2}\int_{\Lambda e^{-\delta\ell}}^{\Lambda}\frac{d^{d-2}q_x}{(2\pi)^{d-2}}\int_{-\infty}^{
\infty}\frac{dq_z}{2\pi}\frac{T^2q_x^{-2}}{\Gamma_{\qv}^2},\nonumber\\
\end{eqnarray}
as the main part of the contribution. We then have the second-order contribution from the random positional pinning as
\begin{equation}
\delta R_v^{(2a)}(u)=A\left[\frac{1}{2}R_v''(u)R_v''(u)
 -R_v''(u)R_v''(0)\right]\delta\ell,
 \end{equation}
 and
 \begin{equation}
\hspace{-2 cm}\delta\Delta_f^{(2a)}(u)=-A_2 R_v''(u)R_v^{(4)}(u)\delta \ell.
\end{equation}
For $d=3$, we have
\begin{eqnarray}
A&=&\frac{\pi-2}{8\pi^2B^2\lambda^3\Lambda^3},\\
A_2&=&\frac{\pi-2}{16\pi^2B^2\lambda^3\Lambda^5}=A/(2\Lambda^2).
\end{eqnarray}

The second order contribution of the random orientational pinning $\Delta_f$ in (\ref{HorderExpansion}) is
\begin{widetext}
\begin{eqnarray}
 -\frac{1}{2T}\left\langle H_{\Delta_f}^2\right\rangle_>^c
 &=& -\frac{1}{32T^3}\sum_{\alpha_1,\beta_1,\alpha_2,\beta_2} \int_{x,z}\int_{x',z'} \bigg\langle\Delta_f\left[u_0^{\alpha_1}(x,z)-u_0^{\beta_1}(x,z)\right]\left|\partial_x\left[u_0^{\alpha_1}(x,z)-u_0^{\beta_1}(x,z)\right]\right|^2\nonumber\\
&&\times\Delta_f\left[u_0^{\alpha_2}(x',z')-u_0^{\beta_2}(x',z')\right]\left|\partial_x\left[u_0^{\alpha_2}(x',z')-u_0^{\beta_2}(x',z')\right]\right|^2\bigg\rangle_>^c,
\end{eqnarray}
\end{widetext}
which after separating the high and low wavevector components becomes
\begin{widetext}
\begin{eqnarray}
&&\hspace{-0.6 cm}\left\langle\Delta_f(u_0^{\alpha_1\beta_1})\left|\partial_x u_0^{\alpha_1\beta_1}\right|^2
 \Delta_f\left(u_0^{\alpha_2\beta_2}\right)\left|\partial_{x'} u_0^{\alpha_2\beta_2}\right|^2\right\rangle_>^c \nonumber\\
&=&\Delta_f\left(u_{0<}^{\alpha_1\beta_1}\right)\Delta_f\left(u_{0<}^{\alpha_2\beta_2}\right)
 \left\langle\left(\partial_x u_{0>}^{\alpha_1\beta_1}\right)^2\left(\partial_{x'} u_{0>}^{\alpha_2\beta_2}\right)^2\right\rangle_>^c \nonumber\\
&& +2\Delta_f\left(u_{0<}^{\alpha_1\beta_1}\right)\frac{\Delta_f''\left(u_{0<}^{\alpha_2\beta_2}\right)}{2}
 \left(\partial_{x'} u_{0>}^{\alpha_2\beta_2}\right)^2
\left\langle\left(\partial_x u_{0>}^{\alpha_1\beta_1}\right)^2
  \left( u_{0>}^{\alpha_2\beta_2}\right)^2\right\rangle_>^c \nonumber\\
&&+4\Delta_f' \left(u_{0<}^{\alpha_1\beta_1}\right)
  \Delta_f' \left(u_{0<}^{\alpha_2\beta_2}\right)
 \partial_x u_{0<}^{\alpha_1\beta_1}\partial_{x'} u_{0<}^{\alpha_2\beta_2}
 \left\langle u_{0>}^{\alpha_1\beta_1} u_{0>}^{\alpha_2\beta_2} \partial_x
 u_{0>}^{\alpha_1\beta_1}\partial_{x'} u_{0>}^{\alpha_2\beta_2} \right\rangle_>^c+\cdots,
\label{Delta_fDelta_f}
\end{eqnarray}
\end{widetext}
where for simplicity of notation we defined $[u_0^{\alpha_1}(x,z)-u_0^{\beta_1}(x,z)]=u^{\alpha_1\beta_1}_0$.
The average in the first term of (\ref{Delta_fDelta_f}) is given by (neglecting the three replica contribution)

\begin{eqnarray}
&&\hspace{-0.6 cm}\left\langle\left(\partial_x u_{0>}^{\alpha_1\beta_1}\right)^2\left(\partial_{x'} u_{0>}^{\alpha_2\beta_2}\right)^2\right\rangle_>^c\nonumber\\
&=&4\left[C_{T,\partial}^>\left(\delta x,\delta z\right)\right]^2\left(\delta_{\alpha_1\alpha_2}\delta_{\beta_1\beta_2}-2\delta_{\alpha_1\beta_2}\delta_{\beta_1\beta_2}\right),\nonumber\\
\end{eqnarray}
in which the correlation of the $x$ derivative of the layer fluctuation, $C_{T,\partial}^>$, is
\begin{equation}
C_{T,\partial}^> \left(\delta x,\delta z\right)=\left\langle \partial_xu_{0>}(x,z)\partial_x u\left(x+\delta x,z+\delta z\right)\right\rangle_>.
\end{equation}
Similarly to (\ref{Delta_vDelta_vcontribution}), this term provides the second-order contribution as
\begin{eqnarray}
\delta R_v^{(2b)}(u)&=&A_3\left[\frac{1}{2}\Delta_f(u)\Delta_f(u)-\Delta_f(u)\Delta_f(0)\right]\delta\ell,\nonumber\\
\\
\delta\Delta_f^{(2b)}(u)&=&-A_4\Delta_f(u)\Delta_f''(u)\delta\ell,
\end{eqnarray}
where
\begin{equation}
A_3\delta\ell=\frac{1}{2T^2}\int_{\delta x,\delta z}
 \left[C_{T,\partial}^>(\delta x,\delta z)\right]^2,
\end{equation}
and
\begin{equation}
A_4\delta\ell=\frac{1}{4T^2}\int_{\delta x,\delta z}(\delta x)^2
 \left[C_{T,\partial}^>(\delta x,\delta z)\right]^2.
\end{equation}

The average in the second term of (\ref{Delta_fDelta_f}) is
\begin{eqnarray}
&&\hspace{-0.6 cm}\left\langle\left(\partial_x u_{0>}^{\alpha_1\beta_1}\right)^2
 \left( u_{0>}^{\alpha_2\beta_2}\right)^2\right\rangle_>^c\nonumber\\
&=&4\left[C_{T,\partial/2}^>\left(\delta x,\delta z\right)\right]^2(\delta_{\alpha_1\alpha_2}\delta_{\beta_1\beta_2}-\delta_{\alpha_1\beta_2}\delta_{\beta_1\beta_2}\nonumber\\
&& -\delta_{\alpha_2\beta_1}\delta_{\beta_1\beta_2}),
\end{eqnarray}
where the correlation $C_{T,\partial/2}^>$ is defined through
\begin{equation}
C_{T,\partial/2}^>(\delta x,\delta z)=\left\langle u_{0>}(x,z)\partial_x u(x+\delta x,z+\delta z)\right\rangle_>.
\end{equation}
It is easy to see that this term contributes to the random orientational pinning nonlinearity
\begin{eqnarray}
\delta\Delta_f^{(2c)}
&=&-A_5\Big[\Delta_f(u)\Delta_f''(u)-\Delta_f(u)\Delta_f''(0)\nonumber\\
&&-\Delta_f(0)\Delta_f''(u)\Big],
\end{eqnarray}
with
\begin{equation}
A_5=\frac{1}{2T^2}\int_{\delta x,\delta z}
 \left[C_{T,\partial/2}^>(\delta x,\delta z)\right]^2.
\label{A_5}
\end{equation}

The average in the third term of (\ref{Delta_fDelta_f}) is
\begin{eqnarray}
&&\hspace{-0.6 cm}\left\langle u_{0>}^{\alpha_1\beta_1} u_{0>}^{\alpha_2\beta_2} \partial_x
 u_{0>}^{\alpha_1\beta_1}\partial_{x'} u_{0>}^{\alpha_2\beta_2}\right\rangle_>^c\nonumber\\
&=& 4\left\{C_{T}^>(\delta x,\delta z)C_{T,\partial}^>(\delta x,\delta z)+\left[C_{T,\partial/2}^>
  (\delta x,\delta z)\right]^2\right\}\nonumber\\
 &&\times (\delta_{\alpha_1\alpha_2}\delta_{\beta_1\beta_2}-2\delta_{\alpha_1\beta_2}\delta_{\beta_1\beta_2}),
\end{eqnarray}
which contributes to the random orientational pinning nonlinearity through
\begin{equation}
\delta\Delta_f^{(2d)}(u)=-A_6\left[\frac{1}{2}\Delta_f'(u)\Delta_f'(u)-\Delta_f'(u)\Delta_f'(0)\right],
\end{equation}
with
\begin{equation}
A_6=\frac{4}{T^2}\int_{\delta x,\delta z}\left\{C_{T}^>(\delta x,\delta z)C_{T,\partial}^>(\delta x,\delta z)
+\left[C_{T,\partial/2}^>(\delta x,\delta z)\right]^2\right\}.
\end{equation}

The contribution to the Hamiltonian from the cross term, the third term in (\ref{Delta_fDelta_f}), is
\begin{widetext}
\begin{eqnarray}
-\frac{1}{T}\left\langle H_{\Delta_f}H_{\Delta_v}\right\rangle_>^c
&=&\frac{1}{8T^3}\sum_{\alpha_1,\beta_1,\alpha_2,\beta_2}
\int_{x,z}\int_{x',z'} \bigg\langle \Delta_f\left[u_0^{\alpha_1}(x,z)-u_0^{\beta_1}(x,z)\right]
 \left|\partial_x\left[u_0^{\alpha_1}(x,z)-u_0^{\beta_1}(x,z)\right]\right|^2\nonumber\\
&&\times R_v\left[u_0^{\alpha_2}(x',z')-u_0^{\beta_2}(x',z')\right]\bigg\rangle_>^c.
\end{eqnarray}
\end{widetext}
Similarly, we can separate the long- and short-scale field and keep only the leading-order contribution, thus
\begin{widetext}
\begin{eqnarray}
&&\hspace{-0.6 cm}\left\langle\Delta_f\left[u_0^{\alpha_1}(x,z)-u_0^{\beta_1}(x,z)\right]
 \left|\partial_x\left[u_0^{\alpha_1}(x,z)-u_0^{\beta_1}(x,z)\right]\right|^2
 R_v\left[u_0^{\alpha_2}(x',z')-u_0^{\beta_2}(x',z')\right]\right\rangle_>^c\nonumber\\
&=&\frac{1}{4}\Delta_f''\left(u_{0<}^{\alpha_1\beta_1}\right)
 R_v''\left(u_{0<}^{\alpha_2\beta_2}\right)\left(\partial_xu_{0<}^{\alpha_1\beta_1}\right)^2
 \left\langle \left(u_{0>}^{\alpha_1\beta_1}\right)^2
 \left(u_{0>}^{\alpha_2\beta_2}\right)^2\right\rangle_>^c \nonumber\\
&& +\frac{1}{2}\Delta_f\left(u_{0<}^{\alpha_1\beta_1}\right) R_v''\left(u_{0<}^{\alpha_2\beta_2}\right)
\left\langle \left(\partial_x u_{0>}^{\alpha_1\beta_1}\right)^2\left(u_{0>}^{\alpha_2\beta_2}\right)^2\right\rangle_>^c+\cdots.\nonumber\\
 \label{Delta_fDelta_v}
\end{eqnarray}
\end{widetext}
The average in the first term of (\ref{Delta_fDelta_v}) is easy to calculate as
\begin{eqnarray}
&&\hspace{-0.6 cm}\left\langle\left(u_{0>}^{\alpha_1\beta_1}\right)^2\left( u_{0>}^{\alpha_2\beta_2}\right)^2\right\rangle_>^c\nonumber\\
&=&4\left[C_{T}^>\left(\delta x,\delta z\right)\right]^2
\big(\delta_{\alpha_1\alpha_2}\delta_{\beta_1\beta_2}-\delta_{\alpha_1\beta_2}\delta_{\beta_1\beta_2}\nonumber\\
&& -\delta_{\alpha_2\beta_1}\delta_{\beta_1\beta_2}\big),
\end{eqnarray}
and it contributes
\begin{eqnarray}
\delta\Delta_f^{(2e)}(u)&=&A\Big[\Delta_f''(u) R_v''(u)-\Delta_f''(0)R_v''(u)\nonumber\\
&& -\Delta_f''(u) R_v''(0)\Big]\delta \ell,
\end{eqnarray}
with constant $A$ given in (\ref{A_1}).

The average in the second term of (\ref{Delta_fDelta_v}) is
\begin{eqnarray}
&&\hspace{-0.6 cm}\left\langle \left(\partial_x u_{0>}^{\alpha_1\beta_1}\right)^2\left(u_{0>}^{\alpha_2\beta_2}\right)^2\right\rangle_>^c\nonumber\\
&=&4\left[C_{T,\partial/2}^>(\delta x,\delta z)\right]^2
\big(\delta_{\alpha_1\alpha_2}\delta_{\beta_1\beta_2}-\delta_{\alpha_1\beta_2}\delta_{\beta_1\beta_2}\nonumber\\
&& -\delta_{\alpha_2\beta_1}\delta_{\beta_1\beta_2}\big),
\end{eqnarray}
and it contributes
\begin{eqnarray}
\delta R_v^{(2f)}&=&-A_5\Big[\Delta_f(u)R_v''(u)-\Delta_f(0)R_v''(u)\nonumber\\
 && -\Delta_f(u)R_v''(0)\Big]\delta\ell,
\end{eqnarray}
with $A_5$ given in (\ref{A_5}).

These second- and lower-order contributions sum up
into flow equations of the non linearities given as Eqs.~(\ref{Delta_fFullFlow}) and (\ref{Delta_vFullflow})
in Sec.~\ref{sec:FRG_calculation_smectic}.

\section{Full behavior of long scale correlation function}
\label{app:full_long_scale_correlation}
The RG and matching analysis in Sec.~\ref{sec:smectic_FRG} predicts that the long-scale smectic
phonon correlation is dominated by random orientational pinning, with strength $\Delta_f(T)=\Delta_f\delta_f(T)$.
The correlation is given by
\begin{eqnarray}
C(x,z)
&=&2\int\frac{dq_z}{2\pi}\frac{d q_x}{2\pi}
\frac{\Delta_f(T)q_x^2}{2B^2\lambda}\frac{\sqrt{\lambda^2q_x^4+q_z^2}-\lambda q_x^2}{q_z^2(\lambda^2q_x^4+q_z^2)}\nonumber\\
&&\times\left(1-e^{iq_x x+iq_zz}\right),
\label{full_long_scale_correlation_integral}
\end{eqnarray}
in which the integrations can be safely extended to infinity with our focus on long scale behavior.
Given $A$ as a positive number, it is easy to observe a relation whereby
\begin{equation}
C(\sqrt{A}x,Az)=\sqrt{A}C(x,z),
\end{equation}
which is consistent with the results shown in Eq.~(\ref{large_scale_correlation}), and, on the $x$ or $z$ axis,
\begin{equation}
C(x,0)\propto x, \mbox{ and, } C(0,z)\propto \sqrt{z}.
\end{equation}
The correlation thus can be characterized by
\begin{equation}
C(x,z)\propto xf\left(\frac{\lambda z}{x^2}\right),
\end{equation}
in which, at different limits of the ratio $t=\lambda z/x^2$,
\begin{equation}
\left\{\begin{array}{ll}
f(t)=const.,&\mbox{ for } t\rightarrow 0,\\
f(t)\propto \sqrt{t},&\mbox{ for } t\rightarrow \infty.
\end{array}\right.
\end{equation}

Redefining variables
\begin{equation}
(u,v)=(x^2q_x^2,zq_z),
\end{equation}
so $(q_x,q_z)=(\sqrt{u}/x,v/z)$, Eq.~(\ref{full_long_scale_correlation_integral}) then becomes
\begin{eqnarray}
C(x,z)
&=&\frac{\Delta_f(T)}{\pi^2B^2\lambda}\int_0^{\infty}dq_z\int_0^{\infty}d q_x
q_x^2\frac{\sqrt{\lambda^2q_x^4+q_z^2}-\lambda q_x^2}{q_z^2(\lambda^2q_x^4+q_z^2)}\nonumber\\
&&\times \left[1-\cos{(q_x x)}\cos{(q_zz)}\right] \nonumber\\
&=&\frac{\Delta_f(T)x}{4\pi^2B^2\lambda^3} 2 t^2\int_0^{\infty}du\int_0^{\infty}d v
\sqrt{u}\frac{\sqrt{t^2u^2+v^2}-tu}
{v^2(t^2u^2+v^2)}\nonumber\\
&&\times\left[1-\cos{(\sqrt{u})}\cos{(v)}\right] \nonumber\\
&=&\frac{\Delta_f(T)}{4\pi^2B^2\lambda^3}xf(t),
\end{eqnarray}
with
\begin{eqnarray}
f(t)&=&2t^2\int_0^{\infty}du\int_0^{\infty}d v\sqrt{u}\frac{\sqrt{t^2u^2+v^2}-tu}
{v^2(t^2u^2+v^2)}\nonumber\\
&&\times\left[1-\cos{(\sqrt{u})}\cos{(v)}\right].
\end{eqnarray}
Writing $1-\cos{(\sqrt{u})}\cos{(v)}$ as $[1-\cos{(\sqrt{u})}]+\cos{(\sqrt{u})}[1-\cos{(v)}]$, it is easy
to obtain the full analytic expression and limits of $f(t)$ as given in Eq.~(\ref{f_t}) and plotted in
Fig.~\ref{fig:f_t}.

\bibliography{refs}		

\begin{thebibliography}{52}
\bibitem [{com({\natexlab{a}})}]{commentRM}%
  \BibitemOpen
  \href@noop {} {} {\natexlab{a}} \bibinfo {note} {Because here we are
  focusing on thermotropic smectics whose period $a$ is comparable to quenched
  disorder correlation length $\xi_0$ (both on the order of molecular scale
  $\sim 10 nm$), we neglected the (otherwise) intermediate random manifold
  regime
  \cite{DSFisherFRG,GiamarchiLedoussalFluxLattice_PRB,GiamarchiLedoussalFluxLattice_PRL,LedoussalWiese,BalentsFisher,GiamarchiLedoussalFluxLattice_PRB}.
  In contrast in lyotropic (e.g. surfactant systems) we expect $a\ll \xi_0$
  thereby allowing random manifold. We leave analysis of this more involved
  case to future study.}\BibitemShut {Stop}%
\bibitem [{com({\natexlab{b}})}]{commentXray}%
  \BibitemOpen
  \href@noop {} {} {\natexlab{b}} \bibinfo {note} {This is a unlikely to be
  a physically realizable regime because for weak pinning the penetration
  length into the bulk is expected to be much longer than the X-ray wavelength.
  This case requires numerical analysis that we leave for the
  future.}\BibitemShut {Stop}%
\bibitem [{com({\natexlab{c}})}]{comment_positional_nonlinear}%
  \BibitemOpen
  \href@noop {} {} {\natexlab{c}} \bibinfo {note} {Similarly, we find for
  the positional pinning $\xi_{NL,x}\sim
  (K^2/\Delta_v)^{1/(6-d)}$.}\BibitemShut {Stop}%
\end{thebibliography}

\end{document}